\documentclass[usenatbib]{mn2e}
\usepackage{graphicx}
\usepackage{rotating}
\usepackage[section] {placeins}
\usepackage{amsmath, amssymb}
\usepackage{color}
\usepackage{multirow}
\usepackage{makeidx}
\usepackage{xspace}
\usepackage{float}

\newcommand{\gtsim}{\mbox{{\raisebox{-0.4ex}{~$\stackrel{>}{{\scriptstyle\sim}}$~}}}}
\newcommand{\ltsim}{\mbox{{\raisebox{-0.4ex}{~$\stackrel{<}{{\scriptstyle\sim}}$~}}}}

\newcommand{\m}[1]{{\bf{#1}}}
\newcommand{\M}[1]{{\bf{#1}}}

\newcommand{\kms}[0]{\ensuremath{km~s$^{-1}$}\xspace}

\def\Msun{$\textrm{M}_{\odot}$}

\def\farcs{\hbox{$.\!\!^{\prime\prime}$}}
\def\fp{\hbox{$.\!\!^{\reset@font\scriptscriptstyle\r@mn{p}}$}}
\def\arcmin{\hbox{$^\prime$}\xspace}
\def\arcsec{\hbox{$^{\prime\prime}$}\xspace}

\def\d{{\rm d}}

\long\def\crap#1{}

\def\GH{Gauss-Hermite\xspace}

\def\kms{km~s$^{-1}$\xspace}

\def\log{\textrm{log}}

\def\p{\ensuremath{^\prime}\xspace}

\def\n#1{\textrm{\tiny#1}}
\def\<{\langle }
\def\>{\rangle }
\def\bi#1{\emph{\bf#1}\xspace}
\def\nHST{43\xspace}
\def\Gyr{Gyr\xspace}
\def\Mag{mag\xspace}

\def\apmag{\ensuremath{\emph{m}}\xspace}
\def\absmag{\ensuremath{\mathcal{M}}\xspace}
\def\sex{{\sc SExtractor}\xspace}
\def\micron{\ensuremath{\mu}\textrm{m}\xspace}
\def\erf{\ensuremath{\textrm{erf}}}

\def\avDeltaBetawBCG{0.23 \Mag}
\def\avDeltaBetawoBCG{0.22 \Mag}
\def\DeltaBeta{$0.22\pm{0.11}$ \Mag}
\def\nsigdisagree{$2\sigma$\xspace}
\def\ZComa{$z=0.024$\xspace}
\def\DComa{$103\pm10$Mpc\xspace}
\def\A1689{Abell~1689\xspace}
\def\LXA1689{20.74$\times 10^{44}$ erg s$^{-1}$\xspace}
\def\LXComa{7.21$\times 10^{44}$ erg s$^{-1}$\xspace}
\def\dage{2.26 \Gyr} \def\ZA1689{z=0.183\xspace}

\def\CMRscatterA1689{$0.054\pm0.004$ \Mag}
\def\CMRage{$5.5 < $ age (\Gyr) $ < 9.5$\xspace}
\def\CMRagelowlim{$>5.5$ \Gyr \xspace}
\def\CMRformz{$0.55 < z < 1.55$\xspace}
\def\CMRformzlowlim{$z>0.55$\xspace}

\def\ageALLMOD{$> 6.0$ \Gyr \xspace}
\def\ageBC03SUN{$8.6 \pm_{1.8}^{4.9}$ \Gyr\xspace}
\def\ageM05SUN{$11.4\pm_{3.9}^{>3.6}$ \Gyr (models only run to 15 \Gyr)\xspace}
\def\AGEav{10 \Gyr}
\def\deltaageB03M05SUN{2.8 \Gyr}

\def\formzBC03SUN{$1.2\pm_{0.4}^{\infty}$\xspace}
\def\formzM05SUN{$2.9\pm_{0.9}^{\infty}$\xspace}

\def\age{$10.2\pm{3.3}$ \Gyr} 
\def\formz{$z=1.8\pm_{0.9}^{\infty}$\xspace}

\def\LumEvolSUNHST{0.25 \Mag} \def\LumEvolSUNrprime{0.23 \Mag} 
\def\nhst{43\xspace}
\def\nsample{38\xspace}
\def\ngmos{531\xspace}
\def\nspec{77\xspace}
\def\nspeccluster{71\xspace}

\title[Data and 2D scaling relations for galaxies in Abell 1689: a hint of size evolution at z$\sim$0.2]{Data and 2D scaling relations for galaxies in Abell 1689: a hint of size evolution at z$\sim$0.2}
\author[R. C. W. Houghton, Roger L. Davies, E. Dalla Bont\`a, R. Masters]{R. C. W. Houghton$^{1}$\thanks{Email: rcwh@astro.ox.ac.uk}, Roger L. Davies$^{1}$, E. Dalla Bont\`a$^{2,3}$, R. Masters$^{1}$\\
$^{1}$ University of Oxford, Denys Wilkinson Building, Keble Road, Oxford, OX1 3RH \\
$^{2}$ Dipartimento di Fisica e Astronomia, Universit\`a degli Studi di Padova, Vicolo dell'Osservatorio 3, I-35122, Padova, Italy \\
$^{3}$ INAF Osservatorio Astronomico di Padova, Vicolo dell'Osservatorio 5, I-35122, Padova, Italy\\}

\begin{document}

\date{}

\pagerange{\pageref{firstpage}--\pageref{lastpage}} \pubyear{2011}

\maketitle

\label{firstpage}

\begin{abstract}

We present imaging and spectroscopy of \A1689 (\ZA1689) from the GMOS multi-object spectrograph on the Gemini-North telescope and the Advanced Camera for Surveys (ACS) on the Hubble Space Telescope (HST). We measure integrated photometry from the GMOS g\p and r\p images (for \ngmos galaxies) and surface photometry from the HST F625W image (for \nhst galaxies) as well as velocities and velocity dispersions from the GMOS spectra (for \nspeccluster galaxies). We construct the Kormendy, Faber--Jackson and colour-magnitude relations for early-type galaxies in \A1689 using this data and compare them to those of the Coma cluster. We measure the intrinsic scatter of the colour--magnitude relation in \A1689 to be $\sigma_\n{CMR}=$\CMRscatterA1689 which places degenerate constraints on the ratio of the assembly timescale to the time available ($\beta$) and the age of the population. Making the assumption that galaxies in \A1689 will evolve into those of Coma over an interval of \dage breaks this degeneracy and limits $\beta$ to be $>0.6$ and the age of the red sequence to be \CMRagelowlim (formed at \CMRformzlowlim). Without corrections for size evolution but accounting for magnitude cuts and selection effects, the Kormendy and Faber--Jackson relations are inconsistent and disagree at the \nsigdisagree level regarding the amount of luminosity evolution in the last \dage. However, after correcting for size evolution the Kormendy and Faber--Jackson relations show similar changes in luminosity (\DeltaBeta) that are consistent with the passive evolution of the stellar populations from a single burst of star formation \age ago (\formz). Thus the changes in the Kormendy, Faber--Jackson and colour-magnitude relations of \A1689 relative to Coma all agree and suggest old galaxy populations with little or no synchronisation in the star formation histories. Furthermore, the weak evidence for size evolution in the cluster environment in the last \dage places interesting constraints on the possible mechanisms at work, favouring harassment or secular processes over merger scenarios.

\end{abstract}


\section{Introduction}
\label{sec:intro}

Scaling relations for early type galaxies (ETGs) are one of the simplest tools available to study the formation and evolution of the present day population. They are particularly useful for observing changes in the stellar populations as a function of galaxy size, velocity dispersion or mass and allow us to test the conventional view that ETGs are composed of old populations which formed in a single burst of star formation some 12 -- 16 \Gyr \citep{Baade58,TinsleyGunn76,Bruzual83,Hamilton85}.
The Faber-Jackson relation \citep[FJR,][]{FaberJackson76} relates the central stellar velocity dispersion ($\sigma$) to the absolute magnitude or luminosity of a galaxy,
\begin{equation}
\label{eq:intro:fjr}
M = \alpha_\n{FJR} \log \sigma + \beta_\n{FJR}.
\end{equation}
Likewise, the Kormendy relation \citep[KR,][]{Kormendy77,HamabeKormendy87} relates the effective radius ($R_{e}$) to the surface brightness of the galaxy ($\mu_{e}$),
\begin{equation}
\label{eq:intro:kr}
\<\mu\>_{e} = \alpha_\n{KR} \log R_{e} + \beta_\n{KR}.
\end{equation}
The original Kormendy relation was defined between the surface brightness \emph{at} $R_{e}$ and $R_{e}$ itself. It is now conventional to use the \emph{average} surface brightness \emph{within} $R_{e}$ against $R_{e}$. For a de Vaucouleurs light profile \citep{deVaucouleurs53}, these are related by a constant factor, but for a S\'ersic profile \citep{Sersic63} this is not the case and the conversion depends on the S\'ersic index $n$. 

Both the FJR and the KR are projections of the Fundamental plane of ETGs \citep[FP:][]{Dressler87,DjorgovskiDavis87}, linking $R_{e}$, $\sigma$ and $\<\mu_{e}\>$ (sometimes expressed in $\rm{L}_{\sun} \rm{pc}^{-2}$ as $\<I\>_{e}$).
\begin{equation}
\log R_{e} = \alpha_\n{FP} \log \sigma + \beta_\n{FP} \<\mu\>_{e} + \gamma_\n{FP}
\end{equation}
The FP is tilted compared to the Virial Theorem prediction which is thought to be caused by a variation in mass-to-light (M/L) with mass \citep[][]{Faber87,RenziniCiotti93}.

The colour-magnitude relation \citep[CMR,][]{Sandage72} is a tight relation between the (red) colour of ETGs and their magnitude (or luminosity). Defining generic variables $\apmag_{b}$ and $\apmag_{r}$ for the magnitudes in the blue and red filters respectively,  
\begin{equation}
\label{eq:intro:cmr}
m_{b} - m_{r} = \alpha_\n{CMR} \absmag_{r} + \beta_\n{CMR}.
\end{equation}
where $\absmag$ refers to the absolute magnitude.

\subsection{Evolution in slope and intercept of scaling relations}
\label{sec:intro:slopegrad}
Studying scaling relations at earlier cosmic times allows us to study the evolution of the stellar populations. Under the assumption that  all galaxies evolve uniformly in brightness from the changes in stellar populations, one should observe a change in the zeropoint, $\beta$. For the FJR, the non-local relation becomes
\begin{equation}
\label{eq:intro:fjrz}
\absmag_{z} = \alpha_\n{FJR} \log \sigma_{z} + \beta_\n{FJR} + \Delta\beta_\n{FJR}.
\end{equation}
where $\absmag_{z}$ and $\sigma_{z}$ are the non-local measurements and $\Delta\beta_\n{FJR}$ represents the difference in luminosity from the local sample. Similarly for the KR,
\begin{equation}
\label{eq:intro:krz}
\<\mu\>_{e,z} = \alpha_\n{KR} \log R_{e,z} + \beta_\n{KR} + \Delta\beta_\n{KR}.
\end{equation} 
Studies of the KR at intermediate redshift show changes in $\Delta\beta_\n{KR}$ consistent with ETGs becoming more luminous with lookback time and the slope generally appears to be unchanged \citep{Barrientos96,Pahre96,Schade96,Schade97,Ziegler99,LaBarbera03,Fritz05,Holden05b}. Similar conclusions are drawn from the evolution of the FJR \citep{Bender96,Ziegler01,Ziegler05,Fritz05} and the FP \citep{vanDokkumFranx96,vanDokkum98,Treu01,Kelson00,vanDokkum01,vanDokkumEllis03,Fritz05,Holden05a,vanderWel05,Moran05,Jorgensen06,Barr06,vDvdM07,vanderWel06,vanderWel07,Fritz09,Holden10,Saglia10}. 

An important use of scaling relations was demonstrated by \citet{KA97} who broke the age--metallicity degeneracy \citep{Worthey94} by showing that the CMR slope is the same in clusters at $z\sim0.2$--$0.4$ and must originate from a variation in metallicity with luminosity, rather than age (in that respect, the cluster ETGs were found to be coeval and old). 

Recently there has been debate over the observation of downsizing \citep{Cowie96,Thomas05} and the role of selection effects: the study of \citet{Holden10} at $z\sim0.8$ found no evidence that the tilt of the high-z FP changes from its local value, which implies that the findings of \citet{Jorgensen06}, \citet{Fritz09} and \citet{Saglia10} are biased by selection effects.

We emphasise the importance of updating local scaling relations with the latest technology, techniques and cosmology in relation to observations at higher redshift -- the vast majority of KR, FJR and FP studies compare high redshift observations to the 15 yr old Gunn r-band observations of \citet{Jorgensen96} and observations in different bands are even older \citep[see the review by][]{Donofrio06}. Although potentially a good reference, the SDSS FP parameters \citep{Bernardi03} have not been used due to their disparity with other work. 

\subsection{Scatter in the CMR}
\label{sec:intro:cmrscatter}
\citet[][hereafter BLE92]{BLE92} first used the U-V Colour--Magnitude Diagram (CMD) of the Coma cluster to place constraints on the star formation histories (SFHs) of the galaxies. They estimated the scatter in the red sequence (RS) of the Coma CMD to be $\sim$0.04 \Mag, which can then be used to infer that either: a) all the galaxies in Coma formed at high redshift (z$>$2) but randomly over the preceding interval, or b) the galaxies in Coma formed more recently but were highly coeval, thus demonstrating a remarkable degree of synchronisation. Given that if all the galaxies formed at z$\sim$1, the star formation would have been apparent with moderately deep surveys and had not been observed, the authors concluded that the ETGs in Coma formed at z$>$2 with little synchronisation in the SFHs.

By considering the effects of more complex stellar populations (CSPs), \citet[][hereafter BKT98]{BKT98} confirmed the original conclusion of BLE92. They also investigated the effects of merging on the RS and found that the small scatter of the CMR requires that the cluster galaxies formed in mass \emph{sub-units} not much smaller than half their present day mass.

\citet{Stanford95}, \citet{Ellis97}, \citet{Stanford98} and \citet{Mei09}, have studied many clusters with $z\ltsim1.3$ to address evolution of the CMR scatter. They all found remarkably constant scatter, generally consistent with that of Coma with values $<0.1$ \Mag. This implies that the bulk of the stars in these cluster ETGs are very old and formed at $z>2$ with little synchronisation in the SFHs, although \citeauthor{Stanford98} and \citeauthor{Mei09} highlight the possibility of progenitor bias. \citeauthor{Mei09} also found evidence that the scatter is less for more luminous galaxies, suggesting more massive galaxies are slightly older (by $\sim0.5$ \Gyr). We note that \citet{Stanford98} discuss a gap in the estimates of the CMR scatter between $0.1<z<0.3$ which, together with their slightly higher CMR scatter at $z>0.3$, led them to suggest possible evolution of the CMR scatter during this epoch. 

\subsection{This study}
\label{sec:intro:thisstudy}

The evolution of the KR, FJR, FP and CMR appears to be consistent with passive fading of the stellar population from a single burst of star formation at $z>1$. Similarly, the current evidence from analysis of the CMR is also greatly in favour of the simple passive fading of ETGs. However, simultaneous study of multiple scaling relations is rarely performed and it is unclear if they are all \emph{quantitatively} consistent. Furthermore, it is important to account for magnitude cuts and selection effects given the results of \citet{Holden10}.

In this paper we present data and analysis techniques for the cluster \A1689. We compare the 2D scaling relations (CMR, KR and FJR) for \A1689 to those of the local cluster Coma, accounting for magnitude cuts and selection effects, and specifically testing for consistency: if the evolution of different scaling relations is governed solely by luminosity evolution of the stellar populations, they should all agree. Note that we will present the FP of \A1689 in a future paper. \A1689 is a massive cluster at \ZA1689 \citep{A1689distance} with an X-ray luminosity of \LXA1689 \citep{Ebeling96} making it one of the most X-ray luminous galaxy clusters known \citep[the X-ray luminosity for Coma is \LXComa,][]{Ebeling96}. The KR, FJR, FP and CMR have not been previously studied in regard to the passive fading of ETGs despite it being one of the main targets for HST lensing surveys. We present colours and magnitudes for \ngmos galaxies, as well as spectroscopic data for \nspeccluster galaxies and surface photometry for \nhst galaxies with which we investigate the CMR, KR and FJR.

Throughout this work, we adopt a WMAP7 Cosmology \citep{WMAP7COSMO}; specifically, we use $H_{0}=71$\kms Mpc$^{-1}$, $\Omega_\n{m}=0.27$ and $\Omega_{\Lambda}=0.73$. All quoted uncertainties are standard ($1\sigma$, 68\%) unless otherwise stated. The structure of this paper is as follows. In \S\ref{sec:red} we discuss the data reduction of the GMOS imaging, GMOS spectra and HST imaging as well as the data analysis techniques for the photometry and the kinematics and end with a discussion of how we fit the 2D scaling relations. Then in \S\ref{sec:res} we present our results and discuss the use of stellar population models to interpret them in \S\ref{sec:spmodels}. In \S\ref{sec:disc} we present the discussion and conclude in \S\ref{sec:conc}. We also present Appendices regarding the data reduction and analysis: Appendix A discusses our curve-of-growth (COG) technique and the associated error analysis; Appendix B discusses how we matched the spectral profile of the stellar library to that of the GMOS-N instrument, and Appendix C presents the techniques used to fit and compare the scaling relations of \A1689.

\section{Data Reduction and Analysis}
\label{sec:red}

We now describe the reduction and analysis techniques used to measure integrated photometry (from GMOS-N \emph{g\_G0301} and \emph{r\_G0303} imaging, hereafter referred to as g\p and r\p), surface photometry (from HST/ACS F625W imaging), and kinematics (from GMOS-N MOS).

\subsection{GEMINI/GMOS imaging}
\label{sec:red:gmosim}

\begin{figure*}
   \centering
   \includegraphics[width=0.9\textheight,angle=90]{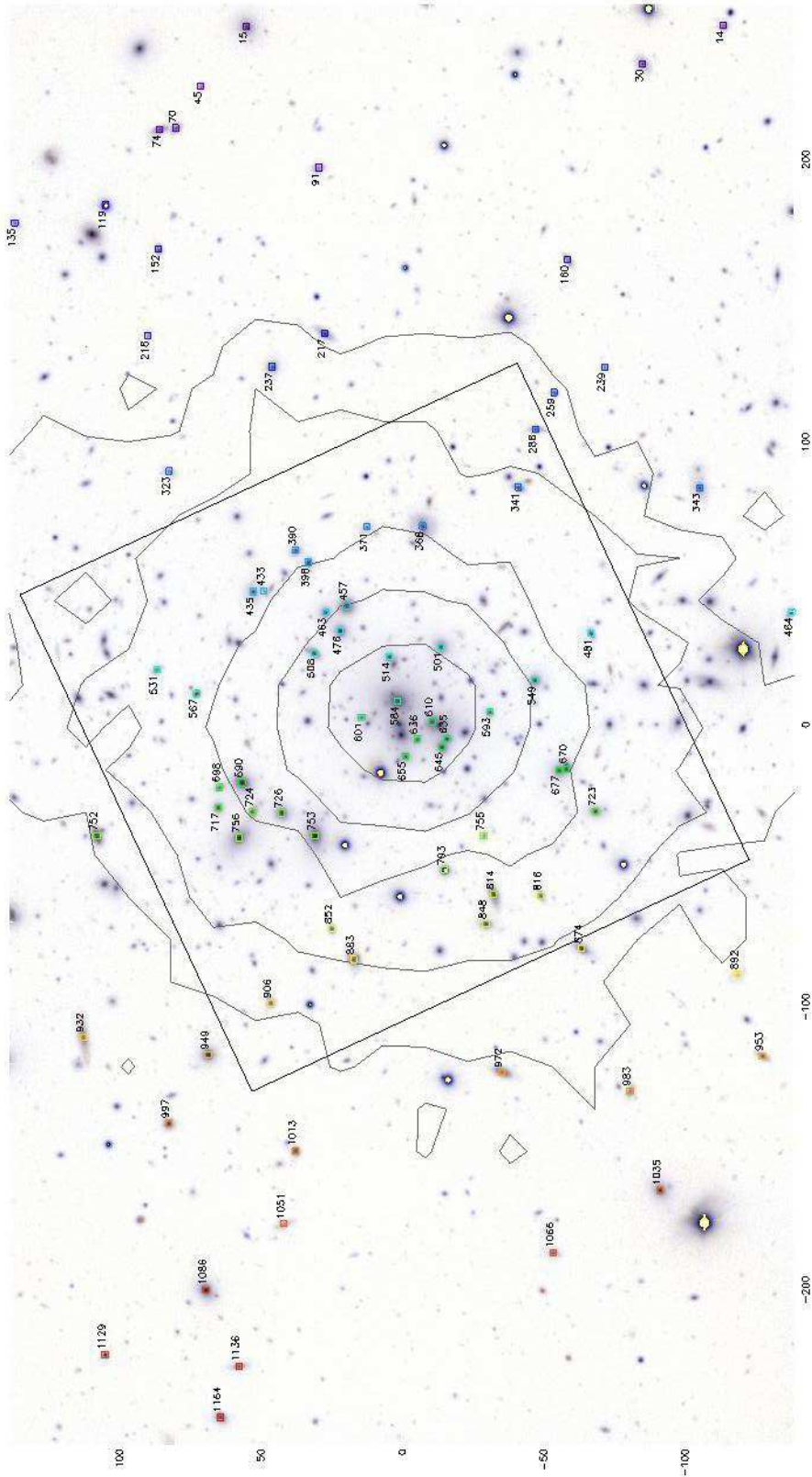}    
   \caption{A negative image of the galaxy cluster Abell 1689 observed using the g\p- and r\p-band filters with the GEMINI/GMOS imager. The g\p-band image has been degraded to match the resolution of the r\p-band image. Overlaid is the HST footprint (solid black square) and the ROSAT X-ray map (grey, solid lines; contours were shifted by \{-0.008,-0.006\} degrees in \{RA,DEC\} to align with the cluster core). Coloured squares mark the galaxies in the spectroscopic sample, apart from \#119, \#793, \#1129 which are guiding/alignment stars. Note that \#160, \#457, \#892, \#932, \#983 and \#1051 are not in the cluster potential according to their recession velocities (see Table \ref{tab:res:kin}). }
   \label{fig:A1689GMOSim}
\end{figure*}

Details of the seeing limited images of Abell 1689 are given in Table \ref{tab:red:imdata}. The g\p-band images have a plate scale of 0\farcs1454 per pixel (2x2 binning) while the r\p-band images have a plate scale of 0\farcs0727 (no binning).
The reduction of the GMOS imaging was based on that of \citet{Jorgensen05} and \citet{Barr05} but with a few differences. We use custom modified versions of the Gemini IRAF data reduction software (v1.14) and for completeness, we outline our method below.

\subsubsection{Preliminary stages: bias subtraction, mosaicing and interpolation}

We first processed the GMOS CCD images by performing the bias subtraction using the overscan region. A flat field image was created using a series of scaled twilight exposures (of differing exposure times) which was then applied to the Abell 1689 images. The individual chips were mosaiced onto a single image using `nearest' interpolation to avoid correlating pixels (the error in nearest interpolation is at most half the pixel size which in this case is less than 10\% of the seeing width, see also \S\ref{sec:red:gmosspec}). All these processing s.ps were performed with {\sc gireduce}, which makes use of {\sc gprepare}. 
\subsubsection{Scattered light removal}
\label{sec:red:scatteredlight}

Before mosaicing the images with {\sc gmosaic} and subtracting the residual scattered light, the background (sky and scattered light) levels on each chip were estimated (using the modal value) and equalised (individual chip levels subtracted, mean level added). However, both the g\p- and r\p-band images show a non-uniform scattered light pattern across the three CCD chips. 

We produced a stacked scattered light frame following the recipe described in \citet{Jorgensen05}, but to prevent removal of the intra-cluster light (ICL) a 4th order polynomial surface was fit to the frame, one chip at a time, to provide a smooth representation of the scattered light while leaving the ICL present. As in \citet{Jorgensen05}, we scaled the scattered light frame (a polynomial surface in our case) by between 0.8 and 1.2 (in 0.05 intervals) and determined the best subtraction by eye for each frame. 

The scattered light corrected frames were then aligned, scaled and combined using a modified version of the {\sc imcoadd} routine  (allowing multiple pointings to be combined into a single mosaic). The default intensity scaling of individual images by {\sc imcoadd} (given by the header keyword {\sc RELINT}) was found to be unreliable; instead, we fitted a Moffat function to a star present in the overlapping pointings of all images, and used the flux within 3 FWHM to scale each of the images to a common standard. 

\subsubsection{Flux calibration}
\label{sec:red:gmosim:fluxcal}

Flux calibration of the g\p-band images followed the prescription in \citet{Jorgensen09} and final zeropoints are listed in Table \ref{tab:red:photstands}. We used standards from \citet{Landolt92Phot} or \citet{Landolt07Spec} to estimate the zero points (ZP) for each night. 

The g\p-data was all taken during photometric conditions and with good seeing ($\sim0\farcs6$ in the combined image); consequently, the images required little relative scaling. The photometric standard was taken minutes after the last exposure, minimising systematic error from atmospheric changes. We scaled all images to the last image of the night, which was nearest in time to the observation of the photometric standard and also had one of the highest relative throughputs of all the images (see Table \ref{tab:red:photstands}). 

The r\p-band data were taken over two consecutive nights, the first of which was not photometric while the second was. From the relative throughput of the images, it is clear that the flux in the photometric images of the second night varies only by a few percent compared to the non-photometric images of the first night, and we scaled all images to the brightest photometric image of the second night (with airmass $<$1.2) and flux calibrated the images using an average of the ZPs derived from all standards. We also scaled the science images to correct for airmass \citep[according to][]{Jorgensen09}. The ZPs in Table \ref{tab:red:photstands} do not include a term for the extinction from the Galaxy. For the coordinates of \A1689, $A_g = 0.106$ and $A_r=0.075$ \citep{extinction_map, extinction_law}: these corrections are applied to all integrated (\S\ref{sec:anal:intphot}) and surface photometry (\S\ref{sec:anal:surphot}) measurements. 

\begin{figure}
   \centering
   \includegraphics[width=0.5\textwidth]{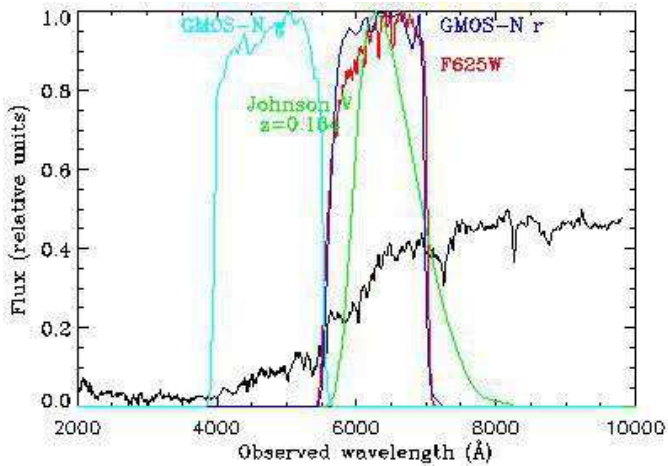} 
   \caption{The GMOS-N filters transmission curves (g\p, r\p) together with the HST/ACS F625W transmission curve are shown together with a galaxy template spectrum \citep{SpecCat} which has been redshifted to \ZA1689. Also shown for comparison is a standard Johnson V-band filter curve \citep[photographic, from]{Bessel90} in the rest frame of the galaxy spectrum.}
   \label{fig:red:imdata:filters}
\end{figure}

\begin{table*}
\begin{tabular}{ccccccccc}
\hline
\textbf{Filename} & \textbf{Airmass}  & \textbf{Moffat} & \textbf{Gaussian} & \textbf{Scaling} & \textbf{Relative}& \textbf{Filter} & \textbf{Obs.} & \textbf{Exp. Time} \\
& \textbf{sec(z)} & \textbf{FWHM} & \textbf{FWHM} & & \textbf{Throughput} & & \textbf{Time (UT)} & \textbf{(s)} \\\hline\hline
N20031224S0065  & 1.327 & 0\farcs56 & 0\farcs619 & 1.059 & 0.971 & g\p & 14:57:29 & 180\\
N20031224S0066  & 1.311 & 0\farcs53 & 0\farcs583 & 1.062 & 0.964 & g\p & 15:01:24 & 180\\
N20031224S0067  & 1.295 & 0\farcs53 & 0\farcs593 & 1.060 & 0.962 & g\p & 15:05:20 & 180\\
N20031224S0068  & 1.281 & 0\farcs52 & 0\farcs585 & 1.059 & 0.960 & g\p & 15:09:14 & 180\\
N20031224S0070  & 1.243 & 0\farcs57 & 0\farcs597 & 1.006 & 1.001 & g\p & 15:20:09 & 180\\
N20031224S0071  & 1.231 & 0\farcs54 & 0\farcs610 & 1.009 & 0.995 & g\p & 15:24:04 & 180\\
N20031224S0072  & 1.219 & 0\farcs52 & 0\farcs600 & 1.000 & 1.000 & g\p & 15:28:00 & 180\\
N20031224S0073  & 1.208 & 0\farcs50 & 0\farcs580 & 1.000 & 0.998 & g\p & 15:31:54 & 180\\\hline\hline
Coadded g\p-band  & 1.219 & 0\farcs56 & 0\farcs607 & 1.000 & 1.000 & g\p & -        & 1440 \\ \hline
N20011223S174	& 1.310 & 1\farcs13 & 1\farcs159 & 1.005 & 1.031 & r\p & 15:01:09 & 300\\ 
N20011223S175	& 1.279 & 1\farcs08 & 1\farcs164 & 1.009 & 1.020 & r\p & 15:09:12 & 300\\
N20011223S176	& 1.252 & 1\farcs07 & 1\farcs147 & 1.007 & 1.015 & r\p & 15:17:14 & 300\\
N20011223S178	& 1.213 & 1\farcs08 & 1\farcs118 & 1.012 & 1.000 & r\p & 15:29:56 & 300\\
N20011223S179	& 1.191 & 1\farcs14 & 1\farcs246 & 1.012 & 0.994 & r\p & 15:38:10 & 300\\
N20011223S180	& 1.172 & 1\farcs02 & 1\farcs066 & 1.009 & 0.992 & r\p & 15:46:14 & 300\\
N20011224S139	& 1.518 & 1\farcs25 & 1\farcs158 & 1.021 & 1.071 & r\p & 14:17:41 & 300\\
N20011224S140	& 1.466 & 1\farcs30 & 1\farcs134 & 1.015 & 1.063 & r\p & 14:25:54 & 300\\
N20011224S149	& 1.167 & 0\farcs85 & 0\farcs864 & 1.000 & 1.000 & r\p & 15:44:19 & 300\\
N20011224S150	& 1.150 & 0\farcs88 & 0\farcs997 & 1.007 & 0.989 & r\p & 15:52:22  & 300\\
N20011224S151	& 1.135 &  - & - & -  &-& r\p &     16:00:28 & 300\\\hline\hline
Coadded r\p-band  & 1.167 & 1\farcs06 & 0\farcs936 & 1.000 & 1.000 & r\p & -        & 3000\\\hline
\end{tabular}
\caption{The GEMINI/GMOS imaging data. The g\p-band data were observed with a plate scale of 0\farcs1454 (2x2 binning) while the r\p-band data were observed with a plate scale of 0\farcs0727 (1x1 binning). The last exposure in the r\p-band was badly affected by twilight and was rejected because the scattered light pattern could not be adequately subtracted off. Images are scaled to the exposure nearest in time to the observation of the photometric standard (see Table \ref{tab:red:photstands})}
\label{tab:red:imdata}
\end{table*}
\begin{table*}
\begin{tabular}{ccccccccccc}
\hline
\textbf{Filename} & \textbf{Object} & \textbf{Airmass}  & \textbf{Exp. Time} & \textbf{Obs. Time} &\textbf{Counts} & \textbf{m$_\textrm{std}$} & \textbf{m$_\textrm{ZP}$} & \textbf{Filter} \\
                  & \textbf{Name}   & \textbf{sec(z)}   & \textbf{(s)}  & \textbf{(HH:MM:SS)} & \textbf{(e$^{-}$/s)} & \textbf{(\Mag)} & \textbf{(\Mag)} & \\\hline\hline
N20031224S0081 & PG1323-086 & 1.241 & 3 & 16:06:09 & 6.17$\times 10^5$ & 13.335$^a$& 27.844& g\p \\ N20031224S0082 & PG1323-086 & 1.239 & 3 & 16:06:58 & 6.38$\times 10^5$& 13.335$^a$,13.283$^c$& 27.881& g\p \\ N20031224S0083 & PG1323-086 & 1.237 & 1 & 16:07:49 & 6.00e5 & 13.335$^a$,13.283$^c$& 27.814 & g\p \\	
N20031224S0084 & PG1323-086 & 1.235 & 1 & 16:08:36 & 6.24e5 & 13.335$^a$,13.283$^c$& 27.856& g\p \\\hline
N20011223S119  & G191B2B    & 1.197 & 5 & 09:42:04 & 24.90$\times 10^5$& 12.044$^b$ & 28.062 & r\p \\ N20011224S153  & PG1323-086  & 1.219 & 1 & 06:09:38 & 4.93$\times 10^5$& 13.663$^a$,13.660$^c$ & 27.919 & r\p \\ N20011224S154  & PG1323-086  & 1.216 & 1 & 06:09:38 & 5.25$\times 10^5$& 13.663$^a$,13.660$^c$ & 27.987 & r\p \\ \end{tabular}
\caption{The GEMINI/GMOS imaging standards used to calibrate the photometry used here. Note that the g\p-band standards were observed straight after the science observations, while the r\p-band standards were observed some hours before hand (see Table \ref{tab:red:imdata}). Zero points (ZP) were calculated according to the prescription of \citet{Jorgensen09}. Magnitudes for the photometric standards ($m_\textrm{std}$) were calculated using the $\textrm{f}(B,B-V)$ transformation in \citet[][]{Smith02} and the data from either \citet[suffix a,][]{Landolt92Phot} or \citet[suffix b,][]{Landolt07Spec}; magnitudes for these standards calculated by \citet{Jorgensen09} are also given (suffix c). The average ZP from the above standards for the g\p-band data is 27.85$\pm$0.03 while the average ZP for the r\p-band is 27.99$\pm$0.07.}
\label{tab:red:photstands}
\end{table*}
\subsection{HST imaging}
\label{sec:red:hstim}
We used {\it HST}/ACS images of Abell 1689 obtained with the Wide
Field Channel (WFC) as part of the ACS Guaranteed Time Observation
program 9289 (P.I. H. Ford). The WFC detector consists of two 4096
$\times$ 2048 SITe CCDs butted together along their long dimension and
separated by a gap corresponding to approximately 50 pixels (2.5
arcsec). The plate scale is 0.050 arcsec pixel$^{-1}$. Each chip uses
two amplifiers to read a single 2048 $\times$ 2048 quadrant.  We
downloaded from the public {\it HST} archive the images employing the
F625W filter, which resembles the SDSS/GMOS $r$\p filter (see
Fig. \ref{fig:red:imdata:filters}). The observations were carried out
on 14 June 2002 and consist of eight exposures (four pointings split
in two to deal with cosmic rays) for a total integration time of 9500
sec. The dithered images were able to cover the gap between the
adjacent chips.

All images were calibrated using the standard reduction pipeline {\sc
  PYRAF/CALACS} maintained by the Space Telescope Science
Institute. Reduction steps include bias subtraction, dark current
subtraction, and flat fielding, as described in detail in the ACS
instrument \citep{may10} and data \citep{pav04} handbooks.
  
We used {\sc PYRAF} task {\sc MULTIDRIZZLE} together with standard
tasks in {\sc IRAF}\footnote{IRAF is distributed by the National
  Optical Astronomy Observatories which are operated by the
  Association of Universities for Research in Astronomy (AURA) under
  cooperative agreement with the National Science Foundation.} to
combine all exposures into a single geometrically corrected image
while rejecting cosmic rays. We let {\sc MULTIDRIZZLE} generate the
inverse-variance weighting map automatically, which is suitable for
\sex \citep{sextractor}. We analyzed the images obtained with
different kernels by comparing the point spread function (PSF) FWHM
derived fitting a two-dimensional Gaussian to a few stars distributed
in the field of view. We chose to use the {\sc square} drizzle
kernel. In addition, for the photometric analysis we need a well
sampled PSF. The most suitable values for scale and {\sc pixfrac}
were found to be 0.030 arcsec pixel$^{-1}$ and 1.0, respectively. Prior
to {\sc MULTIDRIZZLE}, we processed every flat-fielded image by
fitting a sky level in each quadrant separately, as suggested by
\citet{sir05}, because of residual differences after the bias
subtraction.

The final, combined, and geometrically corrected image has a size of
7062 $\times$ 7251 pixel$^2$, with 0.030 arcsec pixel$^{-1}$. The
resulting field of view is approximately 212\arcsec $\times$ 218\arcsec
in the shape of a rhomboid.

The transformation to the AB photometric system follows \citet{sir05}. 
We adopted the photometric keywords tabulated
in the header of the images, as suggested by
\citet{pav04} to ensure up-to-date values based on the throughput
curves of all {\it HST} optical components:
\begin{equation}
r_{\rm AB}= -2.5\,\log f({\rm F625W})+25.9186,
\end{equation}
\noindent where $f$({F625W}) refers to the integrated flux in units of
electrons per second.

We produced a suitable PSF for each sample galaxy included in the {\it
  HST} image. The aim is to obtain realistic PSFs that underwent the
same reduction-steps we performed on the observed images. The method
is similar to the one adopted by \citet{chi09}. For each galaxy we
derived from the flat-fielded images the coordinates in pixel of its
centre. By using the {\sc TINYTIM} package \citep{kri99} we produced
PSFs corresponding to all the exposures. We created {\sc TINYTIM} PSFs
with a diameter of 3 arcsec. The PSFs were added to synthetic images
equal to the flat-fielded ones, but with pixel values set to
zero. These were then combined, aligned, and corrected for geometric
distortion in the same way as the real images. Finally, we extracted
the PSF from the processed image, with a pixel scale of 0.03 arcsec
pixel$^{-1}$.

\subsection{Integrated Photometry}
\label{sec:anal:intphot}

We used \sex on the GMOS g\p- and r\p-band images to calculate the integrated magnitudes in those filters. We selected galaxies using the g\p image (it has better seeing than the r\p image, see Table \ref{tab:red:imdata}) to be above a threshold of 3 times the sky noise and have a minimum area of 8 pixels. A total of 64 de-blending sub-thresholds were permitted with a minimum contrast of $1\times10^{-4}$. No filter was applied for detection.

When running \sex in \emph{dual image mode}, the images must be identically aligned, scaled, sampled and PSF convolved. We re-binned and aligned the r\p-band image to match the g\p-band image and degraded the g\p-band resolution to that of the r\p-band by convolving with a Gaussian (FWHM=0\farcs877\footnote{This number is found by trial and error to give similar Gaussian fits to both the g\p- and r\p-band images.}). A pseudo-colour image of \A1689 using the g\p- and r\p-band images is presented in Fig. \ref{fig:A1689GMOSim}. 

We extract total (MAG\_AUTO: based on the Kron definition) and fixed circular aperture magnitudes (MAG\_AP: diameter of 2\farcs9) for all detected objects. This aperture size was chosen because it is approximately three times the Gaussian FWHM of the seeing disc in the r\p image and exactly 20 pixels. From the aperture magnitudes in g\p and r\p, we calculate the g\p-r\p colour of the galaxies. 

The errors produced by \sex when using just the input science images appeared to be underestimated. Instead, when provided with a variance image to act as a weight map (produced by the GMOS pipeline with stars masked), \sex produced more realistic errors, used here. 

\begin{figure*}
   \centering
   \includegraphics[width=\textwidth, angle=0]{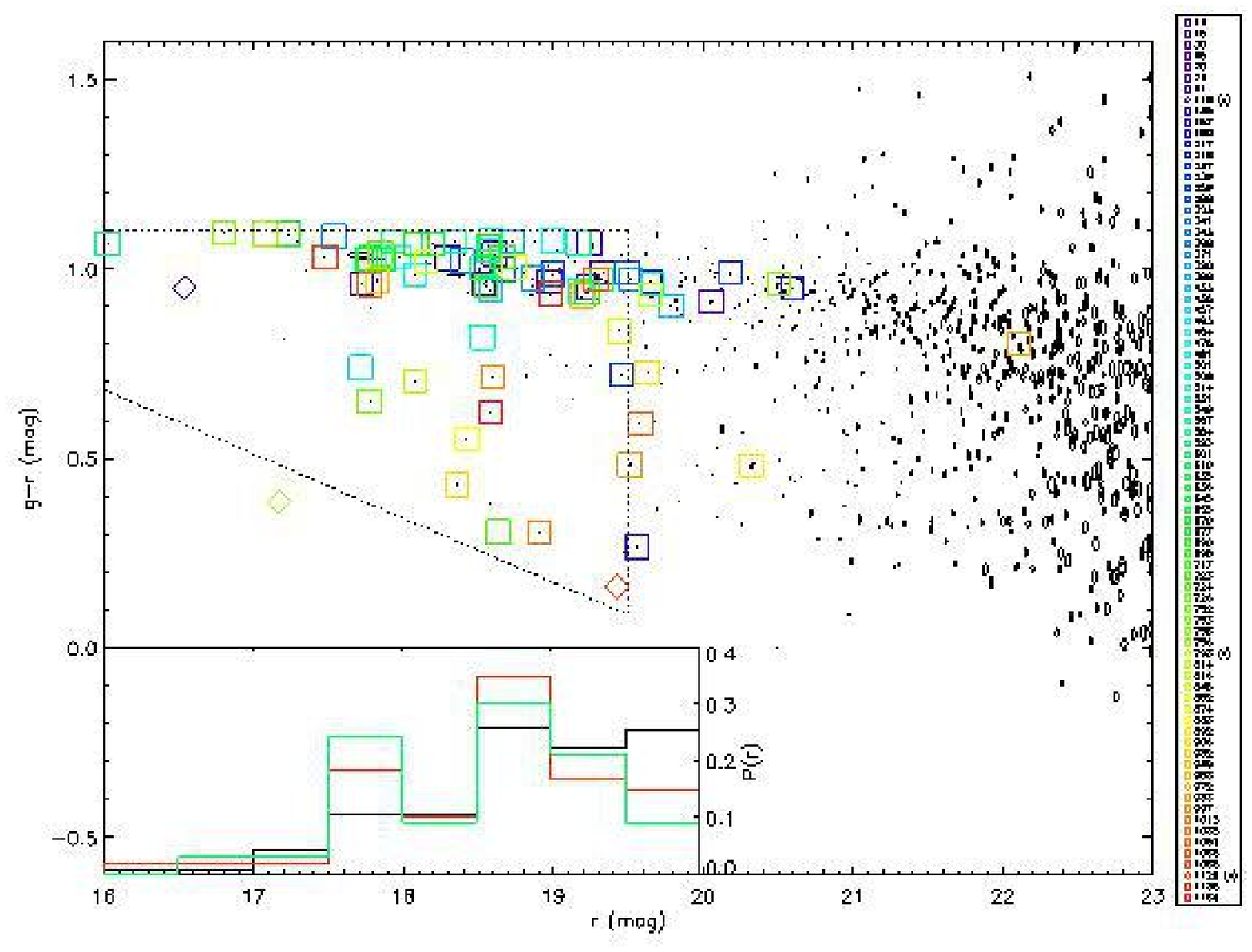} 
   \caption{The Colour-Magnitude diagram for Abell 1689 created using the GEMINI/GMOS photometry with \sex: only data with errors less than 0.1 \Mag are shown. The spectroscopic galaxy sample is highlighted with coloured squares while coloured diamonds represent stars. In the legend, labels followed by a star in parenthesis are guiding/alignment stars (\#119, \#793, \#1129); of the other diamonds, \#972 is reported by \sex to be contaminated (the galaxy is peculiar and appears in the g-band image to have two parts) while \#655 is reported by \sex to be a star but its spectrum confirms it to be a galaxy (as does the HST photometry). The complete sample is shown by plotting the error ellipses (often too small to distinguish from a dot). Selection limits for the spectroscopic sample (\S\ref{sec:red:gmosspec}) are shown as dotted lines (r\p$=19.5$ \Mag represents 50\% completeness) and the inset histogram shows the distribution of all objects brighter than r\p$<20$ for the parent sample (black), for the spectroscopic sample (red) and for the spectroscopic sample inside the HST footprint (green). }
   \label{fig:gmosim:cm}
\end{figure*}

We present the colour-magnitude diagram in Fig. \ref{fig:gmosim:cm}: the r\p magnitude represents the total (Kron) magnitude while the g\p-r\p colour is derived from the aperture magnitudes. To use only reliable photometry of galaxies, we select data where: the errors on magnitudes are less than 0.1 \Mag; total magnitudes are brighter than 25 \Mag in g\p and r\p; \sex has not flagged the photometry for any reason\footnote{other than flags {\sc 1} and {\sc 2} which inform if the photometry was biased by nearby bright objects or originally blended with a neighbour, respectively.}); the object was not identified as a star ({\sc class\_star} $<$ 0.99) and the object has a FWHM$>4$ pixels. Note that the error ellipses neglect the correlation between the r\p-band aperture and total magnitudes. This data is further presented in \S\ref{sec:res:cmd} and discussed in \S\ref{sec:disc:cmd}. We show the constraints for spectroscopic sample selection as dotted lines, which are discussed in \S\ref{sec:red:gmosspec}.

\begin{table*}
\caption{The \sex photometry of the GMOS r\p and g\p images. The \emph{Phot. ID} is the identification from \sex (which we find no need to quote other than in this table), while the \emph{Spec. ID} is the ID in the spectroscopic sample (which \emph{is} quoted in this paper when referring to a galaxy or star). R.A. and DEC. are given in degrees and are calculated from the g\p GMOS image coordinates. The \emph{Red Seq.} column informs us if the object is more likely to be in the red sequence (Y=yes) than not (N=No; i.e. it is more likely to be in the outlier distribution) using the results of the mixture model (see \S\ref{sec:fitting2D}, \S\ref{sec:res:cmd} \& \S\ref{sec:ap:fitting}). The apparent r\p magnitudes and the g\p-r\p colours have been corrected for extinction (atmospheric and Galactic). We refer the reader to the \sex manual for further information on the flag codes (v.2.13, \S9.1, p28). Note that the g\p-r\p colour is an aperture magnitude while the r\p magnitude is a Kron (total) magnitude. Note also that galaxy \#655 does not meet the criterion specified in \S\ref{sec:anal:intphot}, but we include it here for completeness (it \emph{is} included in the KR and FJR). \emph{The full table appears in the online version.}}
\begin{center}
\begin{tabular}{cccccr@{$\pm$}lr@{$\pm$}lcc}
\hline\hline
Phot. & Spec. &  R.A.       & DEC. & Red    &   \multicolumn{2}{|c|}{r\p}     & \multicolumn{2}{|c|}{g\p-r\p} & r\p flag & g\p flag \\
ID    &  ID      &  ($\deg$) & ($\deg$) &Seq.& \multicolumn{2}{|c|}{(\Mag)}& \multicolumn{2}{|c|}{(\Mag)}&               &             \\
\hline\hline
154 & 14 & 197.8058167 & -1.3739161 & Y & 19.2334 & 0.0014 & 0.9509 & 0.0008 & 2 & 2 \\
1364 & 15 & 197.8058014 & -1.3269255 & Y & 17.7165 & 0.0007 & 0.9600 & 0.0004 & 2 & 2 \\
261 & 30 & 197.8095551 & -1.3659689 & Y & 18.5313 & 0.0011 & 0.9600 & 0.0006 & 2 & 2 \\
1317 & 45 & 197.8116913 & -1.3224429 & Y & 20.0542 & 0.0021 & 0.9136 & 0.0013 & 0 & 0 \\
1213 & 70 & 197.8158112 & -1.3199574 & Y & 18.9672 & 0.0009 & 0.9675 & 0.0007 & 3 & 3 \\
1217 & 74 & 197.8159485 & -1.3183731 & Y & 18.6034 & 0.0008 & 1.0388 & 0.0006 & 3 & 3 \\
1508 & 91 & 197.8197632 & -1.3340420 & Y & 19.3286 & 0.0018 & 0.9885 & 0.0011 & 2 & 2 \\
1658 & 135 & 197.8251495 & -1.3040887 & Y & 19.6441 & 0.0014 & 0.9673 & 0.0014 & 0 & 0 \\
1149 & 152 & 197.8277740 & -1.3182346 & Y & 19.2382 & 0.0009 & 1.0686 & 0.0008 & 0 & 0 \\
435 & 160 & 197.8289032 & -1.3585705 & N & 19.5570 & 0.0016 & 0.2652 & 0.0007 & 0 & 0 \\
891 & 217 & 197.8361206 & -1.3346049 & Y & 18.2879 & 0.0006 & 1.0289 & 0.0004 & 2 & 2 \\
1197 & 218 & 197.8362885 & -1.3172021 & Y & 20.5914 & 0.0027 & 0.9489 & 0.0021 & 0 & 0 \\
1430 & 237 & 197.8394165 & -1.3294218 & Y & 18.6605 & 0.0009 & 0.9995 & 0.0006 & 2 & 2 \\
385 & 239 & 197.8395233 & -1.3621749 & Y & 20.1782 & 0.0019 & 0.9885 & 0.0014 & 0 & 0 \\
471 & 259 & 197.8419952 & -1.3572158 & N & 19.4536 & 0.0011 & 0.7208 & 0.0007 & 0 & 0 \\
461 & 286 & 197.8456116 & -1.3554035 & Y & 18.9960 & 0.0009 & 0.9863 & 0.0007 & 2 & 2 \\
1237 & 323 & 197.8496857 & -1.3192796 & Y & 19.4914 & 0.0016 & 0.9888 & 0.0011 & 2 & 2 \\
520 & 341 & 197.8513641 & -1.3536860 & Y & 18.6636 & 0.0007 & 0.9992 & 0.0006 & 3 & 3 \\
... & ... & ... & ... & ... & \multicolumn{2}{|c|}{...} & \multicolumn{2}{|c|}{...} & ... & ... \\
\hline
\end{tabular}
\end{center}
\label{tab:res:intphot}
\end{table*}

\subsubsection{Surface Photometry}
\label{sec:anal:surphot}

There are two common approaches to calculating the effective radius $r_{e}$ and the average surface brightness within $r_{e}$, $\<\mu\>_{e}$ from the flux calibrated images: curve-of-growth fitting \citep[COG, e.g.][]{Dressler87,Jorgensen92} and differential surface fitting \citep[e.g. GASP2D][]{Mendez-Abreu08}. The COG technique was originally used to parameterise the scale and brightness of galaxies, assuming a de Vaucouleurs like profile for all galaxies, where the surface brightness $I(R)$ is defined to be
\begin{equation}
I(R) = I_{0}\textrm{exp}\left( -7.676 [R/R_{e}]^{1/4}\right).
\end{equation}
However, more recently the differential fitting technique has been more widely used in the literature \citep[e.g.][]{vanDokkum96}; furthermore it has become common to fit the more general S\'ersic model (for which the de Vaucouleurs profile is a special case), defined as 
\begin{equation}
I(R) = I_{0}\textrm{exp}\left( -b [R/R_{e}]^{1/n}\right).
\end{equation}
where $b\approx2n-0.324$ over the range $2<n<10$ \citep{Ciotti91}. Recall also that the intensity \emph{at} $R_{e}$ is $I_{e}=I_{0}\exp(-b)$ and the average intensity within $R_{e}$ is $\<I\>_{e}=I_{e}\exp(b)\Gamma(2n)nb^{-2n}$ where $\Gamma$ is the gamma function.

We chose to use the older COG technique for consistency with the majority of earlier work on scaling relations and fit both de Vaucouleurs and S\'ersic profiles, for which the integrated luminosity increases with projected radius as
\begin{equation}
\label{eq:surf:sersiccog}
L(R) = I_{0} R_{e}^{2} \frac{2\pi n}{b^{2n}}\gamma[b(R/R_{e})^{1/n}, 2n]
\end{equation}
where $\gamma$ is the incomplete gamma function \citep{Ciotti91}. In a future paper we will compare these results with full surface fitting and decomposition methods. We developed our own curve-of-growth (COG) software (written in IDL) and implemented this software on the HST/ACS F625W image for the spectroscopic sample. This limited the final sample which could enter our analysis: we observed a total of \nspec galaxies in the spectroscopic sample and while all these galaxies have ground-based g\p- and r\p-band imaging (\S\ref{sec:red:gmosim}), the HST image covers only \nhst of the original \nspec galaxies.

Details of the COG algorithm (including masking of nearby sources, correction of PSF effects) are given in Appendix \ref{sec:ap:cog}. We use Monte-Carlo simulations of the fitting procedure, to estimate the random and systematic uncertainties in $\log R_{e}$ and $\log I_{e}$. These simulations include photon shot noise, typical systematic errors in the subtracted sky level and a suitable range of radii over which we fit model COGs (see Appendix \ref{sec:ap:cog}). We calculate the covariance terms for use when fitting the scaling relations (see \S\ref{sec:fitting2D}) and plotting error ellipses in \S\ref{sec:res}. Both the de Vaucouleurs and S\'ersic COG surface photometry are given in Table \ref{tab:res:surphot}; $\<\mu\>_{e}$ has been corrected for cosmological dimming and $\log R_{e}$ is given in kpc (assuming the cosmology described in \S\ref{sec:intro}).

\begin{table*}
\caption{HST F625W COG surface photometry for \nHST galaxies which are also in the GMOS spectroscopic sample. Both de Vaucouleurs and S\'ersic COGs were used: de Vaucouleurs parameters are headed with \emph{(dV)} while S\'ersic parameters are headed \emph{(S)}; the S\'ersic index \emph{n} only appears in this table for S\'ersic COG fits ($n=4$ for de Vaucouleurs COGs). Values of $\log R_{e}$ are in kpc and values of $\<\mu\>_{e}$ are corrected for extinction (atmospheric and Galactic), Tolman dimming and the bandpass term of the K-correction (assuming the cosmology described in \S\ref{sec:intro}). The apparent magnitudes \apmag are calculated directly from the counts in the COG and have only been corrected for extinction, as described in \S\ref{sec:red:gmosim:fluxcal}. Uncertainties in the parameters ($\sigma$) were calculated from simulation, including the correlation coefficient between $\log R_{e}$ and $\<\mu\>_{e}$ ($\rho$, see \S\ref{sec:anal:surphot} \& \S\ref{sec:cogerrs}). The flags represent the quality of the COG fits: \emph{1} for galaxies whose COG was well represented by the model COG; \emph{2} for galaxies whose COG was not well represented by the model COG; \emph{3} for galaxies whose COG did not asymptotically tend to a constant value and/or had severe contamination from neighbouring galaxies such that, even after masking, were unreliable (see \S\ref{sec:cogcontam}). Only galaxies with flags \emph{1} or \emph{2} enter the fitting process; galaxies \#584 and \#610 are BCGs which were fitted in some cases (c.f. \S\ref{sec:res} and Table \ref{tab:res:fjr&kr}). }
\begin{center}
\begin{tabular}{cr@{$\pm$}l r@{$\pm$}l cr@{$\pm$}lc r@{$\pm$}lr@{$\pm$}lcr@{$\pm$}lcc}
\hline\hline
{Spec.} &  \multicolumn{2}{|c|}{$\log R_{e}$} &  \multicolumn{2}{|c|}{$\<\mu\>_{e}$} & $\rho$ &  \multicolumn{2}{|c|}{\apmag} & {Flag} &  \multicolumn{2}{|c|}{$\log R_{e}$} &  \multicolumn{2}{|c|}{$\<\mu\>_{e}$} & $\rho$ &  \multicolumn{2}{|c|}{\apmag} & n & {Flag} \\
ID & \multicolumn{2}{|c|}{(dV)} & \multicolumn{2}{|c|}{(dV)} & (dV) &  \multicolumn{2}{|c|}{(dV)} & (dV) & \multicolumn{2}{|c|}{(S)} & \multicolumn{2}{|c|}{(S)} & (S) &  \multicolumn{2}{|c|}{(S)} & (S) & (S) \\
\hline\hline
286 & 0.40 & 0.35 & 19.79 & 1.10 & -0.998 & 18.78 & 0.00 & 2 & 0.35 & 0.13 & 19.66 & 0.39 & -1.000 & 18.86 & 0.00 & 2.8 & 2\\
341 & 0.37 & 0.35 & 19.34 & 1.10 & -0.999 & 18.47 & 0.00 & 2 & 0.35 & 0.13 & 19.28 & 0.39 & -1.000 & 18.52 & 0.00 & 3.2 & 2\\
368 & 1.02 & 0.34 & 21.37 & 1.10 & -0.999 & 17.24 & 0.00 & 2 & 0.85 & 0.03 & 20.77 & 0.07 & -0.999 & 17.50 & 0.00 & 2.0 & 2\\
371 & 0.40 & 0.35 & 20.38 & 1.10 & -0.998 & 19.33 & 0.00 & 1 & 0.47 & 0.20 & 20.65 & 0.71 & -1.000 & 19.24 & 0.00 & 5.2 & 1\\
390 & 0.41 & 0.35 & 19.74 & 1.10 & -0.998 & 18.65 & 0.00 & 1 & 0.37 & 0.13 & 19.59 & 0.39 & -1.000 & 18.71 & 0.00 & 3.2 & 1\\
398 & 0.25 & 0.35 & 18.87 & 1.10 & -0.998 & 18.60 & 0.00 & 2 & 0.22 & 0.11 & 18.80 & 0.29 & -1.000 & 18.67 & 0.00 & 2.6 & 1\\
433 & 0.70 & 0.37 & 22.11 & 1.13 & -0.993 & 19.57 & 0.00 & 2 & 0.57 & 0.20 & 21.68 & 0.46 & -1.000 & 19.77 & 0.00 & 2.0 & 2\\
435 & 0.63 & 0.35 & 20.62 & 1.10 & -0.999 & 18.42 & 0.00 & 1 & 0.68 & 0.08 & 20.79 & 0.29 & -1.000 & 18.36 & 0.00 & 4.9 & 1\\
463 & 0.60 & 0.35 & 20.45 & 1.10 & -0.999 & 18.43 & 0.00 & 1 & 0.63 & 0.08 & 20.57 & 0.29 & -1.000 & 18.38 & 0.00 & 4.8 & 2\\
476 & 0.44 & 0.35 & 19.17 & 1.10 & -0.999 & 17.95 & 0.00 & 2 & 0.39 & 0.05 & 19.03 & 0.15 & -1.000 & 18.03 & 0.00 & 2.8 & 2\\
481 & 0.25 & 0.35 & 19.31 & 1.10 & -0.998 & 19.02 & 0.00 & 2 & 0.26 & 0.11 & 19.40 & 0.29 & -1.000 & 19.08 & 0.00 & 2.4 & 1\\
501 & 0.82 & 0.35 & 21.01 & 1.10 & -0.999 & 17.86 & 0.00 & 2 & 0.80 & 0.06 & 20.92 & 0.17 & -1.000 & 17.91 & 0.00 & 3.3 & 2\\
508 & 0.57 & 0.35 & 20.16 & 1.10 & -0.999 & 18.30 & 0.00 & 2 & 0.43 & 0.06 & 19.70 & 0.14 & -0.999 & 18.50 & 0.00 & 2.2 & 2\\
514 & 0.45 & 0.35 & 19.83 & 1.10 & -0.998 & 18.57 & 0.00 & 2 & 0.94 & 0.09 & 21.80 & 0.33 & -1.000 & 18.06 & 0.00 & 9.8 & 2\\
531 & 0.35 & 0.35 & 20.27 & 1.10 & -0.998 & 19.48 & 0.00 & 2 & 0.33 & 0.20 & 20.29 & 0.46 & -1.000 & 19.59 & 0.00 & 1.8 & 2\\
549 & 0.66 & 0.35 & 20.45 & 1.10 & -0.999 & 18.09 & 0.00 & 1 & 0.64 & 0.06 & 20.35 & 0.17 & -1.000 & 18.13 & 0.00 & 3.5 & 1\\
567 & 0.12 & 0.35 & 18.64 & 1.10 & -0.998 & 18.99 & 0.00 & 1 & 0.13 & 0.16 & 18.67 & 0.51 & -1.000 & 19.00 & 0.00 & 3.6 & 1\\
584 & 1.48 & 0.34 & 22.07 & 1.09 & -0.999 & 15.62 & 0.00 & 3 & 0.99 & 0.00 & 20.47 & 0.01 & -0.998 & 16.48 & 0.00 & 1.2 & 3\\
593 & 0.25 & 0.35 & 19.55 & 1.10 & -0.998 & 19.27 & 0.00 & 1 & 0.25 & 0.16 & 19.58 & 0.51 & -1.000 & 19.27 & 0.00 & 3.7 & 2\\
601 & 0.56 & 0.35 & 21.01 & 1.10 & -0.998 & 19.16 & 0.00 & 2 & 0.46 & 0.13 & 20.65 & 0.39 & -1.000 & 19.30 & 0.00 & 2.9 & 1\\
610 & 1.04 & 0.34 & 20.99 & 1.10 & -0.999 & 16.77 & 0.00 & 3 & 1.05 & 0.03 & 21.03 & 0.08 & -1.000 & 16.76 & 0.00 & 3.9 & 3\\
635 & 0.63 & 0.35 & 20.25 & 1.10 & -0.999 & 18.05 & 0.00 & 3 & 0.78 & 0.09 & 20.83 & 0.33 & -1.000 & 17.87 & 0.00 & 5.9 & 3\\
636 & 0.59 & 0.35 & 20.68 & 1.10 & -0.998 & 18.72 & 0.00 & 3 & 0.70 & 0.31 & 21.13 & 1.14 & -1.000 & 18.61 & 0.00 & 6.2 & 3\\
645 & 0.60 & 0.35 & 20.02 & 1.10 & -0.999 & 17.97 & 0.00 & 2 & 0.55 & 0.04 & 19.83 & 0.11 & -1.000 & 18.06 & 0.00 & 2.6 & 1\\
655 & 0.67 & 0.35 & 20.75 & 1.10 & -0.999 & 18.36 & 0.00 & 2 & 0.72 & 0.07 & 20.94 & 0.23 & -1.000 & 18.30 & 0.00 & 4.7 & 2\\
670 & 0.39 & 0.35 & 19.36 & 1.10 & -0.999 & 18.37 & 0.00 & 1 & 0.36 & 0.05 & 19.24 & 0.15 & -1.000 & 18.43 & 0.00 & 3.2 & 1\\
677 & 0.81 & 0.35 & 20.98 & 1.10 & -0.999 & 17.89 & 0.00 & 1 & 0.92 & 0.08 & 21.38 & 0.29 & -1.000 & 17.76 & 0.00 & 5.2 & 1\\
690 & 0.76 & 0.34 & 19.88 & 1.10 & -0.999 & 17.07 & 0.00 & 2 & 0.67 & 0.01 & 19.60 & 0.03 & -0.999 & 17.23 & 0.00 & 2.0 & 1\\
698 & 0.43 & 0.35 & 20.34 & 1.10 & -0.998 & 19.19 & 0.00 & 2 & 0.37 & 0.06 & 20.19 & 0.14 & -0.999 & 19.31 & 0.00 & 2.0 & 2\\
717 & 0.33 & 0.35 & 19.29 & 1.10 & -0.998 & 18.61 & 0.00 & 2 & 0.33 & 0.13 & 19.33 & 0.39 & -1.000 & 18.64 & 0.00 & 3.1 & 1\\
723 & 0.51 & 0.35 & 19.95 & 1.10 & -0.999 & 18.38 & 0.00 & 2 & 0.48 & 0.05 & 19.85 & 0.15 & -1.000 & 18.44 & 0.00 & 3.0 & 1\\
724 & 0.44 & 0.35 & 20.41 & 1.10 & -0.998 & 19.17 & 0.00 & 3 & 0.51 & 0.12 & 20.67 & 0.39 & -1.000 & 19.10 & 0.00 & 4.1 & 3\\
726 & 0.63 & 0.35 & 19.90 & 1.10 & -0.999 & 17.73 & 0.00 & 1 & 0.65 & 0.06 & 20.01 & 0.20 & -1.000 & 17.70 & 0.00 & 4.0 & 2\\
753 & 0.83 & 0.34 & 19.85 & 1.10 & -0.999 & 16.68 & 0.00 & 1 & 0.81 & 0.03 & 19.79 & 0.08 & -1.000 & 16.70 & 0.00 & 3.8 & 1\\
755 & 0.33 & 0.37 & 20.86 & 1.13 & -0.993 & 20.16 & 0.00 & 2 & 0.29 & 0.35 & 20.72 & 1.02 & -1.000 & 20.24 & 0.00 & 2.8 & 1\\
756 & 0.95 & 0.34 & 20.61 & 1.10 & -0.999 & 16.81 & 0.00 & 2 & 1.19 & 0.05 & 21.53 & 0.18 & -1.000 & 16.53 & 0.00 & 6.5 & 2\\
814 & 0.72 & 0.35 & 20.52 & 1.10 & -0.999 & 17.89 & 0.00 & 2 & 0.58 & 0.04 & 20.04 & 0.11 & -1.000 & 18.09 & 0.00 & 2.3 & 2\\
816 & 0.29 & 0.35 & 19.75 & 1.10 & -0.998 & 19.27 & 0.00 & 1 & 0.27 & 0.13 & 19.69 & 0.39 & -1.000 & 19.33 & 0.00 & 2.9 & 1\\
848 & 0.41 & 0.35 & 19.64 & 1.10 & -0.998 & 18.55 & 0.00 & 1 & 0.40 & 0.16 & 19.59 & 0.51 & -1.000 & 18.58 & 0.00 & 3.4 & 1\\
852 & 0.19 & 0.35 & 19.45 & 1.10 & -0.998 & 19.46 & 0.00 & 2 & 0.18 & 0.22 & 19.43 & 0.57 & -1.000 & 19.52 & 0.00 & 2.6 & 2\\
874 & 0.60 & 0.35 & 19.94 & 1.10 & -0.999 & 17.92 & 0.00 & 1 & 0.56 & 0.05 & 19.80 & 0.15 & -1.000 & 17.98 & 0.00 & 3.2 & 1\\
883 & 0.81 & 0.35 & 21.18 & 1.10 & -0.999 & 18.10 & 0.00 & 2 & 0.63 & 0.03 & 20.59 & 0.07 & -0.999 & 18.42 & 0.00 & 1.2 & 2\\
906 & 0.46 & 0.35 & 20.71 & 1.10 & -0.998 & 19.39 & 0.00 & 2 & 0.39 & 0.20 & 20.55 & 0.46 & -1.000 & 19.55 & 0.00 & 1.5 & 2\\
\hline
\end{tabular}
\end{center}
\label{tab:res:surphot}
\end{table*}

\subsection{Photometric accuracy}
We perform a number of cross-checks to ensure the accuracy of our photometry. The results of these checks are discussed below and when (internally) comparing the derived magnitudes of the galaxies we quote the mean difference and the RMS scatter in the relation.

We check the GMOS r\p-band ZP in two ways: firstly we compare the magnitude of an unsaturated star in our field (RA=13:11:33.5, DEC=-01:20:44.7) to the same stars observed with SDSS; we measure an r\p magnitude of 17.12 while SDSS quotes an r \p magnitude of $17.150\pm0.005$ (unfortunately the same star is saturated in the g\p-band due to better seeing). 

We also check how consistent the de Vaucouleurs and S\'ersic COG magnitudes are: a direct comparison in the HST F625W image indicated that the S\'ersic magnitudes were $0.06$ \Mag fainter than the de Vaucouleurs magnitudes (with $0.2$ \Mag scatter; the median difference was also 0.06 \Mag).

Finally we compare the GMOS r\p\ \sex magnitudes to the HST r\p COG magnitudes: we find the \sex Kron magnitudes are fainter than the de Vaucouleurs magnitudes by $0.12$ \Mag (with 0.20 \Mag scatter and median of 0.17 \Mag) and fainter than the S\'ersic magnitudes by $0.08$ \Mag (with 0.22 \Mag scatter and median of 0.13 \Mag).  We estimate that the HST F625W and GMOS r\p images are calibrated to better than 0.1 \Mag (the scatter in the r\p zeropoints in Table \ref{tab:red:photstands} is 0.07 \Mag). We believe the remaining difference is because \sex (Kron) magnitudes are not extrapolated to infinity. In principle, we could generate simulations to test this, but feel it is not central to our analysis because any uncertainty in the GMOS r\p calibration has little bearing on the results, as we use the scatter in the CMR derived from the GMOS r\p\ \sex catalogue, which is independent of the zeropoint.

\subsection{GEMINI/GMOS-N spectroscopy}
\label{sec:red:gmosspec}

\begin{table*}
\begin{tabular}[t]{c c c c c c}
\hline
Mask & Exp. Times & Num of Frames. & Total exposure time & Dates of exposures & Notes \\\hline\hline
03 & 2400 & 6  & 14400 & 10/01/2002, 11/01/2002, & Light clouds on two exposures taken on 10th\\
04 & 2400,2580 & 4 & 10800 & 13/01/2002 & \\
05 & 2400,3000 & 4 & 9960 & 12/01/2002 & \\
06 & 2400,2100,3000 & 6 & 15000 & 14/01/2002, 15/01/2002 & Guiding lost on last 10mins of one exposure \\\hline
\end{tabular}
\caption{Details of the GMOS/Gemini ground based spectra of the galaxies in \A1689.}
\label{tab:specdata}
\end{table*}

The GMOS-N instrument in multi-object spectroscopy (MOS) mode simultaneously provides multiple spectra in the range 0.35\micron\ -- 1.1\micron inside a 5.5\arcmin $\times$ 5.5\arcmin field of view \citep{gmos2}. The spectra of \nspec galaxies in \A1689 and 3 guide/alignment stars were taken with GMOS-N between 10th January 2002 and 15th January 2002 using the B600\_G5303 grating and 0\farcs75 slits, giving a resolution $R\approx1700$ ($\sigma\approx75$\kms) and a (observer-frame) wavelength range of approximately 3500\AA\ to 7000\AA, depending on the position in the field of view (FOV). The spectroscopic sample is bluer than $g\p-r\p=1.1$ \Mag, redder than $g\p-r\p=-0.17r\p + 3.4$ \Mag and 50\% complete to $r\p=19.5$ \Mag (see \S\ref{sec:fit:dblmodel}). Of the \emph{total} number of galaxies matching these limits, around 50\% are in the spectroscopic sample (but note that as a function of magnitude, we are consistent with sampling 60\% of each magnitude bin, see \S\ref{sec:fit:dblmodel}). The locations of the galaxies are shown in Fig. \ref{fig:A1689GMOSim} while the selection criteria are shown on Fig. \ref{fig:gmosim:cm}. The inset histogram in Fig. \ref{fig:A1689GMOSim} compares the distribution of the parent sample to the total spectroscopic sample and the spectroscopic sample that is limited to the HST footprint: the total spectroscopic sample is a good representation of the parent sample, while the sample that lies in the HST footprint is slightly biased towards more luminous galaxies. Note that an apparent magnitude limit of r\p $<$ 19.5 \Mag is equivalent to a rest frame absolute magnitude limit of $\absmag_{V}<-20.04$ \Mag. 

Four masks were used when observing the cluster and for each mask, the data were observed with two different central wavelengths (565 nm and 570 nm) so that the two chip gaps did not cause discontinuity in the data. The number of exposures for each mask varies and details are given in Table \ref{tab:specdata}.
Our reduction of the spectra is initially very similar to that of the imaging (\S\ref{sec:red:gmosim}) but differs in extraction. We again follow the ideas in \citet{Jorgensen05} and \citet{Barr05} but with a few differences: we use custom modified versions of the Gemini IRAF data reduction software (v1.14) and for completeness, we outline our method below.

We process the CCD images in the usual manner, performing bias subtraction using the overscan region and flat field as normal. The individual chips were mosaiced onto a single image using `nearest' interpolation, which retains the statistical independence of the pixels. This has no effect on the final result, as GMOS samples the spectral and spatial PSFs better than the Nyquist limit, by a factor of $\sim5$; in practice the positioning of our data is accurate to 10\% of the PSF FWHM. These processing s.ps were performed with {\sc gsreduce}, which itself makes use of {\sc gprepare}, {\sc gireduce}, {\sc gmosaic} and {\sc gscut}. Furthermore, a cosmic ray rejection step is present in {\sc gsreduce}.

Most of the slits were cut at an angle so it was necessary to rectify the spectra to a uniform space and wavelength grid. This was performed with {\sc gswavelength} using a 4th-order (Legendre) polynomial fit along $\lambda$ and a 2nd-order polynomial fit along the spatial axis. Residuals were typically less than 0.2 \AA. The absolute zero point of the wavelength calibration required correction which was accomplished using the [OI] (5577.34 \AA) and NaD (5889.95 \AA, 5895.92 \AA) sky lines. Our data were reduced to a log-$\lambda$ grid with only one interpolation to help retain the statistical independence of the pixels and minimise correlations. 

The sky was removed using {\sc gsskysub} and a 1D galaxy spectrum extracted using an aperture of diameter 1\farcs4  (4.3 kpc at \ZA1689) tracing the peak flux of the galaxy spectrum. Individual 1D spectra from each exposure were combined by scaling by the median, rejecting bad pixels (including the chip gaps) and clipping the data at $\pm5\sigma$ to remove bad pixels or cosmics that escaped earlier detection.

\subsection{Kinematics}
\label{sec:anal:kin}

To extract the kinematics from the spectra, we assume that the galaxy spectrum $G$ (sampled in $\log \lambda$) is the convolution of a stellar template $T$ with a distribution of stellar velocities $L(v)$, 
\begin{equation}
\label{eq:anal:kin}
G(\log \lambda) = T(\log \lambda) \otimes L(v).
\end{equation}
Furthermore, we assume that $L(v)$ takes the parameterised form of a Gaussian with velocity $V$ and velocity dispersion $\sigma$. 

In order for Eq. \ref{eq:anal:kin} to be valid, the stellar template population must match the galaxy population and both should have the same spectral resolution (i.e. they should both appear to have been observed with the same instrument). In practice, this is achieved by taking a large stellar library \citep[such as the Indo-U.S. Library of Coud\'e Feed Stellar Spectra (CFLIB) used here,][]{cflib} observed at high spectral resolution and degrading it (via convolution) to match the spectral line profile of the instrument;  one then generates a best-fit template in parallel to fitting the kinematics. 

Usually, both the stellar library and the galaxy spectra have spectral profiles closely approximating a Gaussian and so convolving with another Gaussian completes the process of matching the spectral resolutions. However, in the case of GMOS the spectral profile is not Gaussian; thus matching the spectral profiles was somewhat involved and we dedicate an Appendix (\ref{sec:ap:specres}) to explaining our technique. 

\begin{figure}
   \centering
   \includegraphics[height=0.6\textwidth]{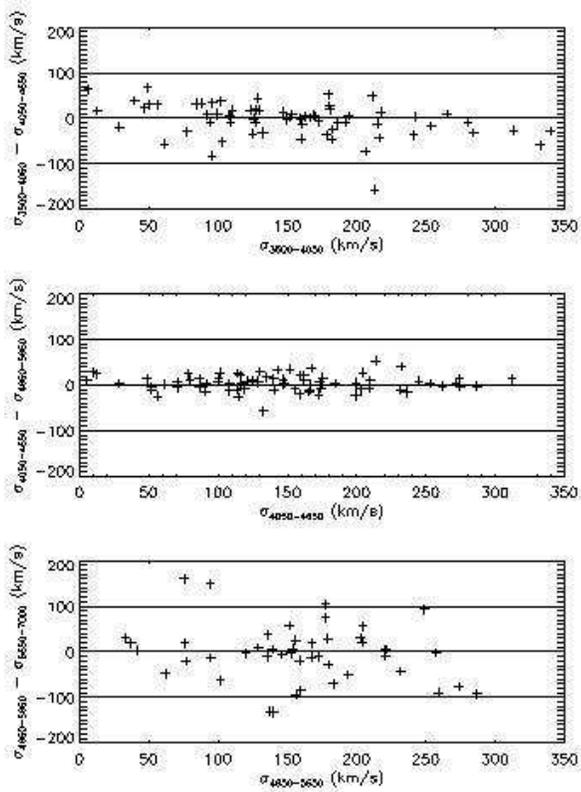} 
   \caption{A comparison of the velocity dispersions $\sigma$ derived from different wavelength ranges: $3500 < \lambda (\textrm{\AA}) < 4050$, $4050 < \lambda (\textrm{\AA}) < 4650$, $4650 < \lambda (\textrm{\AA}) < 5650$ and $5650 < \lambda (\textrm{\AA}) < 7000$. }
   \label{fig:kin:sigmas}
\end{figure}

To calculate the recession velocities $V$ and velocity dispersions $\sigma$ of the individual galaxies, we made use of the freely available {\sc PPXF} (penalised pixel fitting) {\sc IDL} software \citep{ppxf} to fit the parameterised stellar kinematics, and the {\sc GANDALF} (Gas AND Absorption Line Fitting) {\sc IDL} software \citep{SauronV} to fit any emission lines. We divided each input spectrum in four parts and fit them separately: wavelengths less than 4050 \AA; between 4050 \AA\ and 4650 \AA; between 4650 \AA\ and 5650 \AA, and wavelengths greater than 5650 \AA\ (all wavelengths quoted in the rest frame of the cluster). We also excluded regions affected by telluric absorption from the fit, such as the range $6850 < \lambda (\AA) < 6950$. For each wavelength range a new best-fit stellar template was found. Fig. \ref{fig:kin:sigmas} shows the dispersions calculated from the middle two sections (4050 \AA\ -- 4650 \AA\ and 4650 \AA\ -- 5650 \AA) have the least scatter; thus we average the dispersions from these two regions for all subsequent analysis.

Although {\sc PPXF} can determine dispersions less than the instrumental resolution ($\approx 75$ \kms at 0.5 \micron), the errors (both random and systematic) increase. Consequently, we decided that dispersions found to be less than 50 \kms are unreliable and we replace them with an upper limit of 50 \kms and do not use them to fit the scaling relations (see \S\ref{sec:res}). This affects only 4/\nspec of galaxies, two of which could enter the KR and FJR as they have good HST surface photometry (\#883 \& \#906). 

We corrected our velocity dispersion measurements for aperture effects. Each spectrum was extracted from a rectangular aperture of size 1\farcs4 by 1\farcs0 and corrected to the equivalent circular radius following the procedure in \citet{Jorgensen95b}. We also correct our aperture dispersion measurements $\sigma_\n{ap}$ to a velocity dispersion measured within a standard aperture size \citep[1.62 kpc, equivalent 3\farcs4 at the distance of Coma, ][]{Jorgensen95b}. The relation in \citet{Jorgensen95b} is slightly different to the one found by \citet{SauronIV}; however, the effect of this difference on our aperture corrections is small and we persist with the \citeauthor{Jorgensen95b} relation for comparison with the literature.

\begin{table}
\caption{The kinematics of the spectroscopic sample within $\pm6000$\kms of \ZA1689; out of 80 objects targeted, 3 were guide stars leaving \nspec galaxies of which \nspeccluster were found to meet the criterion on $V_\n{rec}$ to be considered part of the cluster. Those galaxies not in the cluster (and not shown here) were \#160, \#457, \#892, \#932, \#983 and \#1051; of these only \#932 has $V_\n{rec}$ relative to \ZA1689 which could be in the cluster (8822\kms). However, none of these galaxies are found to be in the RS according to the mixture model (see Table \ref{tab:res:intphot}). Velocity dispersions measured to be $<50$\kms were deemed unreliable (well below the resolution of the instrument) and replaced with an upper limit of $50$\kms; these points are not used when fitting the scaling relations, but are shown (as diamonds) in Fig. \ref{fig:res:fjr&kr}. Velocity dispersions given here have not been aperture corrected (see \S\ref{sec:anal:kin}). \emph{The full table appears in the online version.}}
\begin{center}
\begin{tabular}{cr@{$\pm$}lr@{$\pm$}l}
\hline\hline
Spec. & \multicolumn{2}{|c|}{$V_\n{rec}$}  &  \multicolumn{2}{|c|}{$\sigma_\n{ap}$} \\
ID      &  \multicolumn{2}{|c|}{(\kms)}      &   \multicolumn{2}{|c|}{(\kms)}     \\ 
\hline\hline
14 & 1672.2 & 2.4 & 83.2 & 3.5 \\
15 & -981.8 & 3.3 & 198.4 & 3.7 \\
30 & 1638.8 & 2.0 & 125.0 & 2.5 \\
45 & 300.8 & 4.0 & 104.3 & 5.4 \\
70 & 1439.5 & 2.7 & 134.8 & 3.3 \\
74 & 2524.8 & 3.5 & 159.1 & 4.2 \\
91 & 2897.4 & 5.2 & 114.8 & 6.6 \\
135 & 948.2 & 2.8 & 85.8 & 4.2 \\
152 & 1291.9 & 2.9 & 158.2 & 3.4 \\
217 & 299.5 & 2.9 & 172.6 & 3.3 \\
218 & 2266.3 & 9.3 & 108.6 & 12.3 \\
237 & -934.2 & 4.2 & 170.8 & 4.8 \\
239 & 1545.2 & 6.3 & 146.8 & 7.6 \\
259 & 1284.4 & 2.3 & 73.6 & 3.7 \\
286 & 2255.1 & 2.5 & 150.7 & 2.9 \\
323 & -83.9 & 2.9 & 90.3 & 4.4 \\
341 & 802.2 & 5.1 & 201.6 & 5.5 \\
343 & -1257.0 & 4.3 & 215.3 & 4.7 \\
...& \multicolumn{2}{|c|}{...} & \multicolumn{2}{|c|}{...} \\
\hline
\end{tabular}
\end{center}
\label{tab:res:kin}
\end{table}

\subsection{The local comparison sample}
\label{sec:localsample}

To draw conclusions about the changes in the properties of the galaxies in \A1689 (at \ZA1689), we need to compare them with local ($z\approx0$) galaxies. For a fair comparison, we must select the local galaxies from a similar environment to the \A1689 galaxies (i.e. from a massive cluster) and the only suitably dense nearby environment is the Coma cluster \citep[\ZComa,][]{Coma_z} at a distance of \DComa (\citealt{Coma_dist1}, \citealt{Coma_dist2} but see also \citealt{Coma_altdist}). 

Fortunately, the ETGs in the Coma cluster have been rigorously investigated in the past. Like previous authors \citep{Ziegler01,LaBarbera03,Barr05}, we start with the data from \citet[][hereafter J99]{Jorg99}: a total of 116 velocity dispersions for galaxies in Coma were compiled from the literature and homogenised in this work. We then compare the sample with the photometry of \citet[][hereafter J95a]{Jorgensen95a} and find that all but 2 galaxies have Gunn r surface photometry available (calculated by fitting de Vaucouleurs curve-of-growth models). We adopt the formal uncertainties given in J95a and J99 but refrain from imposing covariances. Furthermore, we do not attempt to update the Galactic extinction, K-corrections or cosmological corrections but use the data as presented and model them (\S\ref{sec:spmodels}) as though they were observed at $z=0.0$. We determine that this sample is 50\% complete to r\p$=14.8$ (equivalent to $M_{r\p}=-20.28$ ,see \S\ref{sec:fit:dblmodel}). 

\subsubsection{Comparing Coma and \A1689}
\label{sec:ComaVsA1689}

We wish to infer ages and formation timescales of the ETGs in \A1689 and Coma by assuming that galaxies in \A1689 will evolve into those in Coma over \dage. This assumption is not immediately obvious; furthermore, if untrue it could invalidate our analysis. In a hierarchical picture, one expects more massive DM halos to collapse earlier and potentially produce galaxies earlier; thus, if \A1689 were considerably more massive than Coma was at \ZA1689, it is not unreasonable to claim that the \A1689 population would be older, leading to a smaller CMR scatter and $\Delta\beta$, thus biasing our measurements. We now look at these two clusters in detail and discuss how similar or different they are. We address the X-ray luminosities, the cluster masses and the known details of the galaxy populations to gauge how robust our approach is. 

While Coma is relatively X-ray bright (L$_{X}=$\LXComa), its luminosity somewhat dwarfed by the (exceptionally) high X-ray luminosity of \A1689 (L$_{X}=$\LXA1689). On X-ray luminosity evidence alone, one might conclude that the mass of \A1689 is many times the mass of Coma, or that \A1689 has recently undergone a merger \citep[it hasn't, according to the relaxed nature of the X-ray contours,][]{Lemze08}. However, the X-ray luminosity is only part of the picture; indeed, detailed X-ray modelling of Coma and \A1689 indicate that their total masses are quite comparable given the modelling uncertainties \citep[around $5\times10^{14}$ \Msun:][]{Mason&Myers00,Ettori02,Peng09,RiemerSorensen09,Mahdavi08}. The lensing view on the two cluster masses is broadly similar: although \A1689 appears marginally more massive than Coma, there is considerable scatter in the lensing estimates \citep[mass estimates for both are around $1-2\times10^{15}$ \Msun:][]{Kubo07,Gavazzi09,Okabe10,Broadhurst05,Halkola06,Mahdavi08}.

We note (like many others) that the X-ray masses of the inner regions of clusters are generally estimated to be half those of the lensing measurements. \citet{Oguri05} and \citet{Morandi11} were able to resolve this discrepancy with triaxial halos and suggest \A1689 is viewed along its major axis. This is unlikely for Coma as we see the two central BCGs well separated on the sky. Thus certain discrepancies between Coma and \A1689 could in part be down to projection effects. 

In terms of mass evolution, it is unlikely that \A1689 will become much more massive between \ZA1689 and now: according to the work of \citet{Fakhouri10}, for halo masses $>10^{14}$\Msun, the rate of change of mass at $0<z<0.2$ is $<4\times10^{4}$\Msun yr$^{-1}$: therefore, the expected increase in mass over \dage is $<9.2\times10^{13}$\Msun, which is a small fraction of the total mass.

Thus we conclude that, aside from projection effects from possible triaxiallity, the two clusters are well matched in mass and are both reasonably relaxed. 

\subsection{Magnitude conversions, Cosmological corrections and evolution measurements}
\label{sec:fudges}

We cannot compare the Coma data to the \A1689 data directly: the data are in different filter bands using different magnitude systems. Furthermore, the \A1689 data are affected by cosmological dimming (the Coma data has been corrected for this, see \S\ref{sec:localsample}) and evolution of the stellar population (the magnitude of which we wish to determine). Therefore, we must address these differences and where appropriate apply corrections or model the effects (see \S\ref{sec:spmodels}).

\subsubsection{Magnitude systems}
\label{sec:fudges:mags}

The Coma data of JF95a was observed in the Gunn r-band using the Gunn photometric system (relative to a subdwarf F6 star) whereas the HST F625W observations are in the AB system. We convert the Gunn r photometry to the AB system using the corrections listed in \citet{FreiGunn94}. 

\subsubsection{Cosmological Corrections}

When we observe galaxies at significant redshifts, it is necessary to correct for cosmological effects (i.e. expansion and redshift). We adopt the approaches described in \citet{HoggCos} and \citet{HoggKcor} to correct our measurements. We split the K-correction \citep[][]{HoggKcor} into two terms, 
\begin{equation}
\label{eq:fudges:kcor}
K=K_{b}+K_{c}.
\end{equation}
The bandpass term $K_{b}$ is easily corrected in the AB magnitude system by \emph{reducing} the observed brightness by $(1+z)$. However, the colour term $K_{c}$ depends on the details of the underlying stellar population which change with age, metallicity, initial mass function (IMF), dust content, etc.. J95a assumed no \emph{evolution} of the galaxy SED when calculating the colour term of the K-correction $K_{c}$ for the Coma data, which is a good approximation at low redshift. But at higher redshifts, we do not know the exact SED so we do not apply a $K_{c}$ correction to our measurements of \A1689. In \S\ref{sec:spmodels}, we describe how we model our observations by calculating magnitudes based on stellar population models in blueshifted filter profiles, thus negating the need for $K_{c}$. However, it is necessary to correct for \emph{Tolman} dimming \citep{Tolman30,LubinSandage01c}. Incorporating this and $K_{b}$, the bandpass limited AB surface brightness $\mu_\textrm{obs}$ is related to the bandpass limited rest-frame AB surface brightness $\mu_\textrm{rest}$ by 
\begin{equation}
\mu_\n{rest} = \mu_\n{obs} + 7.5\log(1+z).
\end{equation}
Furthermore, we can calculate apparent magnitudes using
\begin{equation}
\label{eq:mue_re_apmag}
\apmag = \<\mu\>_\n{e,rest}  - 5 \log(R_{e}) - 2.5\log(2\pi) + 7.5\log(1+z) 
\end{equation}
and (bandpass corrected) absolute magnitudes using
\begin{equation}
\absmag = \apmag + 2.5\log(1+z) - 5\log(D_{L}) + 5
\end{equation}
where $R_{e}$ is measured in arcseconds and $D_{L}$ is the luminosity distance in pc.

\subsubsection{Size evolution}
\label{sec:fudges:sec}

Recent literature suggests that galaxies (both disks and spheroids) were more compact in the past \citep{Mo98,Ferguson04,Bouwens04,Trujillo07,vdWel08,Hopkins09}. The details are still very much a topic of debate: the degree of compactness could depend on galaxy mass \citep{Barden05,McIntosh05,Trujillo07}. We measure $R_{e}, \<\mu\>_{e}$ and $\sigma$ (for Coma and \A1689) with and without correcting size evolution. The corrections we apply assume that the effective radius now, $R_{e}$, is related to the effective radius at non-zero redshift, $R_{e,z}$, by a simple power-law scaling,  
\begin{equation}
\label{eq:fudges:sec:re}
\log(R_{e}) = \log(R_{e,z}) + \zeta \log(1+z)
\end{equation}
and therefore, assuming no other changes, 
\begin{equation}
\label{eq:fudges:sec:mue}
\<\mu\>_{e} = \<\mu\>_{e,z} + 5 \zeta \log(1+z).
\end{equation}
The value of $\zeta$ is still debated: for massive, high concentration galaxies, \citet{Bouwens04} find $\zeta=1.05\pm0.21$ over the range $2.5<z<6$ which agrees with \citet{vdWel08} who found $\zeta=0.98\pm0.11$ between $0<z<1$ for morphologically selected ETGs. However, \citet{Trujillo07} found slightly stronger evolution for massive spheroid-like (highly concentrated) galaxies: a factor of $4\pm0.4$ since $z=1.5$, equivalent to $\zeta=1.6$. We adopt a value $\zeta=1.0$ in this work.

We also calculate the equivalent velocity dispersion now, $\sigma$, from the velocity dispersion measured at higher redshift, $\sigma_{z}$, to be 
\begin{equation}
\label{eq:fudges:sec:sigma}
\log(\sigma) = \log(\sigma_{z}) - 0.5 \eta \log(1+z).
\end{equation}
with $\eta=1.0$. In dynamical models of S\'ersic-like galaxies, the projected line-of-sight velocity dispersion changes in this way when $R_{e}$ is scaled, if the S\'ersic index is unchanged \citet{Ciotti91}. However, \citet{Hopkins10} suggest the S\'ersic index \emph{does} change. The effect of size evolution on the velocity dispersion is difficult to measure and not well constrained \citep{vdWel08,Cappellari09,vanDokkum09,Hopkins10}. \citet{CenarroTrujillo09} report a value of $\eta\approx0.6$ while the single observation by \citet{vanDokkum09} would suggest $\eta\approx2$. In the absence of a clear measurement of the change in the S\'ersic index, we adopt $\eta=1.0$.

\subsection{Fitting 2D scaling relations}
\label{sec:fitting2D}

There are many different techniques available to fit (thereby represent) scaling relations. The disadvantages and limitations of common techniques are discussed by \citet{HoggFitting} who advocate Markov Chain Monte Carlo (MCMC) methods. We follow these guidelines and use MCMC methods to \emph{explore the posterior distributions of the parameters in each assumed model}; details are given in Appendix \ref{sec:ap:fitting}. 
We investigate three different models: one where the data (with uncertainties) are drawn from a linear relation with unknown slope, intercept and intrinsic dispersion and we do not attempt to correct for magnitude cuts or selection effects (hereafter the linear model or LM); a double linear model where two sets of data (with uncertainties) are assumed to have the same slope and intrinsic dispersion but with different intercepts and we apply corrections for magnitude cuts and selection effects (hereafter the double linear model or DLM); and finally, a model where the data are drawn from a mixture of two distributions: one being the linear model as before, and another \emph{outlier} distribution having unknown first and second moments (hereafter the mixture model, or MM). In all cases, we assume normal distributions for the models.

We use the LM and DLM to fit the KRs and FJRs of Coma and \A1689 to measure any offset (i.e. evolution) between them. We use the MM to isolate and measure the intrinsic scatter of the CMR in \A1689.

\section{Results and Analysis}
\label{sec:res}

We present the Faber-Jackson relation, the Kormendy relation and the colour magnitude diagram for \A1689, below. As discussed in \S\ref{sec:intro:thisstudy}, we will present and analyse the FP in a future paper. 

\subsection{g'- r' Colour Magnitude Diagram}
\label{sec:res:cmd}

The raw g\p-r\p colour magnitude diagram (CMD) for \A1689 was presented earlier in Fig. \ref{fig:gmosim:cm}, showing only the data where the uncertainties were less than 0.1 \Mag. The RS is clearly visible, along with a blue cloud. We also over plot (dotted lines) the selection limits for the GMOS-N spectroscopic sample and show as an insert the luminosity distribution of the parent sample and the spectroscopic sample. 

\subsubsection{Measured scatter of the CMR in \A1689}

We use a mixture model (see \S\ref{sec:fitting2D} \& Appendix \ref{sec:ap:fitting}) to measure the slope, intercept and intrinsic scatter of the CMR (Eq. \ref{eq:intro:cmr}). We restrict the galaxies in the sample to be brighter than r\p$<22$ \Mag as well as the constraints described \S\ref{sec:anal:intphot}. We do not apply constraints on the recession velocities or morphologies of the galaxies, which can only increase the measured scatter. 

The results of fitting the MCMC mixture model are given in Table \ref{tab:res:fjr&kr}; in particular, the intrinsic scatter is found to be $\sigma_\n{CMR}=$\CMRscatterA1689. 
 
\begin{figure}
   \centering
   \includegraphics[width=0.5\textwidth]{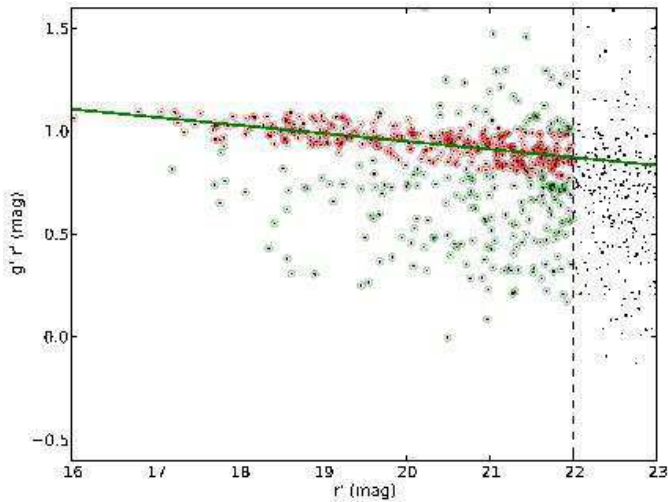} 
   \caption{CMD with RS shown (red) and outliers (green) rejected by the MCMC mixture model.}
   \label{fig:res:CMD_MCMC}
\end{figure}

In Fig. \ref{fig:res:CMD_MCMC} we show the CMD of \A1689 with the galaxies marked according to which of the two distributions they were most likely to be members of. We also show the best fit to the CMR, marginalised over all other parameters. There is an excess of galaxies below the RS, and a few above, which are considered to be \emph{outliers} by the mixture model (points highlighted in green), but the mixture model has successfully isolated the RS (points highlighted in red) \emph{without any cut in colour or clipping} imposed \emph{a priori}. The limit in magnitude (r\p$<22$ \Mag) affects the resulting scatter significantly: brighter (lower) magnitude cuts decrease the scatter while fainter (higher) magnitude cuts increase the scatter. Magnitudes are taken from the entire GMOS r-band image, with dimensions 280\arcsec$\times~$300\arcsec.

BKT98 quote Coma's CMR scatter inside $R <$ 600 kpc to be 0.049 for all galaxies brighter than $\absmag_{V}<-18.2$ \Mag (98 galaxies). We fit the CMR to galaxies brighter than $\absmag_{V}<-17.90$ \Mag inside $R<$570 kpc. The selection effects and resulting scatter in both clusters is very similar; the scatter in Coma is slightly smaller though uncertainties are not given by BKT98. More recently, \citet{Terlevich2001} performed a more extensive study of the CMR in Coma: for all galaxies brighter than $\absmag_{V}<-18.2$ \Mag in roughly a square degree (175 galaxies), they quote an intrinsic scatter of $0.063\pm_{0.01}^{0.008}$ \Mag, which is statistically consistent with our result for \A1689 (although our uncertainties are a factor of 2 smaller). 

We highlight a handful of objects at relatively faint magnitudes (r\p$\sim21$ \Mag) that lie above the CMR. These excessively red objects are found beyond r\p$>20$ \Mag, so they do not enter our spectroscopic sample. Galaxies redder than the RS could be excessively old, excessively metal rich, heavily extincted and/or intermediate or high redshift interlopers. Alternatively, we could be underestimating the scatter in the CMR at this magnitude (particularly because we assume a single scatter across all magnitudes). \citet{Terlevich99} correlated the residuals of the CMR with spectral absorption line indices and showed that the galaxies scattered \emph{blueward} of the mean relation have an increased hydrogen Balmer absorption and are thus younger: if the reverse is true, galaxies redder than the RS could be old relics from the initial cluster population. It is also worth noting that for ages approaching 14\Gyr and metallicities around twice solar, both the BC03 and M05 models predict a maximum g\p-r\p colour of $\ltsim 1.4$. Thus the few objects redder than g\p-r\p=1.4 are unlikely to be part of the cluster unless they are extremely metal rich or extincted; conversely, the majority of the red objects that have g\p-r\p$<1.4$ but are redder than the RS could be cluster members, in principle. 

\begin{figure*}
   \centering
   \begin{minipage}[c]{0.475\textwidth} 
   \begin{minipage}[c]{\textwidth} 
   \includegraphics[width=\textwidth]{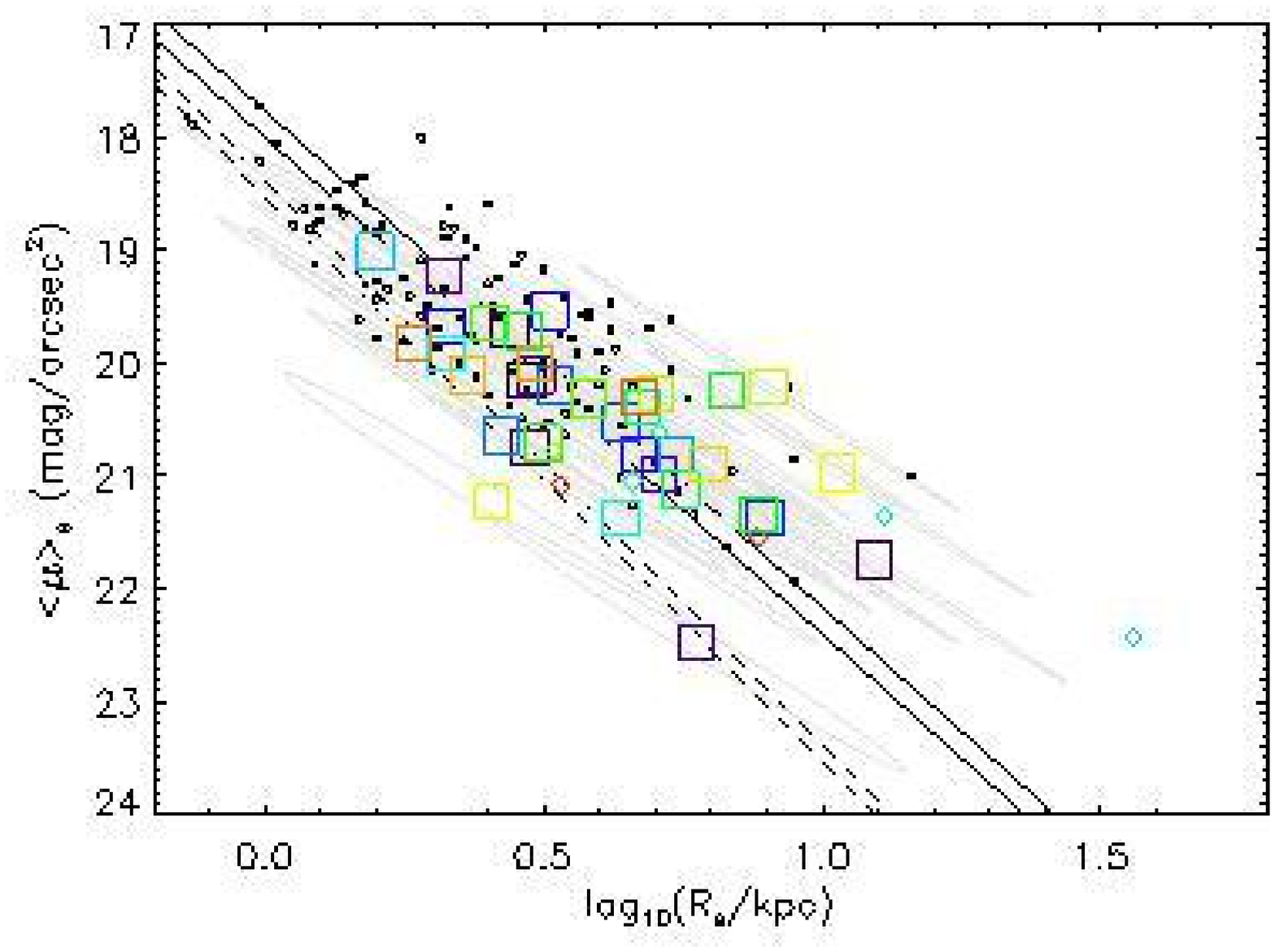} 
   \end{minipage}
   \begin{minipage}[c]{\textwidth}
   \includegraphics[width=\textwidth]{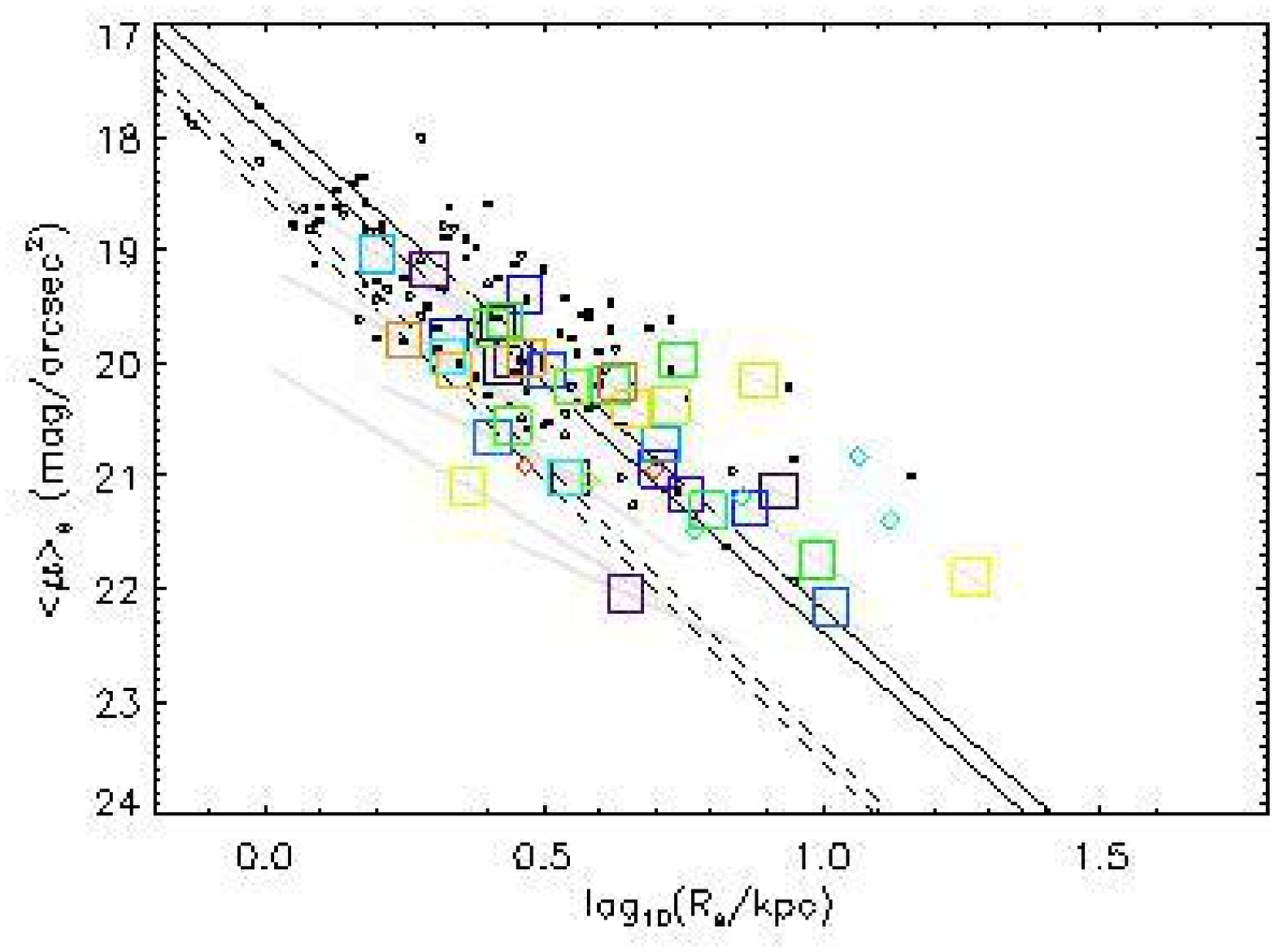} 
   \end{minipage}
   \end{minipage}\begin{minipage}[c]{0.475\textwidth}
   \begin{minipage}[c]{\textwidth}  
   \includegraphics[width=\textwidth]{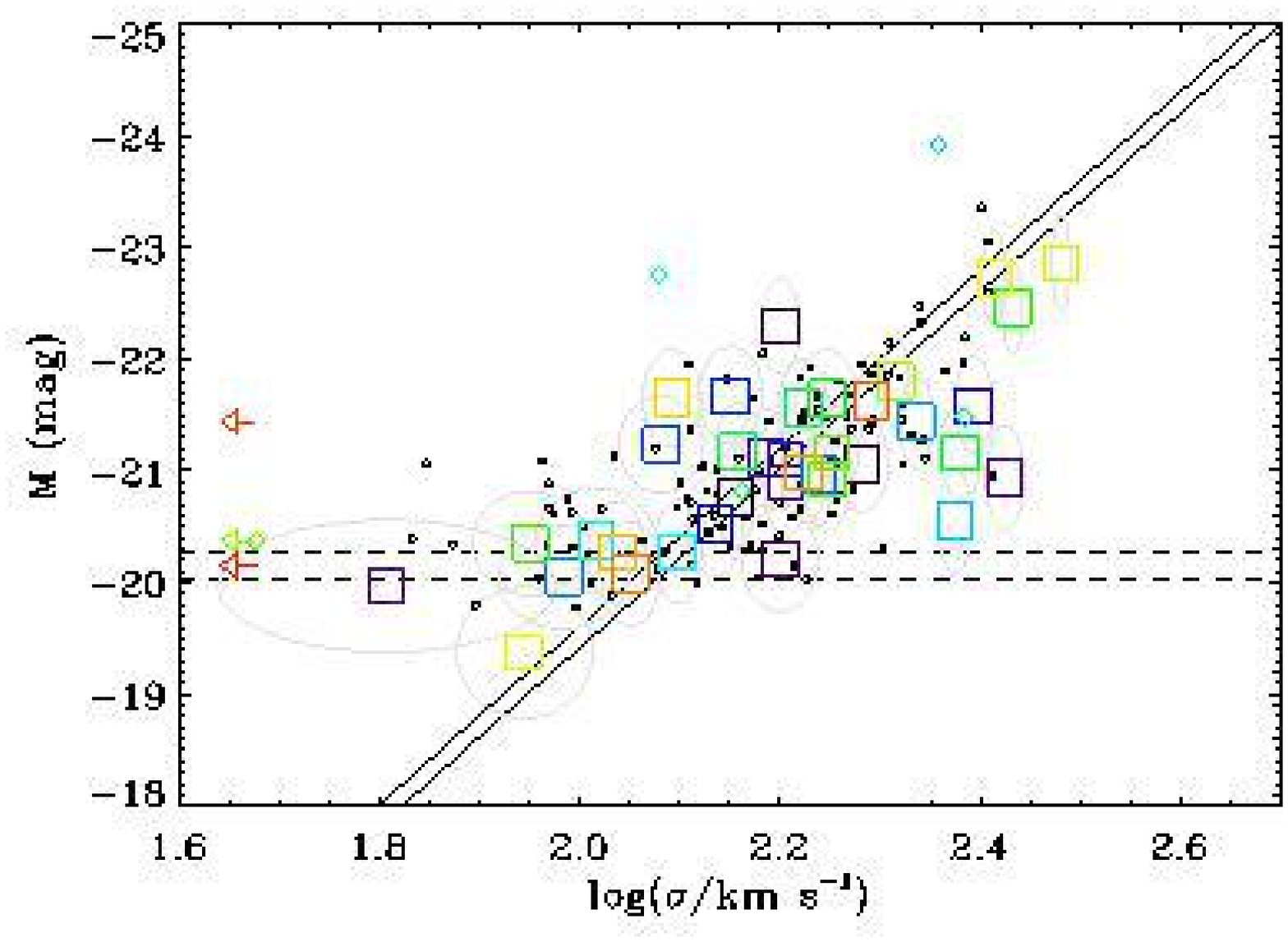} 
   \end{minipage}
   \begin{minipage}[c]{\textwidth} 
   \includegraphics[width=\textwidth]{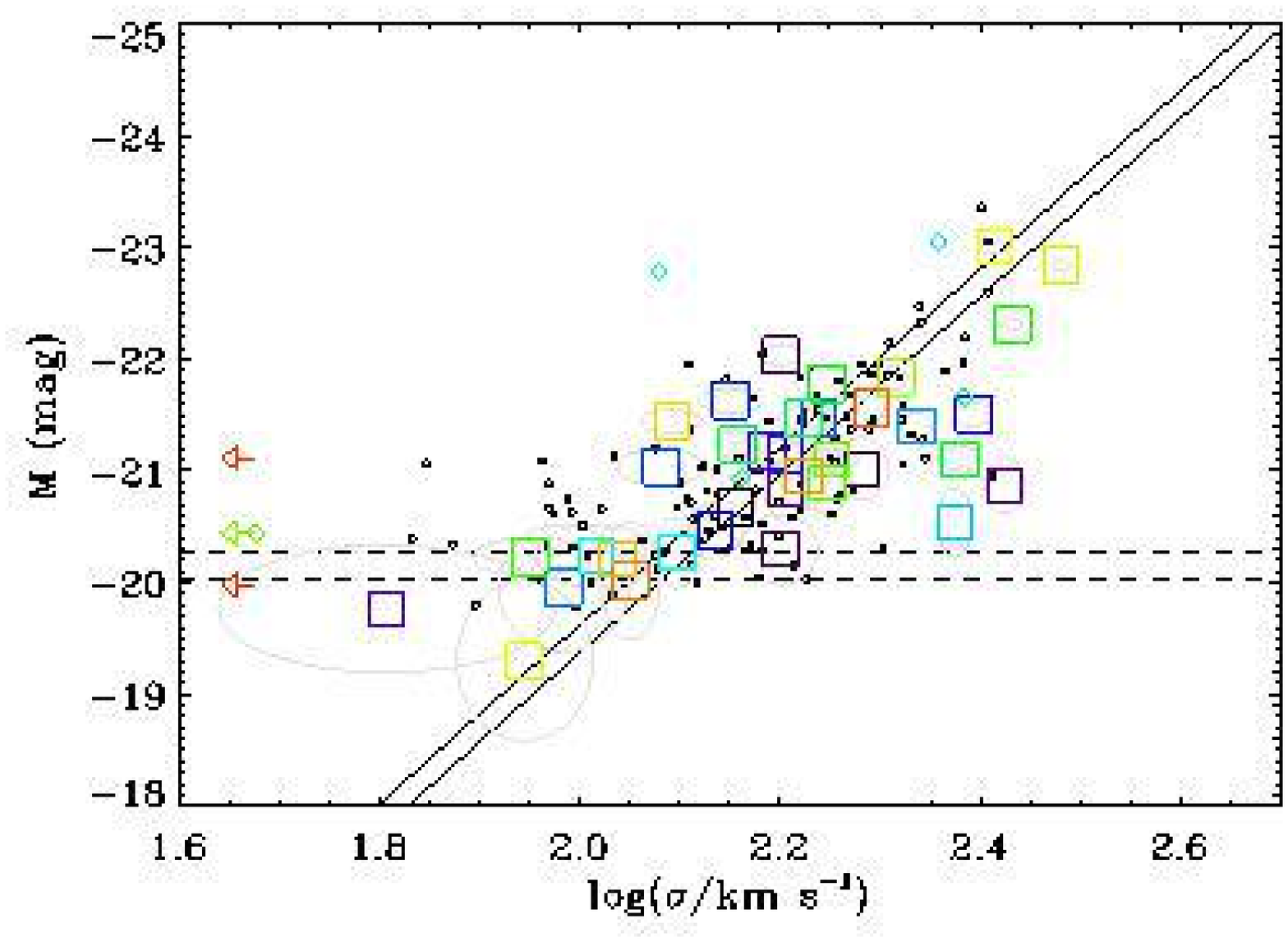} 
   \end{minipage}
   \end{minipage}\begin{minipage}[c]{0.05\textwidth}
   \includegraphics[width=1.7\textwidth]{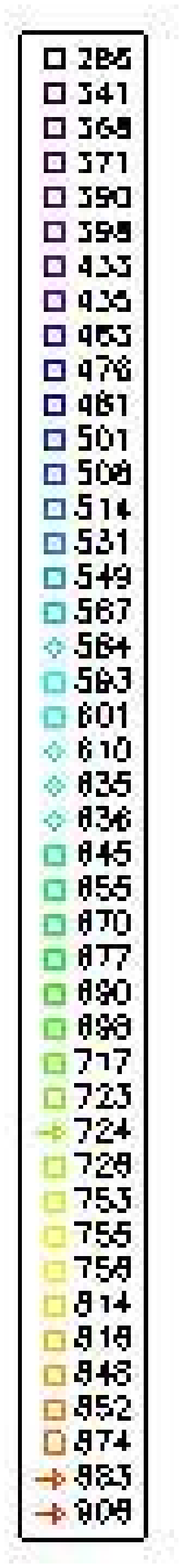} 
   \end{minipage}
   \caption{The de Vaucouleurs (top) and S\'ersic (bottom) Kormendy (left) and Faber-Jackson (right) relations for \A1689. Squares represent good data while diamonds and arrows represent bad data (poor COG fits, or upper limits on velocity dispersions) that were excluded from the fits. Small black squares represent Coma data from \citet[in the r-band, corrected to the AB system]{Jorgensen95a} and \citet{Jorgensen99} while large coloured symbols represent data for \A1689 (from HST/ACS F625W imaging and GMOS-N spectroscopy). Error ellipses on the \A1689 data are shown in light grey and account for correlations between each axis. The error bars are larger for the de Vaucouleurs data because they include the systematic errors associated with fixing the S\'ersic index to $n=4$. Solid lines show the results of fitting the double linear models without BCGs: the light solid line represents the Coma data while the bold solid line represents the \A1689 data. Dashed lines illustrate the magnitude cuts (50\% completeness) for the \A1689 sample (bold dashed) and the Coma sample (light dashed). Note that bad points (diamonds) may lie outside the plotting window. Galaxies for which $\sigma<50$\kms are shown as upper limits in the FJR (\#724, \#884 \& \#906).}
   \label{fig:res:fjr&kr}
\end{figure*}

\begin{table*}
\begin{tabular}[t]{r@{$\pm$}l r@{$\pm$}l r@{$\pm$}l r@{$\pm$}l r@{$\pm$}l l c c c l}
\hline\hline
\multicolumn{15}{|c|}{{\bf Colour-Magnitude Relation}: $(g^\prime-r^\prime) = \alpha_\n{CMR}  (r^\prime-20) + \beta_\n{CMR}$} \\
\hline
\multicolumn{2}{|c|}{$\alpha_\n{CMR}$} & \multicolumn{2}{|c|}{$\beta_\n{CMR}$} & \multicolumn{2}{|c|}{$\sigma_\n{CMR}$} & \multicolumn{4}{|c|}{} & Model & \multicolumn{4}{|c|}{Notes} \\
\hline\hline
-0.039 & $0.003$ & 0.951 & $0.004$ & 0.054 & $0.004$ & \multicolumn{4}{|c|}{} & MM & \multicolumn{4}{|c|}{Fit to sources with $r^{\prime}<22$. } \\
\hline\hline
\multicolumn{15}{|c|}{{\bf Faber-Jackson Relation}: $\absmag = \alpha_\n{FJR} \log (\sigma) + \beta_\n{FJR}$}\\
\hline
\multicolumn{2}{|c|}{$\alpha_\n{FJR}$} & \multicolumn{2}{|c|}{$\beta_\n{FJR}$} & \multicolumn{2}{|c|}{$\sigma_\n{I}$(FJR)} & \multicolumn{2}{|c|}{$\absmag(\log\sigma=2.2)$} & \multicolumn{2}{|c|}{$\Delta \beta_\n{FJR}$} & Model & SEC & Cluster & Profile &  Notes\\
\hline\hline
-8.67 & $_{0.95}^{0.77}$ & -2.1 & $_{1.7}^{2.1}$ & 0.0917 & $_{0.0068}^{0.0076}$ & -21.198 & $_{0.082}^{0.08}$ & \multicolumn{2}{|c|}{-} & LM & N & Coma & dV & all galaxies \\
-10.6 & $_{3.1}^{2}$ & 2.6 & $_{4.5}^{7.1}$ & 0.106 & $_{0.015}^{0.017}$ & -20.70 & $_{0.22}^{0.28}$ & 0.50 & $_{0.24}^{0.29}$ & LM & N & Abell 1689 & dV & all galaxies \\
-10.0 & $_{2.7}^{1.8}$ & 1.2 & $_{4.1}^{6.1}$ & 0.109 & $_{0.013}^{0.016}$ & -20.69 & $_{0.21}^{0.26}$ & 0.51 & $_{0.22}^{0.27}$ & LM & N & Abell 1689 & S & all galaxies \\
{-8.73} & $_{0.73}^{0.59}$ & {-1.6} & $_{1.3}^{1.6}$ & {0.0938} & $_{0.0055}^{0.006}$ & \multicolumn{2}{|c|}{{-}} & {0.42} & $_{0.13}^{0.14}$ & {DLM} & {N} & {Both} & {dV} & {all galaxies} \\
{-8.93} & $_{0.72}^{0.58}$ & {-1.1} & $_{1.3}^{1.6}$ & {0.0980} & $_{0.0051}^{0.0056}$ & \multicolumn{2}{|c|}{{-}} & {0.46} & $_{0.12}^{0.13}$ & {DLM} & {N} & {Both} & {dV+S} & {all galaxies} \\
-8.32 & $_{1.02}^{0.8}$ & -2.9 & $_{1.8}^{2.2}$ & 0.0911 & $_{0.007}^{0.0077}$ & -21.186 & $_{0.08}^{0.079}$ & \multicolumn{2}{|c|}{-} & LM & N & Coma & dV & no BCGs \\
-7.6 & $_{1.7}^{1.2}$ & -3.9 & $_{2.6}^{3.7}$ & 0.081 & $_{0.018}^{0.018}$ & -20.73 & $_{0.15}^{0.17}$ & 0.45 & $_{0.17}^{0.19}$ & LM & N & Abell 1689 & dV & no BCGs \\
-7.9 & $_{1.6}^{1.2}$ & -3.4 & $_{2.7}^{3.7}$ & 0.094 & $_{0.013}^{0.016}$ & -20.69 & $_{0.15}^{0.18}$ & 0.49 & $_{0.17}^{0.2}$ & LM & N & Abell 1689 & S & no BCGs \\
\textbf{-7.92} & $_\textbf{0.58}^\textbf{0.52}$ & \textbf{-3.3} & $_\textbf{1.2}^\textbf{1.3}$ & \textbf{0.0866} & $_\textbf{0.0055}^\textbf{0.0057}$ & \multicolumn{2}{|c|}{\textbf{-}} & \textbf{0.44} & $_\textbf{0.12}^\textbf{0.12}$ & \textbf{DLM} & \textbf{N} & \textbf{Both} & \textbf{dV} & \textbf{no BCGs} \\
\textbf{-8.10} & $_\textbf{0.66}^\textbf{0.55}$ & \textbf{-2.9} & $_\textbf{1.2}^\textbf{1.5}$ & \textbf{0.0906} & $_\textbf{0.0051}^\textbf{0.0054}$ & \multicolumn{2}{|c|}{\textbf{-}} & \textbf{0.50} & $_\textbf{0.1}^\textbf{0.11}$ & \textbf{DLM} & \textbf{N} & \textbf{Both} & \textbf{dV+S} & \textbf{no BCGs} \\
-8.69 & $_{0.93}^{0.79}$ & -2.1 & $_{1.7}^{2}$ & 0.0918 & $_{0.0069}^{0.0076}$ & -21.243 & $_{0.083}^{0.08}$ & \multicolumn{2}{|c|}{-} & LM & Y & Coma & dV & all galaxies \\
-10.6 & $_{3.1}^{2}$ & 2.3 & $_{4.4}^{6.9}$ & 0.106 & $_{0.015}^{0.018}$ & -21.09 & $_{0.21}^{0.23}$ & 0.15 & $_{0.22}^{0.24}$ & LM & Y & Abell 1689 & dV & all galaxies \\
-10.0 & $_{2.7}^{1.8}$ & 0.9 & $_{4}^{5.9}$ & 0.109 & $_{0.014}^{0.016}$ & -21.04 & $_{0.2}^{0.21}$ & 0.20 & $_{0.21}^{0.22}$ & LM & Y & Abell 1689 & S & all galaxies \\
{-8.65} & $_{0.78}^{0.55}$ & {-2.1} & $_{1.2}^{1.7}$ & {0.0937} & $_{0.0054}^{0.0058}$ & \multicolumn{2}{|c|}{{-}} & {0.14} & $_{0.13}^{0.12}$ & {DLM} & {Y} & {Both} & {dV} & {all galaxies} \\
{-8.97} & $_{0.67}^{0.67}$ & {-1.3} & $_{1.5}^{1.4}$ & {0.0980} & $_{0.005}^{0.0055}$ & \multicolumn{2}{|c|}{{-}} & {0.19} & $_{0.13}^{0.12}$ & {DLM} & {Y} & {Both} & {dV+S} & {all galaxies} \\
-8.33 & $_{1.02}^{0.84}$ & -2.9 & $_{1.8}^{2.2}$ & 0.0912 & $_{0.0069}^{0.0078}$ & -21.230 & $_{0.082}^{0.08}$ & \multicolumn{2}{|c|}{-} & LM & Y & Coma & dV & no BCGs \\
-7.6 & $_{1.7}^{1.2}$ & -4.2 & $_{2.7}^{3.7}$ & 0.081 & $_{0.018}^{0.019}$ & -21.01 & $_{0.14}^{0.14}$ & 0.22 & $_{0.16}^{0.16}$ & LM & Y & Abell 1689 & dV & no BCGs \\
-7.9 & $_{1.6}^{1.2}$ & -3.7 & $_{2.7}^{3.6}$ & 0.094 & $_{0.013}^{0.015}$ & -20.98 & $_{0.14}^{0.15}$ & 0.25 & $_{0.16}^{0.17}$ & LM & Y & Abell 1689 & S & no BCGs \\
\textbf{-7.98} & $_\textbf{0.67}^\textbf{0.45}$ & \textbf{-3.5} & $_\textbf{1}^\textbf{1.4}$ & \textbf{0.0868} & $_\textbf{0.0054}^\textbf{0.0057}$ & \multicolumn{2}{|c|}{\textbf{-}} & \textbf{0.19} & $_\textbf{0.11}^\textbf{0.11}$ & \textbf{DLM} & \textbf{Y} & \textbf{Both} & \textbf{dV} & \textbf{no BCGs} \\
\textbf{-7.98} & $_\textbf{0.61}^\textbf{0.49}$ & \textbf{-3.4} & $_\textbf{1.1}^\textbf{1.3}$ & \textbf{0.0905} & $_\textbf{0.0049}^\textbf{0.0054}$ & \multicolumn{2}{|c|}{\textbf{-}} & \textbf{0.25} & $_\textbf{0.1}^\textbf{0.11}$ & \textbf{DLM} & \textbf{Y} & \textbf{Both} & \textbf{dV+S} & \textbf{no BCGs} \\

\hline\hline
\multicolumn{15}{|c|}{{\bf Kormendy Relation }: $\<\mu\>_{e} = \alpha_\n{KR} \log(R_{e}) + \beta_\n{KR}$ }\\
\hline
\multicolumn{2}{|c|}{$\alpha_\n{KR}$} & \multicolumn{2}{|c|}{$\beta_\n{KR}$} & \multicolumn{2}{|c|}{$\sigma_\n{I}$(KR)} & \multicolumn{2}{|c|}{$\<\mu\>_{e}(\log R_{e}=0.5)$} & \multicolumn{2}{|c|}{$\Delta \beta_\n{KR}$} & Model & SEC & Cluster & Profile & Notes\\
\hline\hline
3.84 & $_{0.24}^{0.28}$ & 17.95 & $_{0.13}^{0.12}$ & 0.150 & $_{0.01}^{0.011}$ & 19.870 & $_{0.061}^{0.064}$ & \multicolumn{2}{|c|}{-} & LM & N & Coma & dV & all galaxies \\
3.67 & $_{0.42}^{0.58}$ & 18.10 & $_{0.36}^{0.26}$ & 0.165 & $_{0.025}^{0.026}$ & 19.94 & $_{0.12}^{0.11}$ & 0.07 & $_{0.14}^{0.13}$ & LM & N & Abell 1689 & dV & all galaxies \\
3.99 & $_{0.49}^{0.63}$ & 17.94 & $_{0.38}^{0.3}$ & 0.168 & $_{0.02}^{0.024}$ & 19.93 & $_{0.13}^{0.12}$ & 0.06 & $_{0.14}^{0.14}$ & LM & N & Abell 1689 & S & all galaxies \\
{4.11} & $_{0.2}^{0.22}$ & {18.06} & $_{0.13}^{0.12}$ & {0.1522} & $_{0.0085}^{0.0089}$ & \multicolumn{2}{|c|}{{-}} & {0.218} & $_{0.095}^{0.093}$ & {DLM} & {N} & {Both} & {dV} & {all galaxies} \\
{4.16} & $_{0.21}^{0.23}$ & {18.03} & $_{0.14}^{0.13}$ & {0.1579} & $_{0.0077}^{0.0084}$ & \multicolumn{2}{|c|}{{-}} & {0.211} & $_{0.098}^{0.095}$ & {DLM} & {N} & {Both} & {dV+S} & {all galaxies} \\
4.22 & $_{0.27}^{0.31}$ & 17.83 & $_{0.14}^{0.12}$ & 0.1319 & $_{0.0091}^{0.0104}$ & 19.942 & $_{0.061}^{0.065}$ & \multicolumn{2}{|c|}{-} & LM & N & Coma & dV & no BCGs \\
4.67 & $_{0.69}^{0.76}$ & 17.63 & $_{0.42}^{0.38}$ & 0.106 & $_{0.059}^{0.04}$ & 19.97 & $_{0.13}^{0.13}$ & 0.03 & $_{0.14}^{0.14}$ & LM & N & Abell 1689 & dV & no BCGs \\
4.29 & $_{0.51}^{0.66}$ & 17.83 & $_{0.39}^{0.31}$ & 0.150 & $_{0.019}^{0.023}$ & 19.98 & $_{0.12}^{0.12}$ & 0.04 & $_{0.14}^{0.14}$ & LM & N & Abell 1689 & S & no BCGs \\
\textbf{4.39} & $_\textbf{0.19}^\textbf{0.21}$ & \textbf{17.96} & $_\textbf{0.12}^\textbf{0.12}$ & \textbf{0.1313} & $_\textbf{0.0079}^\textbf{0.0085}$ & \multicolumn{2}{|c|}{\textbf{-}} & \textbf{0.199} & $_\textbf{0.094}^\textbf{0.091}$ & \textbf{DLM} & \textbf{N} & \textbf{Both} & \textbf{dV} & \textbf{no BCGs} \\
\textbf{4.41} & $_\textbf{0.21}^\textbf{0.22}$ & \textbf{17.93} & $_\textbf{0.14}^\textbf{0.13}$ & \textbf{0.1424} & $_\textbf{0.0072}^\textbf{0.0077}$ & \multicolumn{2}{|c|}{\textbf{-}} & \textbf{0.173} & $_\textbf{0.095}^\textbf{0.094}$ & \textbf{DLM} & \textbf{N} & \textbf{Both} & \textbf{dV+S} & \textbf{no BCGs} \\
3.84 & $_{0.25}^{0.28}$ & 17.96 & $_{0.13}^{0.12}$ & 0.150 & $_{0.01}^{0.011}$ & 19.881 & $_{0.06}^{0.063}$ & \multicolumn{2}{|c|}{-} & LM & Y & Coma & dV & all galaxies \\
3.66 & $_{0.41}^{0.59}$ & 18.20 & $_{0.39}^{0.28}$ & 0.165 & $_{0.024}^{0.026}$ & 20.03 & $_{0.14}^{0.12}$ & 0.15 & $_{0.15}^{0.14}$ & LM & Y & Abell 1689 & dV & all galaxies \\
3.98 & $_{0.49}^{0.64}$ & 18.01 & $_{0.43}^{0.34}$ & 0.168 & $_{0.02}^{0.023}$ & 20.00 & $_{0.15}^{0.13}$ & 0.12 & $_{0.17}^{0.15}$ & LM & Y & Abell 1689 & S & all galaxies \\
{4.11} & $_{0.2}^{0.21}$ & {18.13} & $_{0.14}^{0.13}$ & {0.1524} & $_{0.0085}^{0.009}$ & \multicolumn{2}{|c|}{{-}} & {0.276} & $_{0.1}^{0.096}$ & {DLM} & {Y} & {Both} & {dV} & {all galaxies} \\
{4.09} & $_{0.21}^{0.22}$ & {18.13} & $_{0.15}^{0.14}$ & {0.1584} & $_{0.0078}^{0.0084}$ & \multicolumn{2}{|c|}{{-}} & {0.279} & $_{0.103}^{0.097}$ & {DLM} & {Y} & {Both} & {dV+S} & {all galaxies} \\
4.23 & $_{0.27}^{0.31}$ & 17.84 & $_{0.14}^{0.12}$ & 0.1322 & $_{0.0091}^{0.0101}$ & 19.949 & $_{0.061}^{0.065}$ & \multicolumn{2}{|c|}{-} & LM & Y & Coma & dV & no BCGs \\
4.68 & $_{0.69}^{0.78}$ & 17.65 & $_{0.49}^{0.43}$ & 0.107 & $_{0.058}^{0.04}$ & 20.00 & $_{0.16}^{0.14}$ & 0.05 & $_{0.17}^{0.15}$ & LM & Y & Abell 1689 & dV & no BCGs \\
4.30 & $_{0.52}^{0.66}$ & 17.88 & $_{0.43}^{0.34}$ & 0.150 & $_{0.019}^{0.023}$ & 20.03 & $_{0.15}^{0.13}$ & 0.08 & $_{0.16}^{0.15}$ & LM & Y & Abell 1689 & S & no BCGs \\
\textbf{4.40} & $_\textbf{0.2}^\textbf{0.2}$ & \textbf{18.00} & $_\textbf{0.14}^\textbf{0.13}$ & \textbf{0.1312} & $_\textbf{0.0079}^\textbf{0.0085}$ & \multicolumn{2}{|c|}{\textbf{-}} & \textbf{0.238} & $_\textbf{0.098}^\textbf{0.094}$ & \textbf{DLM} & \textbf{Y} & \textbf{Both} & \textbf{dV} & \textbf{no BCGs} \\
\textbf{4.41} & $_\textbf{0.21}^\textbf{0.23}$ & \textbf{17.97} & $_\textbf{0.15}^\textbf{0.15}$ & \textbf{0.1424} & $_\textbf{0.0075}^\textbf{0.0078}$ & \multicolumn{2}{|c|}{\textbf{-}} & \textbf{0.212} & $_\textbf{0.103}^\textbf{0.099}$ & \textbf{DLM} & \textbf{Y} & \textbf{Both} & \textbf{dV+S} & \textbf{no BCGs} \\
\hline
\end{tabular}

\caption{Parameters and marginalised uncertainties for the fits to the Faber-Jackson and Kormendy relations for Coma and \A1689 as well as the CMR relation for \A1689. Model abbreviations are: MM=Mixture model; LM=Linear model; DLM=Double linear model. The column headed SEC shows whether a size evolution correction was applied (\S\ref{sec:fudges:sec}). The Profile column shows if the surface photometry (COG) was calculated using de Vaucouleurs (dV) or S\'ersic (S) profiles. Note that in the double linear fits made to Coma and \A1689 data, the Coma data is archival and was measured using de Vaucouleurs COGs while for the \A1689 data we had the option to fit S\'ersic or de Vaucouleurs profiles. Where BCGs were removed from the fitting procedure, we removed galaxies \#584 and \#610 from the \A1689 sample and GMP2921 and GMP3329 from the Coma sample. For the KR and FJR, results in bold are the ones we consider to be most reliable: note that without size evolution corrections the FJR and KR luminosity evolutions disagree (\nsigdisagree). Taking the average of (i.e. combining the samples for) the KR and FJR results corrected for size evolution while excluding BCGs gives $\Delta\beta=$\DeltaBeta (see \S\ref{sec:res:sumlumevol}) which we compare to stellar population models in \S\ref{sec:spmodels:slopegrad}}
\label{tab:res:fjr&kr}
\end{table*}

\subsection{The Faber-Jackson relation}
\label{sec:res:fjr}

Table \ref{tab:res:fjr&kr} presents the results for fits to the FJRs (Eq. \ref{eq:intro:fjr}) of Coma and \A1689 using individual linear models (LMs) and the double linear model (DLM). 

In \A1689, the calculation of the surface photometry for the central BCGs (brightest cluster or \emph{cD} galaxies: \#584 \& \#610) was compromised by contamination from other galaxies and the ICL. Similar arguments could be made for the two BCGs in Coma (GMP2921 \& GMP3329). Thus we fit the data both with and without the two BCGs and with and without correcting for size evolution (see \S\ref{sec:fudges:sec}).

The values for $\Delta \beta_\n{FJR}$ in Table \ref{tab:res:fjr&kr} were either calculated by subtracting the absolute magnitude of the FJR at $\log\sigma=2.2$ (where the FJRs are best constrained in individual LMs) or directly when fitting the DLM to both the \A1689 and Coma data. For nearly all cases, the galaxies in \A1689 are fainter in the F625W band than the Coma galaxies in the r\p-band. The evolution measured from the LMs and DLM generally agree, suggesting little bias from cuts or selection effects in the FJR.

Correcting for size evolution has a significant effect in decreasing the measured evolution between \A1689 and Coma, decreasing $\Delta \beta$ by around 0.25 \Mag; removing the BCGs increases $\Delta \beta$ by around 0.05 \Mag. 

Fig. \ref{fig:res:fjr&kr} shows the FJR measured for \A1689 in F625W compared to that measured for Coma in the r\p-band. Both de Vaucouleurs and S\'ersic surface photometry are shown for \A1689 but only de Vaucouleurs photometry is available for Coma. We have corrected the data points for size evolution and over plot the best fit DLM (omitting BCGs) in each case. The DLM fits the data well and both the Coma and \A1689 galaxies populate the same $\{\apmag, \sigma\}$ parameter space.

\subsection{Kormendy relation}
\label{sec:res:kr}

Table \ref{tab:res:fjr&kr} presents the results for the fits to the KRs (Eq. \ref{eq:intro:kr}) of Coma and \A1689 using LMs and the DLM, as was done for the FJR above. 

The values for $\Delta \beta_\n{KR}$ in Table \ref{tab:res:fjr&kr}, like $\Delta \beta_\n{FJR}$, were calculated both at $\log R_{e}=0.5$ from individual LMs, and directly from the DLM. As before, we favour the values calculated by the DLM. As for the FJR, galaxies in \A1689 are fainter in the F625W band than Coma galaxies in the r\p-band ($\Delta \beta > 0$). Unlike the FJR, there is a significant difference between the evolution measured from the LMs and the DLM: the KR is sensitive to magnitude cuts and selection effects which is unsurprising given that the cut runs almost parallel to the KR (unlike the FJR). Correcting for size evolution increases $\Delta \beta_\n{KR}$ only by around 0.05 \Mag, while including the BCGs has a negligible effect: clearly the KR is relatively robust to these two factors. 

Fig. \ref{fig:res:fjr&kr} shows the KR measured for \A1689 in F625W compared to that measured for Coma in the r\p-band. Both de Vaucouleurs and S\'ersic surface photometry are shown (for \A1689) and as before, we have corrected the data for size evolution and over plot the best fit DLM (omitting BCGs). The galaxies in \A1689 \emph{do not} populate the same $\{\langle\mu\rangle_{e}, R_{e}\}$ parameter space as the Coma galaxies: there appear to be more bright, compact galaxies in Coma. Recall that we have corrected the \A1689 galaxies for size evolution, thus making them larger and fainter, but we did not apply any cuts in $R_{e}$ or $\<\mu\>_{e}$ \emph{a priori}; the only cut was in absolute magnitude, which is well matched to the absolute magnitude cut in the Coma sample. Indeed, using \sex effective radii as an estimate for size, there is no difference between the size distributions of our spectroscopic sample and the parent sample with r\p$<19.5$.  Thus the different locations of the Coma and \A1689 data along the KR are not due to size selection effects.

\subsection{The effects of size evolution in the KR and FJR}

Without correction for size evolution and excluding BCGs, the measured $\Delta\beta_\n{KR}$ and $\Delta\beta_\n{FJR}$ disagree at the \nsigdisagree level. Although the need to correct for size evolution is at a modest statistical significance, reviewing the effect reveals why the FJR was more affected than the KR.

If we consider both the corrections to $R_{e}$ and $\<\mu\>_{e}$, using our knowledge of how these two are related (Eq. \ref{eq:intro:kr}) we can infer how they shift the whole relation up or down in $\<\mu\>_{e}$, which can be (mistakenly) interpreted as luminosity evolution of the stellar population. Using Eq. \ref{eq:fudges:sec:re} and Eq. \ref{eq:fudges:sec:mue} in Eq. \ref{eq:intro:kr}, we find
\begin{equation}
\mu_{e,z} = \alpha_\n{KR} \log R_{e,z}  + \beta_\n{KR} + \Delta\beta_\n{KR}^{\prime}
\end{equation}
with 
\begin{equation}
\label{eq:fudges:sec:dbetaKR}
\Delta\beta_\n{KR}^{\prime} = (\alpha_\n{KR}-5) \zeta \log(1+z).
\end{equation}
This $\Delta\beta_\n{KR}^{\prime}$ appears as a change in the surface brightness; it arises just from scaling the size of the galaxies and not from changes in the stellar populations. We see that the effects from changing $R_{e}$ and $\<\mu\>_{e}$ counteract with each other in the KR: the typical value for $\alpha_\n{KR}$ is $\sim4$, which with $\zeta=1$, gives $\Delta\beta_\n{KR}^{\prime} = -\log(1+z)$. This tells us that if we were to compare $\<\mu\>_{e}$ vs. $\log R_{e}$ for Coma to $\<\mu\>_{e,z}$ vs. $\log R_{e,z}$ for \A1689, even without any changes in the stellar populations, we would find that the \A1689 galaxies appear \emph{brighter} than the standard KR prediction by $\log(1+z)$.

Repeating this analysis for the FJR relation using Eq. \ref{eq:fudges:sec:sigma} in Eq. \ref{eq:intro:fjr}, we find a similar relation to before
\begin{equation}
\absmag = \alpha_\n{FJR} \log(\sigma_{z}) + \beta_\n{FJR} + \Delta\beta_\n{FJR}^{\prime}
\end{equation}
but this time with 
\begin{equation}
\Delta\beta_\n{FJR}^{\prime} = - 0.5 \alpha_\n{FJR} \eta \log(1+z).
\end{equation}
As with the KR, one could easily misinterpret $\Delta\beta_\n{FJR}^{\prime}$ as luminosity evolution of the stellar populations, but it is in fact just an effect of size evolution. However, unlike the KR, the change is quite substantial. Our typical $\alpha_\n{FJR}$ is $\sim-8$ and with $\eta=1$, gives $\Delta\absmag=4\log(1+z)$. If we were to compare $\absmag$ vs. $\log\sigma$ for Coma to $\absmag_{z}$ vs. $\log\sigma_{z}$ for \A1689 and there was no change to the stellar populations, we would still find \A1689 galaxies to be \emph{fainter} than the standard FJR prediction by $4\log(1+z)$. The canonical value for $\alpha_{FJR}$ is $\sim10$ ($L\sim\sigma^{4}$), which increases the effect to $5\log(1+z)$.

Thus size evolution has opposite effects in the KR and the FJR: it makes the KR of earlier galaxy populations appear \emph{brighter}, the FJR \emph{fainter}, and we observe these changes in $\Delta \beta _\n{FJR}$ and $\Delta \beta _\n{KR}$. 

\subsection{Luminosity evolution from the KR and FJR}
\label{sec:res:sumlumevol}

After correcting for size evolution, the differences between the KRs and FJRs of Coma and \A1689 ($\Delta \beta_\n{KR}$ and $\Delta \beta_\n{FJR}$) agree. To avoid propagating multiple calculations of $\Delta \beta$ any further, we now consider which values are our best estimates. 

The values of $\Delta \beta$ calculated using the DLM account for magnitude cuts and selection effects and are fully marginalised over the other model parameters; the values calculated from separate LMs to Coma and \A1689 are only marginalised over the model parameters fitted to each cluster, depend on where (on the x-axis) we compare the two clusters, and do not account for magnitude cuts or selection effects. We favour calculating $\Delta\beta$ using the DLM for these reasons. However, the DLM, as currently defined, cannot address downsizing. 

We should only consider values of $\Delta \beta$ which were calculated with size evolution corrections: without such corrections, $\Delta \beta_\n{KR}$ and $\Delta \beta_\n{FJR}$ disagree. Including BCGs reduces $\Delta \beta_\n{FJR}$ and has a negligible effect on $\Delta \beta_\n{KR}$, but the changes are small and within the uncertainties. The average $\Delta \beta$ \emph{with} BCGs is \avDeltaBetawBCG while the average without BCGs is \avDeltaBetawoBCG. We somewhat arbitrarily choose to use the values of $\Delta \beta$ calculated without the BCGs (see \S\ref{sec:disc:se}). Finally, $\Delta\beta$ measured using S\'ersic photometry in \A1689 is on average slightly larger than that measured using de Vaucouleurs photometry for \A1689, but the difference is far less than the uncertainties. Consequently, we take the average of the KR and FJR offsets (excluding BCGs) using both S\'ersic and de Vaucouleurs photometry as the overall luminosity evolution. We now seek the uncertainty in this value. The $\<\mu\>_{e}$ and $R_{e}$ data used in the KR are strongly correlated to the magnitudes used in the FJR so we cannot treat the uncertainties in $\Delta\beta_\n{KR}$ and $\Delta\beta_\n{FJR}$ as independent. 
We combine the MCMC samples for $\Delta\beta_\n{KR}$ and $\Delta\beta_\n{FJR}$ from the fits to S\'ersic and de Vaucouleurs surface photometry, thereby assuming equal evidence for all these models and measure the difference in luminosity between the Coma galaxies (measured in rest-frame r\p) and the \A1689 galaxies (measured in the observed-frame F625W) to be \DeltaBeta. We do not attempt to quantitatively justify the assumption of equal evidence as the results from the different models are very similar.

\section{Stellar population models}
\label{sec:spmodels}

As discussed in \S\ref{sec:intro}, changes in the scaling relations can be understood in terms of evolution of the stellar populations. In order to infer such evolution from the observations, it is necessary to make use of stellar population synthesis models which reproduce the SED given some assumptions regarding that population, such as the age, the IMF, the metallicity (Z) and the star formation history (SFH). The latter describes the distribution of star formation over time and in our case is assumed to be a delta function at some previous time (i.e. a simple stellar populations or SSPs), so has a single characteristic age.  

It is beyond the scope of this paper to perform an exhaustive comparison of all the possible models now available. We use two common population synthesis models (\citealt{BC03} and \citealt{M05}, hereafter BC03 and M05) to study the evolution of the scaling relations. Both provide SSP SEDs\footnote{http://www.cida.ve/$\sim$bruzual/bc2003 (BC03) and http://www.icg.port.ac.uk/$\sim$maraston/ (M05)} spanning a variety of ages, metallicities and IMFs, although we only consider models using a Salpeter IMF.  

We use these models together with the filter curves provided by BC03 (SDSS and Gunn bands) and the GEMINI Observatory\footnote{http://www.gemini.edu/sciops/instruments/gmos/} (GMOS-N r\p and g\p) and the Vega and Solar spectra provided by the HST calibration database (CALSPEC\footnote{http://www.stsci.edu/hst/observatory/cdbs/calspec.html, post Feb. 2010 update.}). We make use of the {\sc cosmolopy}\footnote{http://roban.github.com/CosmoloPy/} python library to calculate look back times and redshifts given our cosmological parameters (see \S\ref{sec:intro}). The model magnitudes are calculated so as to be comparable to observations that are K-corrected for bandpass effects ($K_{b}$), but not for colour effects ($K_{c}$).  

Observations of objects at different redshifts equate to observations in different frames of reference (observer, rest), while all the model calculations are performed in the rest frame. This, together with the different magnitude systems (Vega, AB, Gunn) and the different \emph{epochs} at which we observe our galaxies, need particular attention. Consequently, where necessary, we label the specified filter with the magnitude system, the frame of reference and the epoch of the observation in parenthesis. For example, F625W(AB, z=0.0, T) refers to AB magnitudes of local objects in the F625W filter, while F625W(AB, \ZA1689, T-\dage) refers to observations in F625W for an object at \ZA1689 (the look back time for an object at \ZA1689 is \dage for our adopted cosmology). The difference F625W(AB,\ZA1689,T-\dage) - r\p(AB,z=0.0,T) is what we measure when comparing our \A1689 data to the Coma data (i.e. $\Delta\beta$). We also find it useful to quote model values for which we do not have observations: F625W(AB, \ZA1689, T) refers to observations in the blueshifted F625W filter at the current age of the Universe and so F625W(AB, \ZA1689, T)-F625W(AB, \ZA1689, T-\dage) is just the luminosity evolution between now and \dage ago in the blueshifted F625W filter bandpass (approximately equivalent to the V-band). 

These stellar population models can be used in many ways, but we choose two distinct and very different techniques for the CMR analysis and KR/FJR analysis. We present the BC03 and M05 results for modelling the passive evolution of SSPs as a function of age and various metallicities in both cases. 

\subsection{Analysis of the g\p-r\p CMR}
\label{sec:spmodels:cmrscatter}

\begin{figure}
   \centering
   \includegraphics[width=0.5\textwidth]{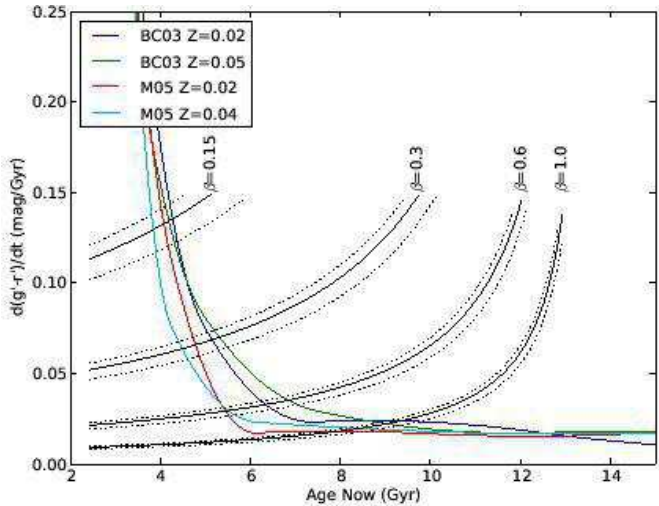} 
   \caption{The rate of change of colour vs. the age of the stellar population for the gmos g'-r' colour using BC03 and M05 models for solar and super-solar metallcities. The x-axis represents $t_\n{f}+$\dage, which is the age \emph{now}, rather than the age at \ZA1689; this is to allow easy comparison with Fig. \ref{fig:res:rr_ekcor}.}
   \label{fig:res:betaplot}
\end{figure}

As discussed in \S\ref{sec:intro}, the intrinsic scatter in the CMR tells us about the SFH of the galaxies. We can perform a similar analysis to BLE92 for \A1689, but using state of the art population synthesis models.  Fig. \ref{fig:res:betaplot} shows the rate of change of (g\p-r\p) as a function of the age of the stellar population for the BC03 and M05 models with solar and super-solar metallicities. Like BLE92 and BKT98, we smoothed the models; the resulting systematic error was no more than 0.1 \Mag.  Our SSP tracks in Fig. \ref{fig:res:betaplot} are broadly similar to those of BLE92 and BKT98, but the M05 models show a more rapid fall in the rate of change of colour before $<$4 \Gyr, followed by very slow or even negligible further reddening beyond $>4$ \Gyr. 

Following BLE92 and BKT98, we model the formation of galaxies with a uniform distribution of random events distributed over a time interval $\Delta t$ with mean age (now) $t_\n{f}$. We parameterise the time interval $\Delta t$ as a fraction, $\beta$, of the time available up to the end of $\Delta t$; when galaxies form between times $t_\n{start}$ and $t_\n{stop}$, $\Delta t=t_\n{stop}-t_\n{start}$, and $\Delta t=\beta t_\n{stop}$ with $0 < \beta < 1$ (when $\beta=1$, galaxy formation is continuous from the start of the Universe to some final time $t_\n{stop}$ so we have the largest possible scatter in ages and no synchronisation). We can write $\Delta t$ in terms of $t_{f}$, 
\begin{equation}
\Delta t =  \frac{t_\n{univ} - t_{f}}{\beta^{-1}-1/2}
\end{equation}
where $t_\n{univ}$ is the age of the Universe now.

As noted by BLE92, if the slope of the CMR is produced by variation in metallicity with luminosity and all galaxies have have identical IMFs, then the RMS scatter of the CMR, $\sigma_\n{CMR}$ is related to the RMS scatter in the formation ages of the RS galaxies, $\sigma_\n{age}$ by 
\begin{equation}
\label{eq:colrate}
\frac{\sigma_\n{CMR}}{ \sigma_\n{age}} \approx \frac{\partial (\textrm{g}^\prime-\textrm{r}^\prime)}{\partial t} = f(\textrm{age})
\end{equation}
where the RMS scatter $\sigma_\n{age}\approx\Delta t / 3.5$ for the uniform distribution considered here (see BLE92 \& BKT98).

The observed scatter in the CMR can therefore tell us about the scatter in the formation times (synchronisation, $\beta$). But as galaxies become redder with time and the rate of change of colour decreases, a small scatter in the CMD could also imply a very old population with no synchronisation. Eq. \ref{eq:colrate} can be written
\begin{equation}
\label{eq:betalink}
\frac{ \partial (\textrm{g}^\prime-\textrm{r}^\prime)}{\partial t} \approx 
3.5 \sigma_\n{CMR} \left( \frac {\beta^{-1}-1/2} {t_\n{univ} - t_{f}}\right)
\end{equation}
For a given $\beta$, there is an age, $t_\text{f}$, which is compatible with the observed scatter $\sigma_\n{CMR}$. In Fig. \ref{fig:res:betaplot}, we plot the LHS of Eq. \ref{eq:betalink} (curves of  $\partial (\textrm{g}^\prime-\textrm{r}^\prime)/\partial t$ derived from stellar population models) and the RHS of Eq. \ref{eq:betalink} (``beta'' curves). Where these curves intersect gives the model age \emph{now} that satisfies Eq. \ref{eq:betalink}. While BLE92 assumed that the variation in the rate of change of colour with metallicity was sufficiently small that they could consider only one metallicity (in effect assuming that the slope of the CMR was zero), we investigate solar and super solar metallicities. Fig. \ref{fig:res:betaplot} shows all these curves for the BC03 and the M05 models. Significant differences only exist for younger ages (i.e. for $\beta \sim 0.6$). We show the 68\% uncertainty limits for individual beta curves as dotted lines (based on the uncertainty in $\sigma_\n{CMR}$).

Fig. \ref{fig:res:betaplot} shows that $\sigma_\n{CMR}$ in \A1689 is compatible with a large range of ages, depending on the degree of synchronisation. BLE92 and BKT98 found the same for Coma and that they argued that younger ages and smaller $\Delta t$ are increasingly unlikely because such high SFRs would be easily visible but are not observed. However, we observe \A1689 some \dage earlier than Coma: combining both results as though we had observed the same cluster twice over a \dage interval adds another constraint: if the scatter in the CMR has not significantly changed over the last \dage, then the rate of change of colour also cannot have changed appreciably over \dage. To try and put an upper limit on how much the rate of change of colour could have changed over these \dage, let us consider a lower limit on the Coma CMR scatter of 0.036 \Mag (the scatter for \emph{just} ellipticals within 600 Mpc in BTK98). The scatter of all RS galaxies \dage earlier (in \A1689) is measured to be 0.054 \Mag; this gives a maximum rate of change of -0.008 \Mag/\Gyr for the CMR scatter, or -0.028 \Mag/\Gyr for the rate of change of colour. In Fig. \ref{fig:res:betaplot}, models that allow for such a small change over \dage have ages $>6$ \Gyr and $\beta>0.6$. According to the exact intersections in Fig. \ref{fig:res:betaplot}, the galaxies in the RS of Coma and \A1689 formed between \CMRformz (have \CMRage), if the \A1689 RS evolves into the Coma RS. The scatter could be overestimated in \A1689 (we do not select galaxies by morphology or recession velocity), so the upper limit on age, like the lower limit on the scatter, could be questioned; thus we are only confident that the galaxies in the RS of Coma and \A1689 are \CMRagelowlim old (formed at \CMRformzlowlim). Similarly, only models with $\beta>0.6$ are consistent with this view.

\subsection{Analysis of the KR and FJR}
\label{sec:spmodels:slopegrad}

\begin{figure*}
   \centering
   \begin{minipage}[c]{0.5\textwidth} 
   \includegraphics[width=\textwidth]{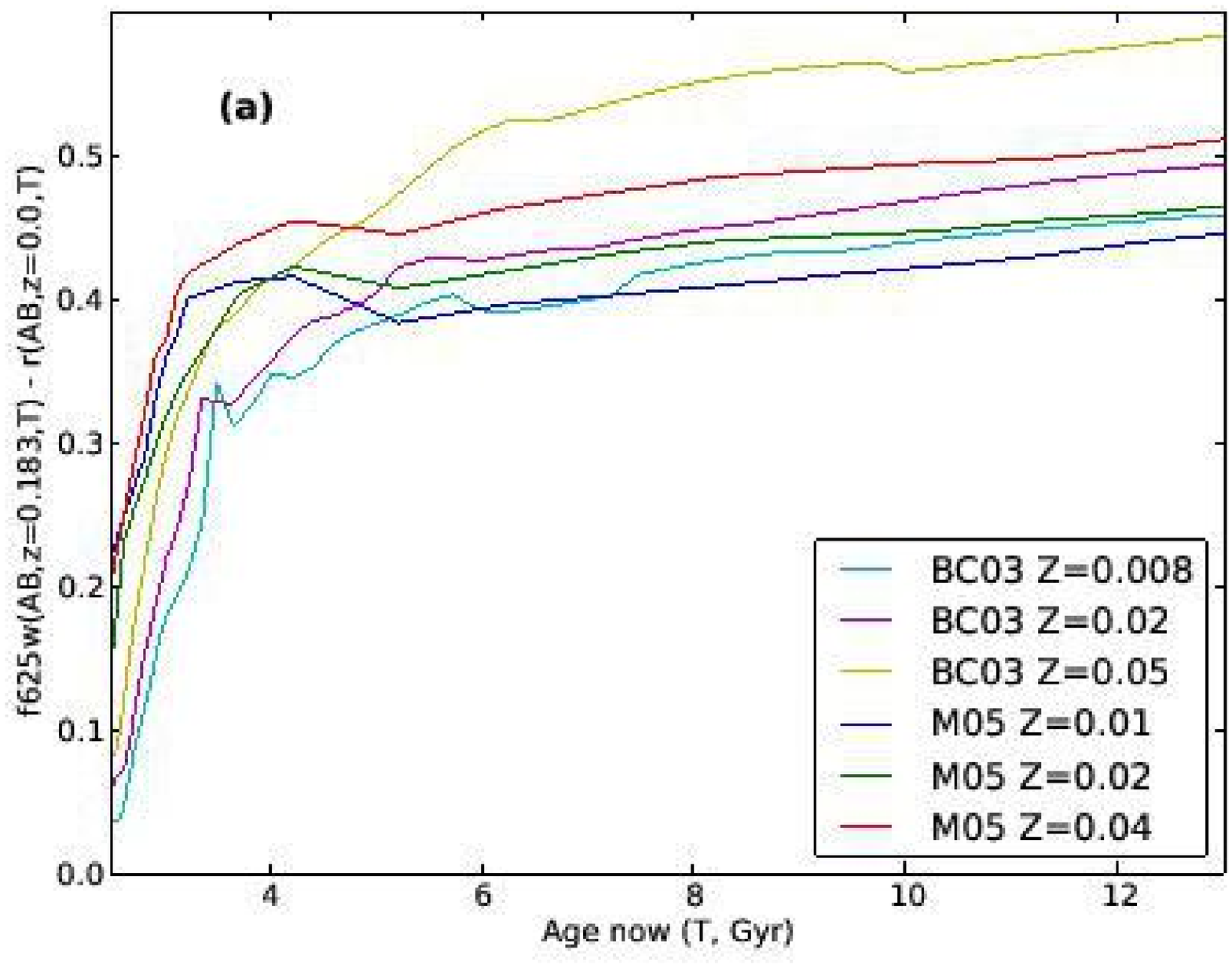} 
   \end{minipage}\begin{minipage}[c]{0.5\textwidth} 
   \includegraphics[width=\textwidth]{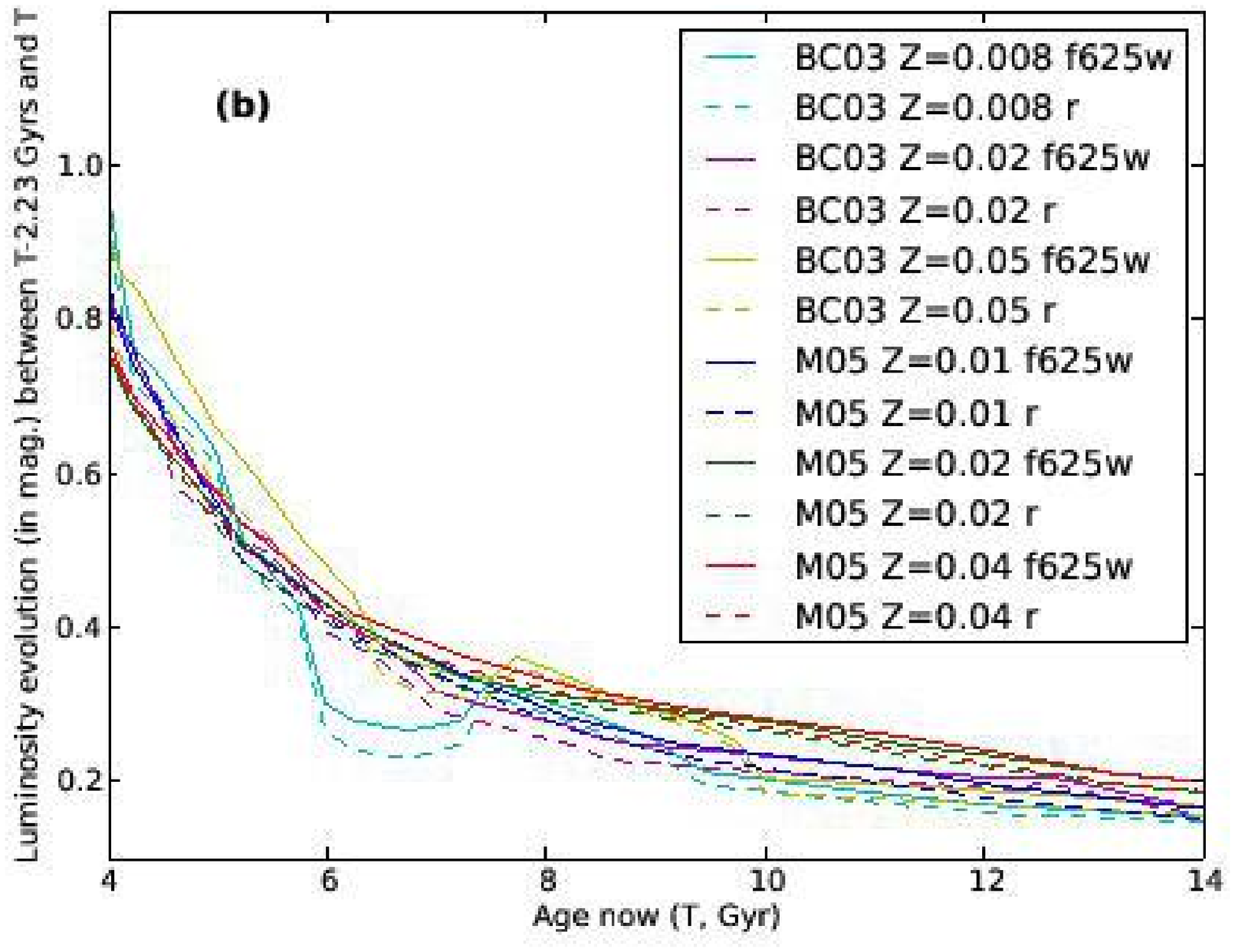} 
   \end{minipage}
   \begin{minipage}[c]{0.5\textwidth} 
   \includegraphics[width=\textwidth]{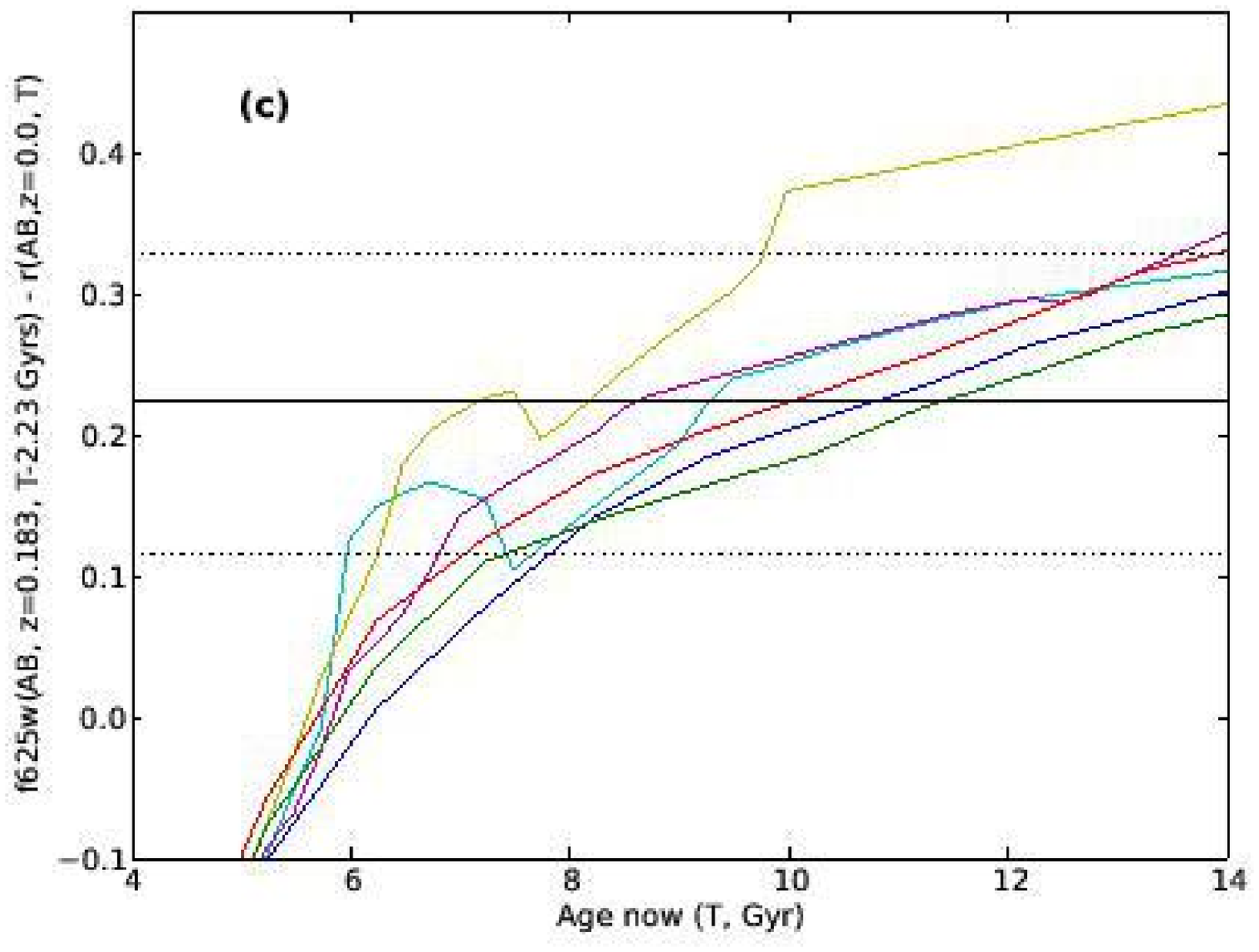} 
   \end{minipage}\begin{minipage}[c]{0.5\textwidth} 
   \includegraphics[width=\textwidth]{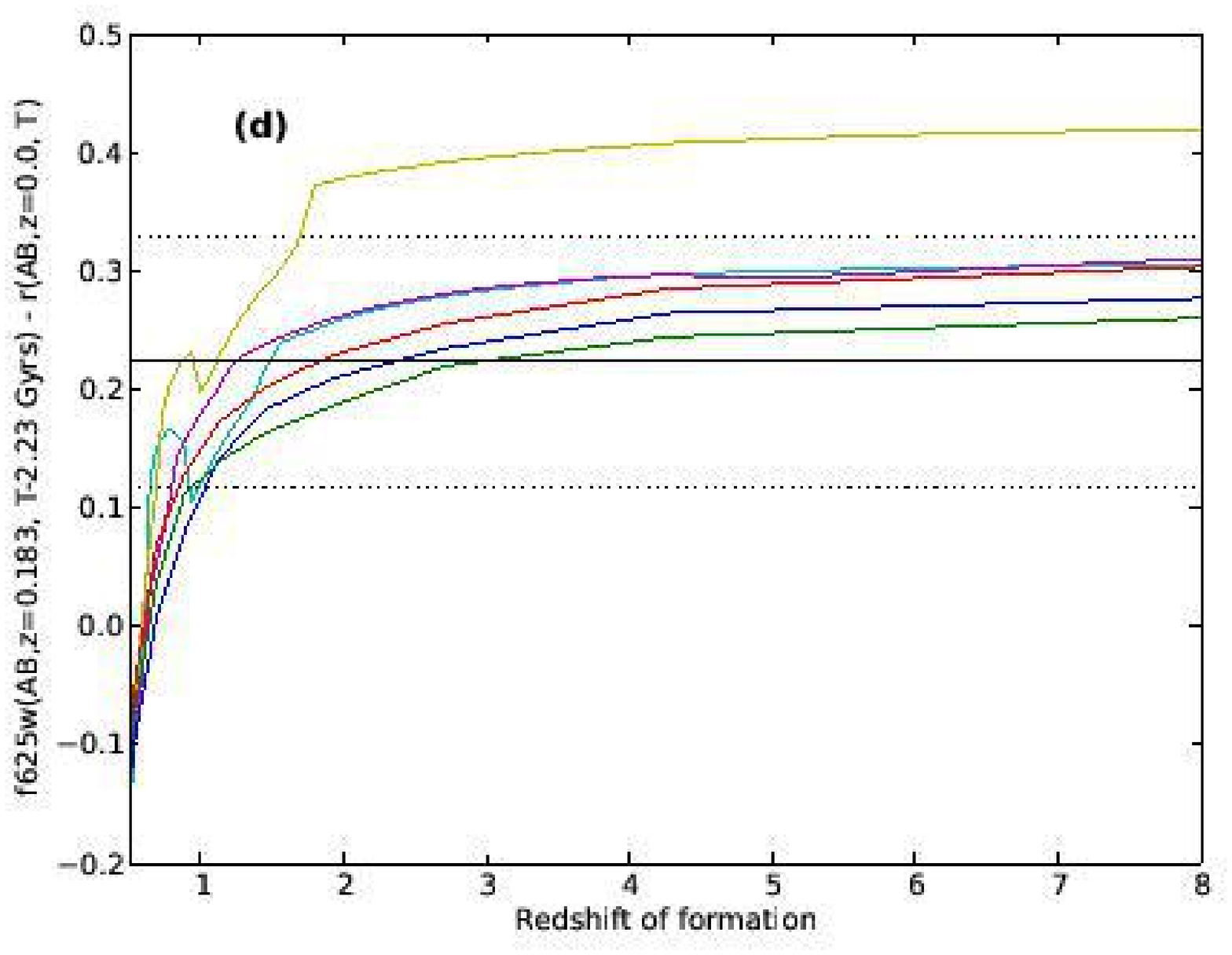} 
   \end{minipage}
   \caption{\bi{(a)}: the change in F625W(AB, \ZA1689, $T$)-r(AB, z=0.0, $T$) colour as a function of age, $T$; \bi{(b)}: the change in luminosity for the r(AB, z=0.0, $T$) and F625W(AB, \ZA1689, $T$) filters as a function of $T$; \bi{(c)}: the difference between F625W(AB, \ZA1689, $T$-\dage) and r(AB, z=0.0, $T$) as a function of $T$, line colours are as in (a), solid and dashed lines show the evolution and associated uncertainty inferred from the FJR and KR; \bi{(d)}: like (c), but as a function of formation redshift, line colours are as in (a). \bi{All}: magnitudes are in the AB system; two different population synthesis models (BC03 and M05) are shown for sub-solar, solar (Z=0.02) and super-solar metallicities. All models are consistent with galaxy ages \ageALLMOD. The BC03 $Z_{\odot}$(=0.02) models suggest ages of \ageBC03SUN while the M05 $Z_{\odot}$ models suggest ages of \ageM05SUN. }
   \label{fig:res:rr_ekcor}
\end{figure*}

Our principle goal is to determine what (if any) luminosity and colour evolution has taken place between now and \ZA1689. We use the BC03 and M05 models to calculate the evolution of F625W(AB, \ZA1689, $T$) - r(AB, z=0.0, $T$) as a function of the age, $T$ (roughly a V-r colour). This is shown in Fig. \ref{fig:res:rr_ekcor}(a) and shows that after $\sim4$ \Gyr, F625W(AB, \ZA1689, $T$)-r(AB, z=0.0, $T$) for the model populations changes slowly from 0.4 to 0.5 \Mag at 13 \Gyr, so any ETG older than 6 \Gyr \emph{now}, would not be much bluer at \ZA1689. Note that this colour is equivalent to the colour term of the K-correction in Eq. \ref{eq:fudges:kcor}, $K_{c}$: there is variation with age and metallicity, which is why we do not try to apply it to our measurements. We calculate the luminosity evolution over \dage for F625W(AB, \ZA1689, $T$) and r(AB, z=0.0, $T$) in Fig. \ref{fig:res:rr_ekcor}(b): here we see the evolution is very similar for all models after 4 \Gyr (to within $\sim$0.1 \Mag) and drops rapidly with age. Finally, we calculate the luminosity evolution we measure when we compare the KRs and FJRs of Coma and \A1689 in Fig. \ref{fig:res:rr_ekcor}(c) and Fig. \ref{fig:res:rr_ekcor}(d): the evolution of F625W(AB, \ZA1689, $T$-\dage) - r(AB, z=0.0, $T$) is dominated by the change in luminosity and not colour after 4 \Gyr, according to panels (a) and (b). After 6 \Gyr, the solar metallicity models show the slowest rate of change and there is significant difference between the BC03 and M05 solar metallicity predictions; because lookback time is not linearly related to redshift, the evolution with respect to \emph{redshift} is almost asymptotically slow in (d).

In Fig. \ref{fig:res:rr_ekcor}(c) and (d) we also over-plot the measured luminosity evolution ($\Delta \beta$) found between the galaxies of Coma and \A1689 (see \S\ref{sec:res:sumlumevol}). The various stellar population models predict different ages for the Coma and \A1689 galaxies from this measurement: considering all models and the $1\sigma$ limits, the luminosity evolution between Coma and \A1689 is consistent with an age \ageALLMOD. Limiting the models to solar metallicities, the BC03 model suggests an age of \ageBC03SUN, while the M05 model suggests an age of \ageM05SUN: there is a discrepancy between the ages inferred from the BC03 and M05 solar metallicity models, the latter being older by \deltaageB03M05SUN, although both lower limits are comparable. In the absence of any reasonable cause to discriminate between the BC03 and M05 models, we average the ages and uncertainties (i.e. combine the samples, assuming equal evidence for the different models) to conclude that the galaxies in Coma and \A1689 are now \age old (i.e. formed at \formz). Note that because the M05 models are only calculated to 15 \Gyr and the $1\sigma$ upper limit on $\Delta \beta$ is greater than the model predictions at that age, we can only quote the upper limit from the BC03 model.

It is customary to quote the luminosity evolution in the same band and our stellar population models allow us to calculate this. The BC03 and M05 solar metallicity models both give a luminosity evolution of \LumEvolSUNHST in the F625W(AB,z=0.0) band and \LumEvolSUNrprime in the r\p(AB,z=0.0) band for their respective ages.

\section{Discussion}
\label{sec:disc}

We now discuss the context of our findings and highlight relevant caveats so that the reliability of our approach can be judged. We discuss the CMD first, followed by the Faber-Jackson and Kormendy relations and summarise the findings at the end.

\subsection{The CMR of \A1689}
\label{sec:disc:cmd}

The scatter in the CMR of \A1689, $\sigma_\n{CMR}=$\CMRscatterA1689 (for all galaxies with M$_{V}<17.9$ \Mag within a projected radius of 570 kpc from the cluster centre) is comparable to the bi-weight scatter quoted by \citet{Terlevich2001} of $0.063\pm_{0.01}^{0.008}$ \Mag. Combining these results, we break the $\beta$--age degeneracy and show that the galaxies in both clusters have an age \CMRagelowlim (and formed at \CMRformzlowlim) and had little or no synchronisation in their formation ($\beta>0.6$). We now ask if this result is in agreement with other clusters at other redshifts. Then we compare Coma and \A1689 to judge if we are making a fair comparison. We also discuss the usefulness of placing \emph{lower limits} on the CMR scatter and finally explain why certain caveats in our approach should not strongly bias the outcome.

\subsubsection{\A1689 in perspective}

As discussed in \S\ref{sec:intro:cmrscatter}, \citet{Stanford95}, \citet{Ellis97}, \citet{Stanford98} and most recently \citet{Mei09} have studied the CMR scatter in clusters up to $z\sim1.3$. The last two authors used a clipped bi-weight estimator to measure the CMR scatter; to our knowledge, we are the first to model and extract the CMR scatter using a mixture model which does not rely on arbitrary cuts, clipping thresholds or parameter tuning. Furthermore, we present a CMR scatter that is marginalised over all reasonable slopes and intercepts, whereas previous work quotes values \emph{given} a measured slope and intercept. That said, the findings of these papers agree with our work: the scatter in the CMR has remained relatively unchanged in the ETG population of massive galaxy clusters since z$\sim1.3$. Both \citeauthor{Stanford98} and \citeauthor{Mei09} found the scatter in their clusters to be nearly always less than 0.1 \Mag with the average being around 0.06 and 0.05 \Mag, respectively. Both authors were also careful to discuss how selection effects may bias the determination of the CMR scatter at higher redshift (so-called \emph{progenitor-bias}): they selected galaxies morphologically to be Es and S0s, so mergers or spirals were excluded from the sample. We know that more (blue) spirals exist in clusters at higher redshift \citep{ButcherOemler84} so it is reasonable to consider that the population of galaxies on the CMR at higher redshift is not the same population observed at lower redshift. Furthermore, clusters grow over time and more field ellipticals will join the cluster later on, producing a similar bias. But the bias is less here, because we do not morphologically select our galaxies and at \ZA1689 we are only looking \dage into the past, which is less time for change in the ETG progenitors compared to Coma. 

\subsubsection{Lower limits on the CMR scatter}

Until now, no one has tried to interpret a \emph{lower} limit on the scatter of the CMR. We do so now tentatively, because we know that that scatter is likely to be smaller for just ellipticals and also for samples limited to more luminous galaxies. Taken at face value, the lower limit of the scatter measured here suggests that, with $\beta=1.0$, the highest possible mean formation age for the RS galaxies of \A1689 is 9.8 \Gyr in the past ($z=1.7$, a $1\sigma$ upper limit). This may be meaningless because we know we have mixed populations (ellipticals, S0s etc.) which increase the scatter, but in future surveys it may be of interest to consider an upper bound on the mean formation redshift if we wish to dedicate effort to witnessing it. If all the stars in the Coma ETGs formed in the first \Gyr after the big bang ($t_\n{f}=13.2$ \Gyr, $\beta=1.0$), the CMR scatter would be $\sim$0.005 \Mag\ -- seven times smaller than that of Coma. It takes time for dark matter (DM) to cluster sufficiently to allow the baryons to start forming stars, but the difference between a scatter of 0.036 \Mag and 0.005 \Mag in the CMR of Coma implies there was a delay of $\sim$3 \Gyr between the big bang and the average SF period for rich clusters (assuming $\beta=1$). This suggests that the majority of stars formed for $z<3$ which is in agreement with the cosmic star formation history \citep{Madau96,Madau98}.

\subsubsection{Caveats}

BKT98 investigated exponentially declining SFHs truncated at different times and also looked at the effects of merging. They found the same conclusions as BKT92, validating the SSP approach used there (and here). We treat the random errors in magnitude and colour as independent, but in reality they will be correlated: noise in the central aperture of the r-band images will affect the colour term as well as the total r-band magnitudes. This will have minimal effect on our scatter measurement, but could slightly affect the uncertainties. We investigated the assumption that \A1689 will evolve into Coma in \S\ref{sec:ComaVsA1689}.

\subsection{The Faber-Jackson and Kormendy relations}

After correcting for size evolution, magnitude cuts and selection effects, analysis of the FJR and KR in \A1689 shows that the \A1689 galaxies in F625W are fainter than the Coma galaxies in rest frame r\p-band by \DeltaBeta, which suggest and age of \age (\formz). We now compare these findings with the literature and discuss the importance of correcting for size evolution.

\subsubsection{Evidence for the passive evolution of ETGs}

The luminosity evolution in the KR and FJR in Coma and \A1689 is consistent with a passively evolving stellar population from single burst \AGEav ago. This is in agreement with the long-standing findings that ETGs both locally and at moderate redshift appear to be dominated by old passively evolving stellar populations that formed at $z>1$. The majority of precision age determinations stem from measurements of absorption line indices: although the broad-band colours of the Coma ETGs have been known to be compatible with old populations for a long time (see \S\ref{sec:intro:slopegrad}), the well documented age--metallicity degeneracy limits this approach for local clusters, but the degeneracy can be reduced by modelling the absorption line indices of the metals. Little work has been done specifically on the galaxy populations of \A1689: \citet{Mieske04} studies the ultra-compact dwarf fraction and luminosity function of the cluster galaxies while \citet{Banados10} study the faint end of the luminosity function. \citet{Carretero07} studied line indices of ETGs in 4 massive clusters at $z\sim0.2$ and found ages of around 10--15\Gyr, but their sample did not include \A1689. Considerably more work has been done on the galaxy populations of Coma, in particular with regard to determining the stellar population ages \citep{Clemens09,Jorgensen99,Mehlert03,SanchezBlazquez06,SanchezBlazquez09,Rakos07,Trager08,Smith09,Harrison10,Harrison11,Price11}. There is still some uncertainty regarding the ages (and presence of multiple populations), but the current evidence suggests ages between 5 and 10 \Gyr for the Coma ETGs. Our age estimate from the broad band photometry of \A1689 and Coma is therefore towards the top of this range, though the BC03 result alone (\ageBC03SUN) is nearer the midpoint and the M05 result at the upper end (\ageM05SUN); intriguingly, the age range implied by just the scatters of the CMRs in \A1689 and Coma (\CMRage) brackets the literature ages very well, even though earlier we expressed doubt regarding this upper limit (see \S\ref{sec:spmodels:cmrscatter}). 

\subsection{The effects of size evolution}
\label{sec:disc:se}
The measured luminosity evolution between \A1689 and Coma agrees with that of other clusters \citep{Barrientos96,Pahre96,Ziegler01,LaBarbera03,Fritz05}, although none of these studies corrected for size evolution. Most did not simultaneously study both the FR and FJR except for \citeauthor{Fritz05} who found that the galaxies appeared brighter in the KR than in the FJR (but the 0.07 \Mag difference is considerably less than the 0.3 \Mag found here). \citet{Saglia10} allow for size evolution in their study of the FP out to $z\sim1$. However, we are not probing a high redshift cluster, but a comparatively local cluster at \ZA1689. The fact that we need size evolution to bring the KR and FJR into agreement suggests that whatever causes it is a gradual process and its effects are not limited to the high-z universe; indeed, the mechanism is probably still in action now and does not appear to have deviated from the high-$z$ $(1+z)^{\zeta,\eta}$ laws. 

While there has been a reasonable passage of time between \ZA1689 and now (\dage), observing the size evolution taking place \emph{in the cluster environment} and \emph{in the local Universe} place important restrictions on models that wish to explain size evolution, which we now discuss.
\begin{description}
\item[\em Merging:] Wet or dry major merger scenarios \citep[e.g.][]{Khochfar&Silk06,Hernquist93,Hopkins09e} are unlikely to occur in the cluster environment because cluster galaxies have a very high relative velocity (larger than the typical escape velocity of two massive galaxies) making it prohibitively difficult to merge galaxies. An exception are central BCGs which would likely merge when multiple clusters merge as they lie at rest at the bottom of the potential. Minor or late accretion of smaller galaxies \citep[e.g.][]{NaabTrujillo06,Maller06,Naab09,Hopkins09d,Hopkins09g} could be more feasible because of the larger number of low-mass satellites, but in general the relative velocities would still be higher than the escape velocity unless the impact was at glancing angle.

\item[\em Accretion:] Accretion of the intra-cluster stars (previously stripped from other galaxies) may be energetically plausible, if not in mass transfer rates. However, while the simulations of \citet{Bois2011} and \citet{Khochfar2011} suggest that ATLAS3D \emph{slow rotators} grow from minor accretions, \citet{Stott2011} finds no evidence of size evolution in BCGs up to $z<1$.

\item[\em Adiabatic expansion:] The action of AGN expelling gas as proposed by \citealt{Fan08} is perhaps unfeasible here because we know that massive reservoirs of gas are stripped when galaxies fall into a cluster \citep[observed for gas rich spirals falling into the cluster potential today,][]{Chung09}. However, the mass lost from evolving stars could be gradually stripped by the ICM. Adiabatic expansion could also occur when central mass (dark or luminous) is stripped or \emph{harassed} to larger radii by high speed encounters, which are inevitable in clusters. 

\item [\em Secular/Cosmological expansion:] There are few secular arguments (the action of bars and/or resonances) for size evolution in the literature at present and perhaps these should be investigated further given our findings: if S0s are believed to be faded spirals \citep[the vast majority of Es show fast rotating disc-like disk kinematics,][]{SAURONIX,ATLAS3DIII}, then whatever action reduces disks to spheroids could also be responsible for the size evolution we observe. Recall also that size evolution in disks was predicted as a consequence of them being truncated to the critical density \citep{Mo98}, which is entirely a secular/cosmological effect. 
\end{description}

\subsubsection{Caveats}
Some caveats to consider for our KR and FJR analysis include not correcting for colour gradients, invoking size evolution, \emph{not} selecting galaxies morphologically or to be on the RS, and using non-independent PDFs as though they were independent. We have already investigated the assumption that \A1689 will evolve into Coma in \S\ref{sec:ComaVsA1689}.

An important difference between the FJR and KR is that the latter is potentially sensitive to internal colour gradients; this may be a source of error in our work given that our F625W observations of \A1689 are approximately rest-frame V-band compared to the Coma r-band observations. The colour gradients in ETGs cause them to be redder towards the centre; this makes $R_{e}$ larger at bluer wavelengths. Using the Coma data from \citet{Jorgensen95a}, we find that $R_{e}$ in Gunn g is a mean of $\sim5\pm15$\% (or a median of 2\%) larger than $R_{e}$ in Gunn r. However, if one considers the combined effects in the KR, we find that they cancel to first order. Using the same s.ps that gave rise to Eq. \ref{eq:fudges:sec:dbetaKR}, increasing $R_{e}$ by 5\% leads to an \emph{apparent} luminosity evolution of $\log(1.05)=0.02$ \Mag in the KR. If we calculate the difference between the central colour ($\<\mu\>_{e}(\rm{g})-\<\mu\>_{e}(\rm{r})$) and the global colour (g-r) from the \citet{Jorgensen95a} data, we find a mean $[\<\mu\>_{e}(\rm{g})-\<\mu\>_{e}(\rm{r})] - [g-r]$ = of $0.05\pm0.35$ \Mag (median of 0.07 \Mag). Allowing for the large scatter, this agrees with the above and is a negligible correction. Therefore we conclude that it is not necessary to correct for colour gradients.

We have invoked the existence of size evolution to make sense of the different luminosity evolution seen in the KR and FJR. However, we have chosen a prescription for how the size and internal kinematics of the galaxies change with time based on the evidence to hand (\S\ref{sec:fudges:sec}). Although this is the simplest approach, one could try to measure this prescription from the data by making the simple assumption that variation in the mass-to-light is determined only by passive evolution of the stellar population, is independent of galaxy luminosity (or stellar/dynamical mass) and is the same for the KR and the FJR. 

Readers may have noticed that we do not select \A1689 galaxies in the KR and FJR to be morphological E/S0s or to be on the RS (even though we have that information from the CMR analysis). This is deliberate: we do not want to bias ourselves away from including (blue, spiral) progenitors of ETGs in Coma. However, blue spirals tend to exclude themselves from our sample because they have velocity dispersions below our spectral resolution (most likely because the luminosity is dominated by light from central dynamically cold star forming regions). Of the \nsample galaxies that have sufficient quality spectra and HST imaging to enter the fitting procedures, only one is found \emph{not} to be on the RS (\#435). Thus although we tried not to select just RS galaxies, we end up being significantly biased towards them.

When calculating the difference between the evolution seen in the KRs and the FJRs, we treated the PDFs as though they were independent normal distributions. Neither of these assumptions is true (the Coma and \A1689 data used in both are obviously correlated), but it is the best approach we are reasonably able to take.

\renewcommand{\labelenumi}{(\roman{enumi})}

\section{Conclusions}
\label{sec:conc}
We have presented the Kormendy, Faber--Jackson and colour-magnitude relations for ETGs in \A1689 using HST/ACS imaging, GEMINI/GMOS imaging and GEMINI/GMOS spectroscopy and conclude: 
\begin{enumerate}
\item The intrinsic scatter in the colour--magnitude relation of \A1689 places degenerate constraints on $\beta$ (the ratio of assembly timescale and time available) and the age of the population: specifically, it is consistent with the galaxies in the colour--magnitude relation having either formed randomly at high redshift ($z\gtsim2$), or at lower redshift with increasing synchronisation (smaller assembly timescales). However, assuming the intrinsic scatter of Coma and \A1689 is that of the same cluster observed twice over an interval of \dage breaks this degeneracy and limits $\beta$ to be $>0.6$ (little or no synchronisation) and the age of the red sequence to be \CMRagelowlim (formed at \CMRformzlowlim).

\item After accounting for size evolution effects, the F625W Kormendy and Faber--Jackson relations both show a similar change in luminosity compared to the rest frame r\p-band relations of Coma: \DeltaBeta. This is consistent with passive evolution of the stellar populations from a single burst of star formation long ago: the galaxies in Coma and \A1689 have an SSP age of \age (i.e. formed at \formz), which agrees with the CMR analysis above. However, not accounting for size evolution causes the Kormendy and Faber--Jackson relations to be inconsistent and they then disagree about the amount of luminosity evolution at the \nsigdisagree level.

\item We therefore find weak evidence that size evolution appears to have taken place in the cluster environment in the last \dage; if true, this places interesting constraints on the models, favouring harassment or secular mechanisms over major or minor merger scenarios.
\end{enumerate}

\appendix

\section{The Curve of Growth (COG) Technique and Associated Errors}
\label{sec:ap:cog}

The COG method we use is in essence very simple: locate the centre of the galaxy via some means (in our case, via ellipse fitting); integrate out in radius to create a curve-of-growth function; fit an analytical curve-of-growth (Eq. \ref{eq:surf:sersiccog}), varying the model parameters (apparent magnitude, $\apmag$; effective radius, $R_{e}$ and S\'ersic index $n$) to minimise the square of the residuals ($\chi^{2}$):
\begin{equation}
\chi^{2} = \sum_{i=1}^{N} \left[ D_{i} - M_{i}\right]^{2}
\end{equation}
where the sum is performed over samples $i$ for the observed COG $D_{i}$ and model COG $M_{i}$. We sample the COG at the plate scale of the input image.

We now discuss how we deal with: PSF effects, contamination and masking of other nearby galaxies, and estimating our uncertainties.

\subsection{PSF effects}
\label{sec:ap:cog:psf}

\begin{figure}
   \centering
   \includegraphics[width=0.5\textwidth]{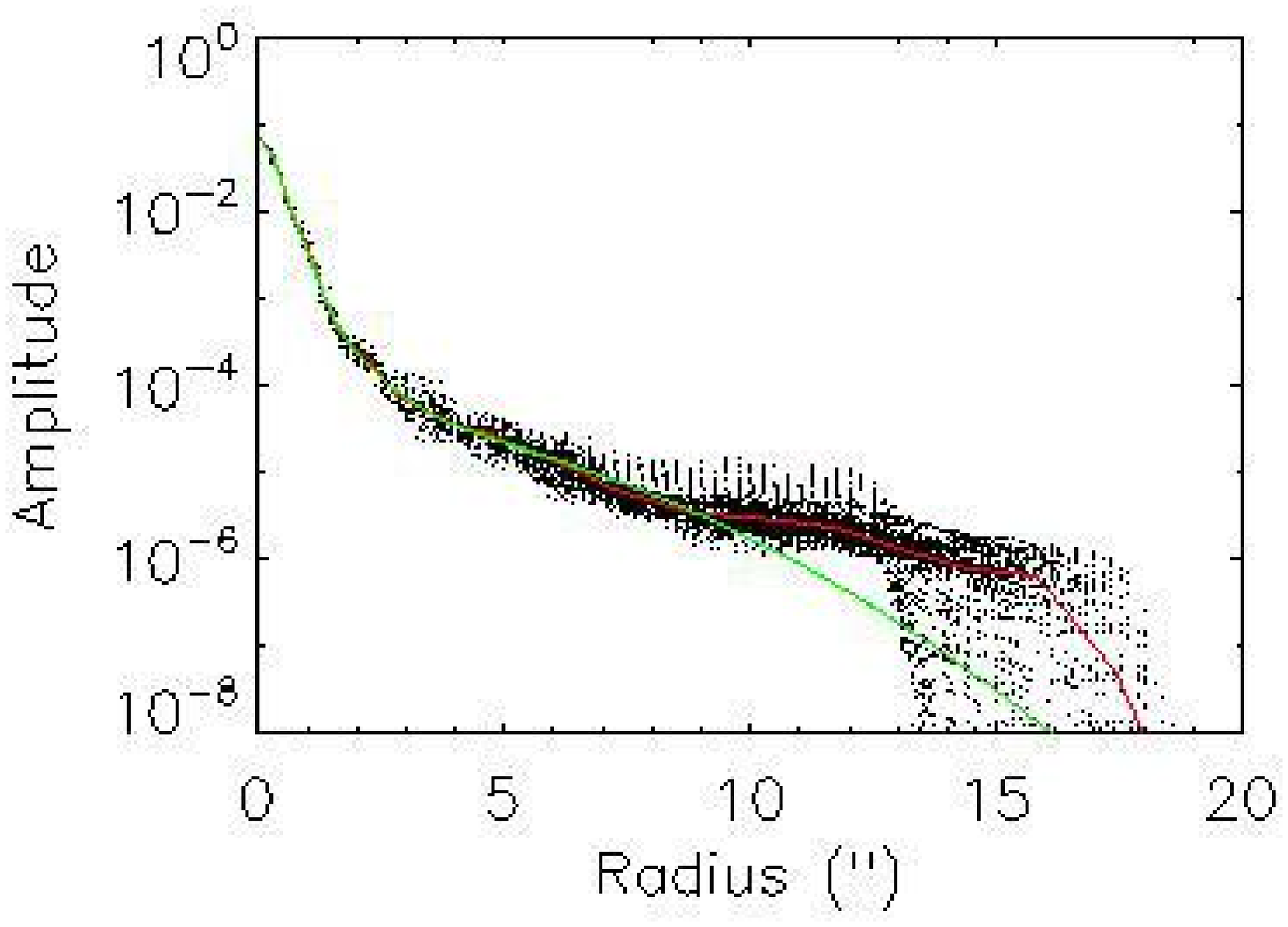}
   \caption{An azimuthal average of multiple HST/ACS PSFs (black points); radially binned points (red) and MGE approximation (green). This MGE approximation is used with Eq. \ref{eq:anal:surf:sersicgaussconvol} to account for PSF effects when fitting COGs.}
   \label{fig:anal:surf:m.psf}
\end{figure}

The instrumental PSF of the image creates a systematic difference between the observed and true COG. \citet{Saglia93}, \citet{Trujillo01a} and \citet{Trujillo01b} describe methods to correct for this. However, the HST/ACS PSF is not well approximated by a single Gaussian or Moffat function, for which these methods were developed. We approximate the ACS PSF with a multi-Gaussian expansion \citep[MGE,][]{Bendinelli91}. We azimuthally average (4-fold reflect) multiple PSFs from across the FOV (see \S\ref{sec:red:hstim}), bin radially, and then use a 1D MGE fitting routine to calculate a radial approximation \citep[made available by M. Cappellari][]{Cappellari02}. Fig. \ref{fig:anal:surf:m.psf} shows an example. 

We make use of the well known result that the convolution of the S\'ersic function $I(R, R_{e}, n)$ with a Gaussian $G(r,\sigma)$ is $I^{\prime}(R, R_{e}, n, \sigma)=$
\begin{equation}
\label{eq:anal:surf:sersicgaussconvol}
\frac{1}{\sigma^{2}} \int_{x=0}^{x=\infty} I(x, R_{e}, n) \exp\left[\frac{1}{2}\left(\frac{x^{2}+R^{2}}{\sigma^{2}}\right)\right] B\left(\frac{xR}{\sigma}\right)
\end{equation}
where $B$ is a modified Bessel function of the first kind, zeroth order \citep[][]{Moffat69,Bendinelli82}. Summing this according to the amplitudes of the MGE then provides a good approximation to the full PSF convolution. We do not account for ellipticity in the fitting process. A Levenberg-Marquardt algorithm  ({\sc MPFIT\footnote{http://www.physics.wisc.edu/$\sim$craigm/}}) minimised $\chi^{2}$ to fit $R_{e}$, $I_{e}$ and $n$ (we fix $n=4$ for a de Vaucouleurs profile). 

We find good agreement between this 1D approach and S\'ersic profiles convolved with the full 2D HST/ACS PSFs: systematic errors in recovered parameters ($R_{e}, I_{e}, \apmag, R_{e}I_{e}^{0.8}$) are always $<$1\% over a wide range of parameter space.

\subsection{Contamination and Masking}
\label{sec:cogcontam}
\begin{figure}
  \centering
  \begin{minipage}[c]{0.5\textwidth} 
    \includegraphics[width=\textwidth]{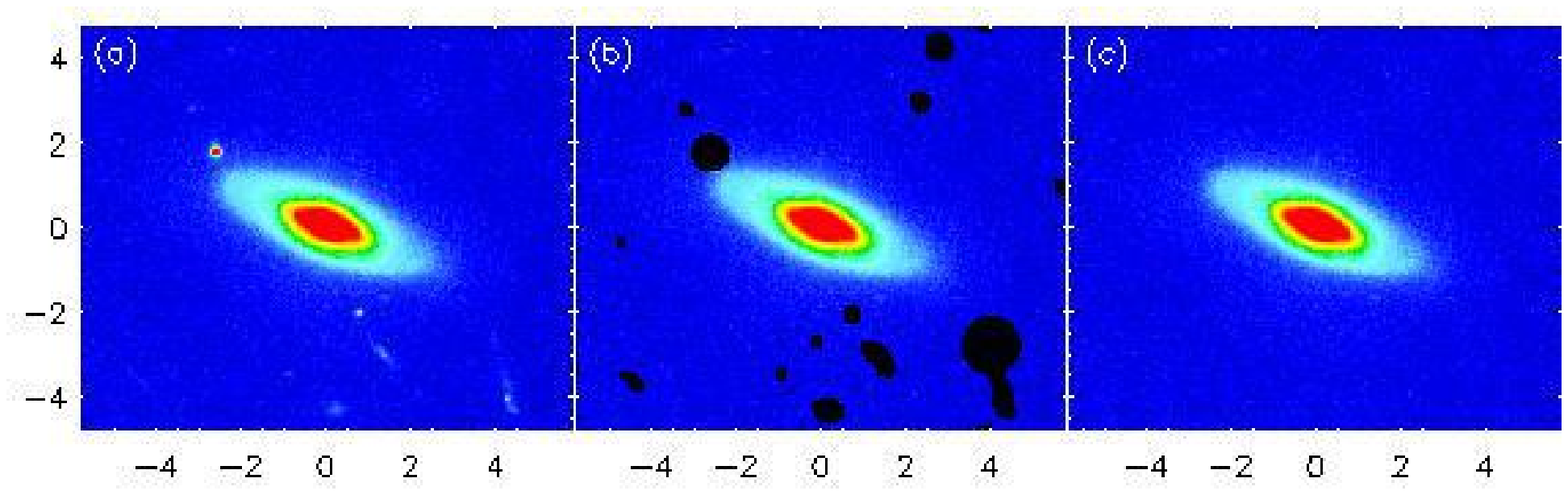}
  \end{minipage}
  \begin{minipage}[c]{0.5\textwidth} 
    \includegraphics[width=\textwidth]{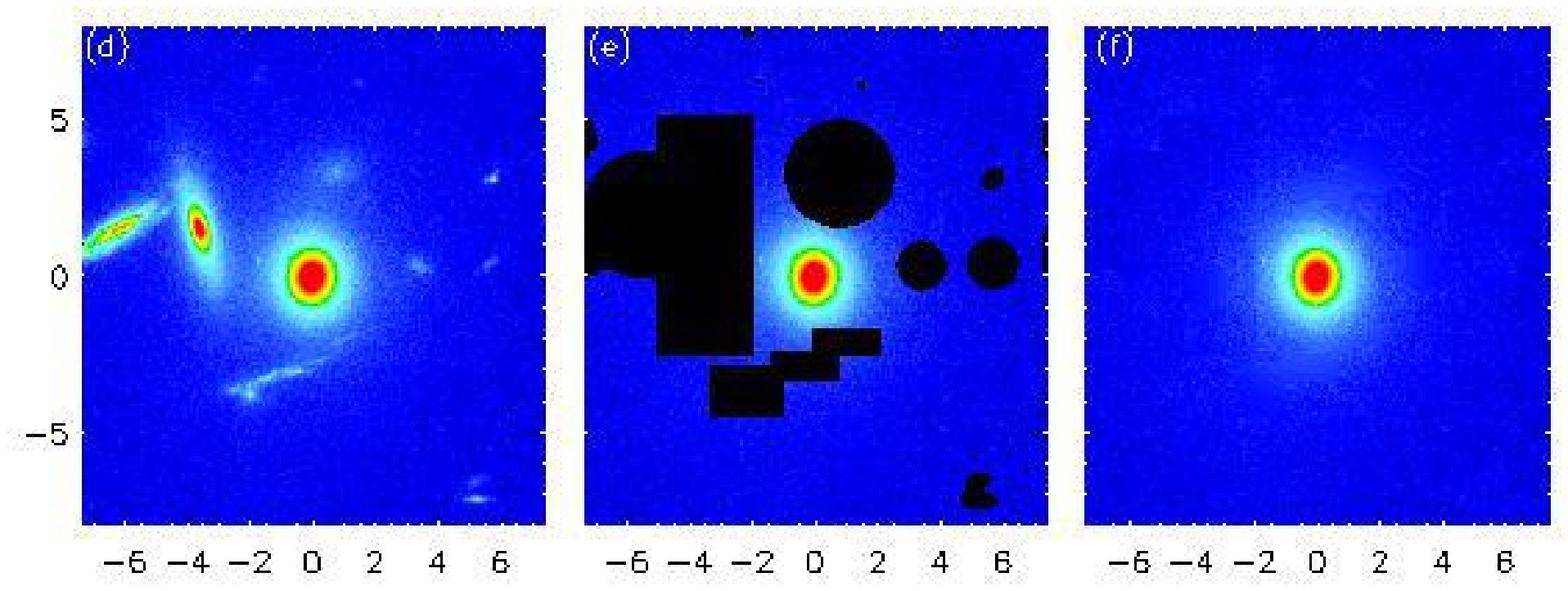}
  \end{minipage}
  \begin{minipage}[c]{0.5\textwidth} 
    \includegraphics[width=\textwidth]{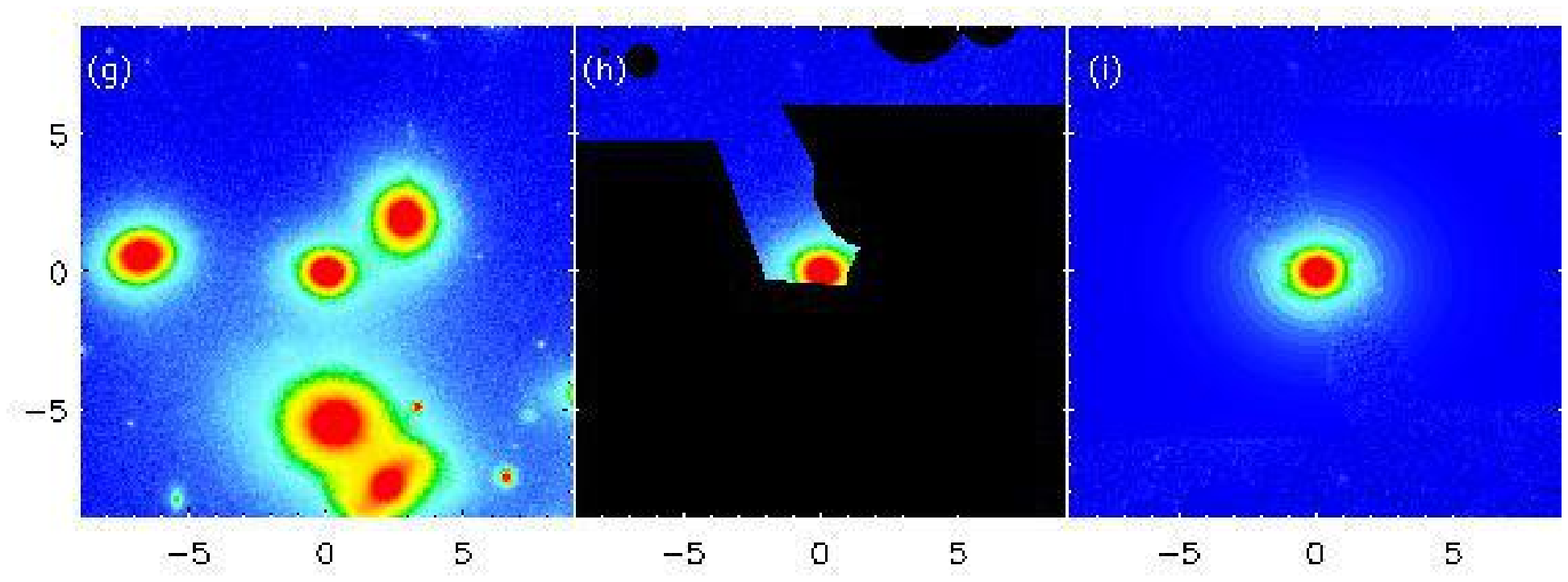}
  \end{minipage}
  \caption{Examples of the masking procedure before fitting a curve of growth model. \textbf{(a)} HST/ACS F625W image of galaxy \#286 which is relatively uncontaminated; \textbf{(b)} shows the mask applied to this image (black pixels) and \textbf{(c)} shows the result of the image reconstruction from replacing masked pixels with a 2-fold reflection. \textbf{(d)} HST/ACS F625W image of galaxy \#435 which has considerable contamination; \textbf{(e)} shows the mask for this image and \textbf{(f)} shows the reconstructed image, produced by replacing pixels with a 2-fold reflection or azimuthally averaged values. \textbf{(g)} galaxy \#635 is so badly contaminated that a reliable measure of the COG is almost impossible; \textbf{(h)} shows the mask (leaving enough information to provide an azimuthal average) and \textbf{(i)} shows our best reconstruction. Although the reconstruction in (i) and the associated COG appear convincing, such results always score a 3rd class rating because of the large uncertainties. Scales on the x- and y-axis are in arcseconds, centred on the galaxy. The greyscale is logarithmic and fixed throughout, to highlight faint structure.}
  \label{fig:anal:surf:cogs}
\end{figure}

The galaxies in Abell 1689 are densely packed and many overlap in projection. To remove this contamination, we initially bin azimuthally averaged profiles of the chosen galaxy and sigma-clip from the mode of each bin. If necessary, we further manually mask the images, ensuring identical masks for both the de Vaucouleurs and the S\'ersic COG fits to prevent systematic differences. To replace masked pixels, a simple 2-fold reflection about the galaxy centre was usually sufficient. When masking was more severe, after a 2-fold reflection, we azimuthally binned the image, averaged the modes of bins at each radius, and replaced masked pixels with this average. This reconstructed strongly contaminated images and gave good COG fits. However, as a method of quality control when the galaxy image was minimally contaminated and the model COG fitted the measured COG to within 2\%, we rated the fit 1st class; when the contamination was again relatively minor but the model COG didn't fit the measured COG to within 2\%, the fit was rated 2nd class; finally, when contamination was severe, such that we had little confidence in the measured COG and thus the model fit, we rated it 3rd class (shown in Table \ref{tab:res:surphot}). Examples of masking and image reconstruction are shown in Fig. \ref{fig:anal:surf:cogs}. 

\subsection{Estimating errors (random and systematic)}
\label{sec:cogerrs}
We use Monte-Carlo simulations to estimate uncertainties in the COG parameters. Generating S\'ersic profiles of various $R_{e}$, $I_{e}$ and S\'ersic index $n$ convolved with a typical 2D HST/ACS PSF (see \S\ref{sec:red:hstim}), we add random sky offsets and read and shot noise to simulate real observations. We use 100 realisations for each parameter combination ($R_{e}$,  \apmag and $n$) over the parameter space of the obervations. We also vary the radius at which we stop the COG fitting according to the statistics of the real data. This approach allows us to estimate random and systematic uncertainties (the latter always being $<1\%$). 

The simulations show that $\log R_{e}$ and $\log I_{e}$ are very strongly correlated. Decomposing the covariance matrix into principle components, the first component always accounts for $>$98\% of the variation: $R_{e} I_{e}^{0.6 \pm 0.1}$ is best constrained while $R_{e}^{-0.6\pm0.1}I_{e}$ is most poorly constrained. Knowing the covariance matrix as a function of input $R_{e}$ and \apmag allows us to use it when fitting the scaling relations; treating $R_{e}$ and $I_{e}$ as independent significantly overestimates the error in the direction perpendicular to the KR which is important when minimising perpendicular residuals (see \S\ref{sec:ap:fitting}).

\section{Matching spectral resolutions}
\label{sec:ap:specres}
The GMOS arc and sky lines appeared top-hat like because the slit width, not the grating, dictated the spectral resolution. Matching spectral resolutions in this case is difficult: firstly, the size or shape of the objects being observed in the GMOS slits can define the resolution and spectral PSF if they are smaller than the slit width; secondly, the kernel required to match the spectral PSF of the stellar library with the GMOS spectra will not be Gaussian. Fortunately, the seeing FWHM (estimated by Gaussian fits to the profiles of guiding/alignment stars) is greater than the 0\farcs75 slit width (varying between 0\farcs74 and 1\farcs49 with an average of 1\farcs06). Thus the illumination of the slits was roughly uniform and similar to that of the arc or sky lines, which enables us to use arc or sky lines as a reference for the spectral PSF of the galaxy spectra. 
An additional complication in matching spectral resolutions is that the width of the GMOS arc lines decreases over the wavelength range used to extract kinematics; if one matches local stellar templates to the observer-frame arc spectra then when later redshifting the templates to the galaxy rest-frame, the non-uniform spectral resolution would introduce a systematic difference between the stellar and galaxy resolutions. 

We developed a technique to account for all these effects by finding a \emph{transfer function} which, when convolved with the spectral profile of the stellar library, reproduces the spectral profile of the galaxy spectra (which is equivalent to reproducing line profiles of a \emph{blueshifted} arc spectrum). We found the GMOS arc lines were well described by a truncated Gauss--Hermite (GH) expansion \citep{Gerhard93,vdM&Franx93}:
\begin{equation}
   \label{eq:GHseries}
   T(\lambda) = {\gamma\over\sqrt{2\pi}\sigma} \exp\left[-\left({{(\lambda-\lambda_{0})\over{2\sigma}}}\right)^2\right]
  \sum_{i=0}^N h_iH_i\left(\omega\right),
\end{equation}
which is a Gaussian centred on $\lambda_{0}$ with scale factor~$\gamma$, dispersion~$\sigma$ weighted by a sum of Hermite polynomials~$H_i$. Taking $h_{0} = 1$, we quantify deviations from a Gaussian using the truncated series $\{h_{2},h_{4},h_{6}\}$ (odd \GH\ moments are not required to describe a symmetric shape). Furthermore, given that positive values of $h_{i}$ (where $i$ is even) provide more top-hat like profiles, whereas negative values give profiles with large wings and sharp peaks, only positive $h_{i}$ values are required.

\begin{figure}
   \centering
   \includegraphics[width=0.5\textwidth]{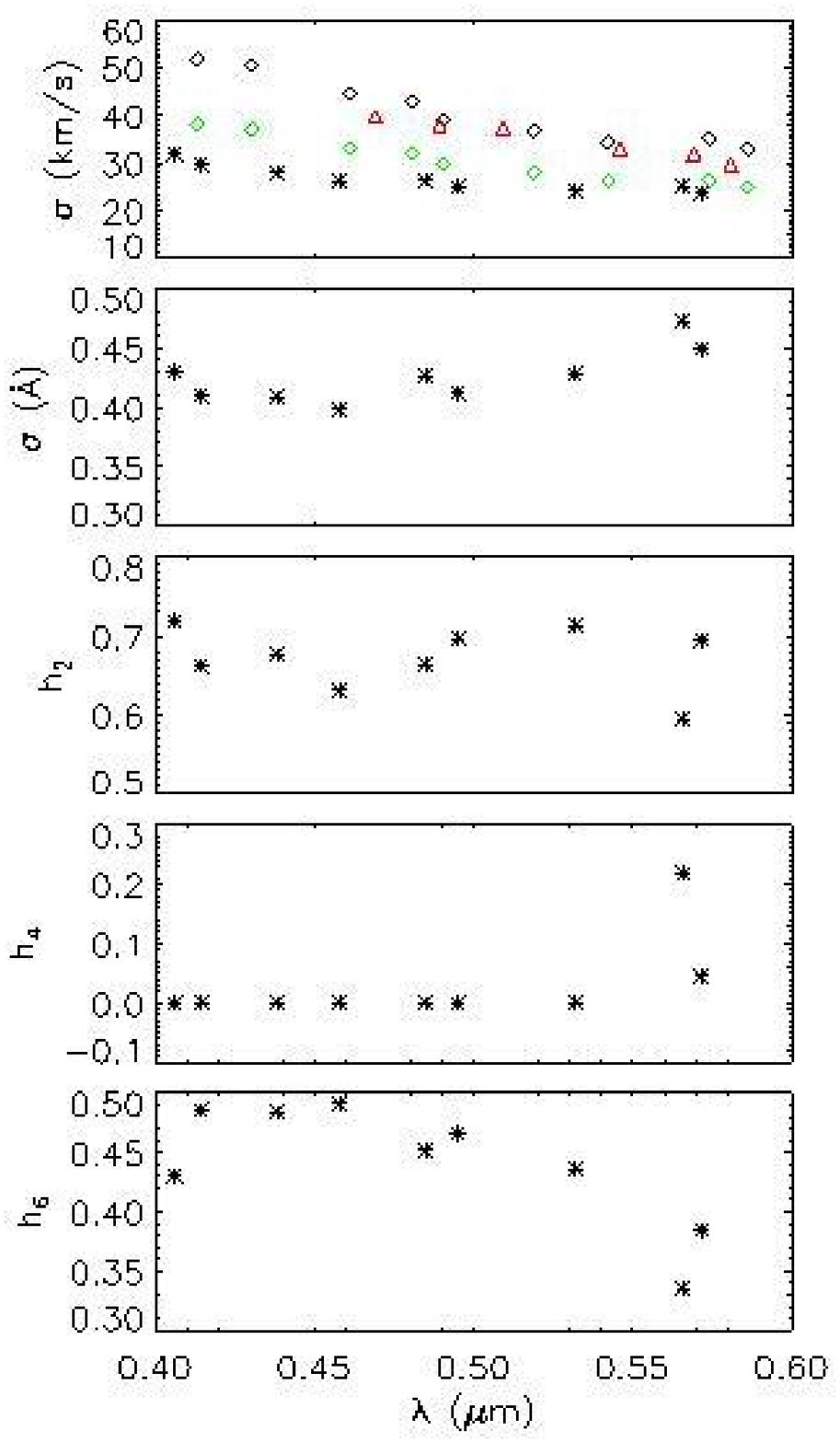} 
   \caption{The best-fit transfer function, matching the spectral response of the stellar library to GMOS arc emission lines at various wavelengths. Parameters for the best-fit transfer functions (Eq. \ref{eq:GHseries}) to blueshifted arc spectra are approximately uniform for all wavelengths (asterisks) and are well approximated by a single function with parameters $\{\sigma, h_{2}, h_{4}, h_{6}\} = \{0.428\textrm{\AA}, 0.676, 0.0, 0.435\}$. The widths of the GMOS arc lines ($\sigma$, black diamonds) are not constant (in $\lambda$ or $\log-\lambda$ space), which can introduce a systematic error when stellar templates are redshifted to the recession velocity of the cluster. The correct transfer function parameters, accounting for this effect are shown as green diamonds (previously the black asterisks fitted to blueshifted arc spectra, but now redshifted back). The widths of the transfer functions if we had not corrected for the redshift effect are shown as red diamonds: these appear roughly equal to the raw arc widths (black diamonds) but this is bad because the transfer function still needs to be convolved with the response of the stellar library; after this, the stellar templates would have a larger spectral resolution than the galaxy spectra, introducing a systematic error into the kinematics (dispersions would be too low). 
}
   \label{fig:kin:CONarcfit}
\end{figure}

\begin{figure}
   \centering
   \includegraphics[width=0.5\textwidth]{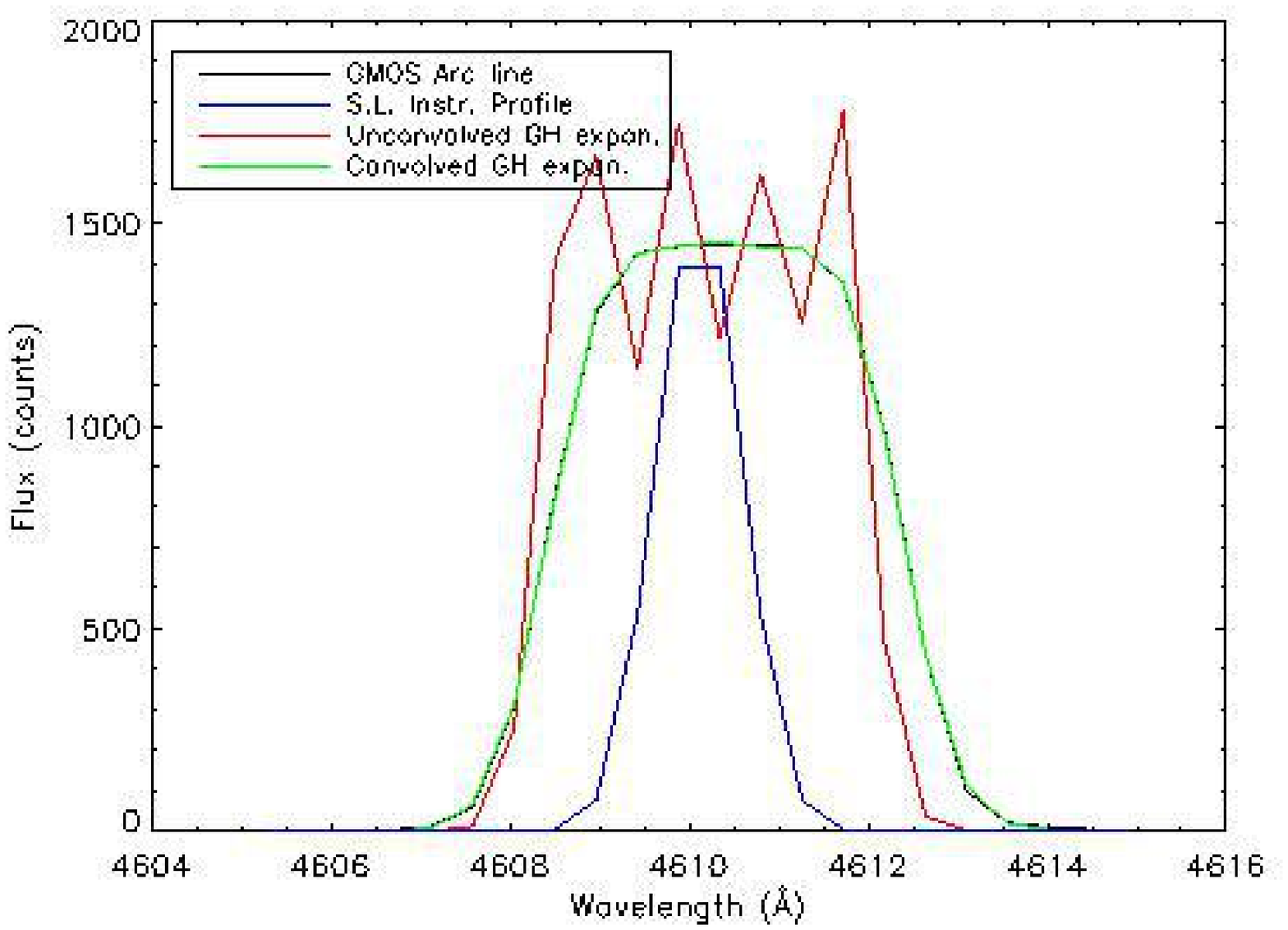} 
   \caption{A fiducial GMOS arc line (black) with the raw stellar library instrument profile (blue), best-fit \GH transfer function (red) and matched stellar library profile (green) over plotted. Note that the matched stellar library profile is virtually identical to the GMOS arc line profile.}
   \label{fig:kin:CONarcprof}
\end{figure}

We then saught a \GH expansion which, when convolved with another Gaussian of FWHM equal to that of the stellar library resolution (1\AA), matched the \emph{blueshifted} arc lines. We accomplished this by minimising the $\chi^{2}$ difference given by the expression
\begin{equation}
\chi^{2} = \sum_{i=1}^{N} \left[ A^{\prime}_{i} -  \left(S_{i}(\sigma\p) \otimes T_{i}(\sigma,h_{j}) \right)\right]^{2}
\end{equation}
where $A^{\prime}$ is the (blueshifted) arc line being fit, $S$ is the spectral profile of the stellar library (width $\sigma\p$), $T$ is the \GH transfer function (Eq. \ref{eq:GHseries}) and the sum is performed over pixels. Fig. \ref{fig:kin:CONarcprof} illustrates the parametric results of this process at various wavelengths (and what would happen if we were not to account for the redshift effect). It is evident that a single transfer function with parameters $\{\sigma, h_{2}, h_{4}, h_{6}\} = \{0.428\textrm{\AA}, 0.676, 0.0, 0.435\}$ is sufficient to degrade the spectral resolution of the stellar library to the blueshifted GMOS arc spectrum. We did occasionally experience a problem of multiple minima when fitting the convolved \GH transfer functions but careful choice of starting parameters solved the problem. Note that because the width (\AA) and properties of the transfer function are uniform in $\lambda$ space, we convolved the stellar library with $T$ \emph{before} resampling it to $\log-\lambda$ space. Fig. \ref{fig:kin:CONarcprof} compares the profile of the matched stellar library to a typical profile for a (blueshifted) GMOS arc line: they are barely distinguishable from each other, confirming we are able to match the spectral profile of the stellar library to that of the galaxy spectra.

\section{Fitting Scaling relations using an MCMC approach}
\label{sec:ap:fitting}
In \S\ref{sec:fitting2D}, we briefly discussed the different models we use to fit the scaling relations in Coma and \A1689. Here we discuss those models in more detail. We have taken on board the recent remarks made by \citet{HoggFitting} and we turn to Markov Chain Monte Carlo (MCMC) methods to `fit' the data.  We use three different models; each one is discussed below.

\subsection{The (Simple) Linear model}
\label{sec:fit:linmodel}

This model is described by \citet{HoggFitting} in their Eq. 35. We assume that our data $\{x_{i}, y_{i}\}$ with (normally distributed) uncertainties $\{\sigma_{xi}, \sigma_{yi}\}$ and covariance $\sigma^{2}_{xyi}=\rho_{xyi}\sigma_{xi}\sigma_{yi}$ come from a linear relation $y = \alpha x + \beta$, with \emph{intrinsic} scatter $\sigma_\textrm{I}$. Given a data point $x_{i}$ with uncertainty $\sigma_{xi}$ we can define the expected frequency\footnote{see \citet{HoggFitting}, note \#13} of observing the corresponding data point in the range $[y_{i},y_{i}+\d y]$ with uncertainty $\sigma_{yi}$ given some particular model parameters $\{\alpha, \beta, \sigma_\textrm{i}\}$ as
\begin{equation}
p(y_{i}|x_{i}, \sigma_{xi}, \sigma_{yi}, \alpha, \beta, \sigma_\textrm{I}) = \frac{1}{\sqrt{2 \pi (\sigma^{2}_\textrm{di} + \sigma^{2}_\textrm{I})}}\exp\left[-\frac{\Delta_{i}^{2}}{2(\sigma^{2}_\textrm{di}+\sigma^{2}_\textrm{I})}\right]
\end{equation}
where 
\begin{eqnarray}
\Delta_{i} & = &\m{v}^\textrm{T}.\m{Z}_{i} - \beta^{\prime}
\end{eqnarray}
measures the perpendicular distance between the linear model and a data point because
\begin{eqnarray}
\m{Z}_{i}^\textrm{T} & = & [x_{i}, y_{i}] \\
\m{v}^\textrm{T} & = & [\cos \theta, \sin \theta] \\
\beta^{\prime} & = & \beta\cos \theta \\
\alpha & = & \frac{\sin \theta}{\cos \theta}
\end{eqnarray}
while the combined projected variance perpendicular to the linear model is
\begin{eqnarray}
\sigma^{2}_\textrm{di} & = & \m{v}^\textrm{T}.\Sigma_{i}.\m{v}
\end{eqnarray}
and $\m{\Sigma}_{i}$ is just the covariance matrix
\begin{eqnarray}
\m{\Sigma}_{i} & = & \left[
   \begin{matrix}       \sigma_{x,i}^{2} & \sigma_{xy,i}^{2} \\
      \sigma_{yx,i}^{2} & \sigma_{y,i}^{2} \\
   \end{matrix}\right]
\end{eqnarray}

Equally, we could have introduced the rotated coordinate system of the linear model, $Z\p = [x\p, y\p]$ which describe coordinates parallel and perpendicular to the regression line. They are related to the non-rotated coordinates by $[x\p,y\p]=\M{R}.[\m{x},\m{y}] - \beta\p$ where $\M{R}$ is the usual rotation matrix which depends only on $\theta$. There is an obvious equality between $\Delta_{i}$ and $y\p_{i}$. There is no need to persist with this rotated coordinate system here, but it will be useful in the next section.

The \emph{likelihood} of observing all our data points given a particular choice of linear model is then the product of the frequency distributions,
\begin{equation}
\mathcal{L}_\n{LM}=\prod_{i=1}^{N} p(y_{i}|x_{i}, \sigma_{x,i}, \sigma_{y,i}, \alpha, \beta, \sigma_\textrm{I}).
\end{equation}
Standard fitting procedures based on minimising $\chi^{2}$ or `maximising the likelihood' then vary the model parameters to maximise this expression and find the mode of the likelihood distribution. However, the Bayesian view on this is to say that the \emph{posterior} probability distribution of the model is
\begin{equation}
p(\alpha,\beta, \sigma_\textrm{I} | \{y_{i}\}_{i=1}^{N}, Q) = \frac{\mathcal{L} . p(\alpha,\beta | Q)}{p(\{y_{i}\}_{i=1}^{N} | Q)}
\end{equation}
where $Q$ is shorthand for all the other (not always prior) knowledge of the problem, such as the $x_{i}$, $\{\sigma_{x,i}, \sigma_{y,i}\}$ etc. Readers familiar with Bayes Theorem will recognise all the terms above, but for the unacquainted, $p(\alpha,\beta | Q)$ contains additional information about the problem (and is referred to as the prior) and $p(\{y_{i}\}_{i=1}^{N} | Q)$ is the normalising constant we evaluate by insisting that our final posterior distribution is normalised to one. Note that maximising the likelihood is equivalent to maximising the posterior given uniform priors on all parameters.

MCMC samplers can probe the above likelihood with different model parameters and provide a set of sample points that are distributed according to the posterior probability distribution (PDF) because they were drawn from it. This is our preferred method here, for two reasons. Firstly, it relaxes the standard assumption that the parameters are normally distributed; while normality may be a good approximation for $m$ and $b$, it is certainly not so for $\sigma_\textrm{I}$ because $\sigma_\textrm{I} \ge 0$, providing a hard (asymmetric) edge to the allowed distribution which we can set as a prior. Secondly, using MCMC here in a relatively simple model prepares us for more complicated models. 

In terms of implementation, we use the {\sc PYMC} library \citep{pymc} to sample parameters $\theta, \beta^{\prime}, \sigma_\textrm{I}$ using the standard Metropolis-Hastings MCMC algorithm. We choose relatively unrestrictive priors that are flat but always over a finite range: we let $\theta$ range between $\theta_{0}\pm\pi/2$ (where $\theta_{0}$ is a good initial guess; this avoids searching for solutions near the boundaries);  the intercept ($\beta^\prime$) is limited to a finite range because we can't sample to $\pm\infty$; the intrinsic scatter ($\sigma_\textrm{I}$) is not allowed to be negative and also has some sensible upper limit. 

Whenever we quote a \emph{best-fit} or \emph{most probable} value for one of the parameters, we quote the median of the resulting samples (which by default is marginalised over all other parameters). The MCMC usually requires reasonably good initial guesses to converge rapidly and such a guess can be provided by the `maximum likelihood' estimates or by eye.

As highlighted by \citet{HoggFitting}, outliers can cause havoc in this simple linear model and \S\ref{sec:fit:mixmodel} discusses our implementation of the solution proposed by those authors. 

\subsection{The (Complex) Double Linear Model}
\label{sec:fit:dblmodel}

It is often the case when studying scaling relations that one wishes to compare the intercept or offset of one relation to another. A common solution is to calculate the offset at a particular value along the x-axis, but such a result is dependent on the different slopes assumed for the two relations (as well as their intercepts). Alternatively, if there is good reason to believe the slope of the relations hasn't changed, one might fit the slope of the local relation (which usually has more data) and then impose this fixed slope on the high redshift data \citep[e.g.][]{Barr06}; then at least there is no error created by imposing two \emph{different} slopes, although there is \emph{still} a strong dependence on the slope itself, the intercepts and any internal scatter being derived. 

To circumvent these problems, we propose a model consisting data drawn from two populations, $Z_{1,i}^{T}=[x_{1,i}, y_{1,i}]$ and $Z_{2,i}^{T}=[x_{2,i}, y_{2,i}]$, each of which is described by a linear relation (as in \S\ref{sec:fit:linmodel}) but we constrain the linear relations to have the same slope $\alpha$ and internal scatter $\sigma_\n{I}$; we then let the first population have intercept $\beta_{1}$ while the second population with intercept $\beta_{2}$ is offset to this by $\Delta \beta$, such that $\beta_{2}=\beta_{1}+\Delta \beta$. In this way, the two linear models share the same properties, except for the intercept and we have parameterised the variable we are most interested in as $\Delta \beta$. The likelihood for this model thus far would be
\begin{equation}
\label{eq:dlmlike_nocut}
\mathcal{L}_\n{DLM}= \sum_{i} p_\n{1,i}  \times \sum_{j} p_{2,j}
\end{equation}
where 
\begin{eqnarray}
p_\n{1,i} & = & \frac{1}{\sqrt{2 \pi \sigma_{1,i}^{2}}}\exp\left[-\frac{\Delta_{1,i}^{2}}{2\sigma^{2}_\textrm{1,i}}\right]  \\
p_\n{2,j} & = & \frac{1}{\sqrt{2 \pi \sigma_{2,j}^{2}}}\exp\left[-\frac{\Delta_{2,j}^{2}}{2\sigma^{2}_\textrm{2,j}}\right]
\end{eqnarray}
give the frequency distributions of the individual data points for the two populations: $\Delta_{1,i}$=$\m{v}^\textrm{T}.\m{Z_{1,i}}-\beta\p_{1}$ defines the residuals for the first population and $\Delta_{2,j}$ similarly defines the residuals for the second population; the expressions
\begin{eqnarray}
\sigma_\n{1, i}^{2} & = & \sigma^{2}_\textrm{1,di} + \sigma^{2}_\textrm{I} \\
\sigma_\n{2, j}^{2} & = & \sigma^{2}_\textrm{2,dj} + \sigma^{2}_\textrm{I}
\end{eqnarray}
give the total variances perpendicular to the relation. Hopefully the definitions of $\sigma_\n{I}$, $\sigma_\n{1,di}$ and $\sigma_\n{2,dj}$ are obvious to the reader from \S\ref{sec:fit:linmodel}. 

One would also like to account for magnitude cuts and selection effects which are known to cause bias in the derived parameters \citep{Willick94, Teerikorpi97, Saglia01, LaBarbera03, Butkevich05}. In the KR and FJR, magnitude cuts have the effect of masking a region of parameter space from being studied. In both cases, the division between the unsampled and sampled regions can be defined by a linear relation of the form $y=\alpha_\n{cut} x + \beta_\n{cut}$. For the KR, 
\begin{eqnarray}
\alpha_\n{cut}&=&5\\
\beta_\n{cut} &=& \apmag_\n{cut} + 2.5\log(2\pi) \\ \nonumber
& &- 7.5\log(1+z) - 5\log\left(\frac{\textrm{kpc}}{\textrm{arcsec}}\right)
\end{eqnarray}
while for the FJR,
\begin{eqnarray} 
\alpha_\n{cut}&=&0\\
\beta_\n{cut}&=&\apmag+2.5\log(1+z)-5\log D_{L}+5
\end{eqnarray}

The ratio of the spectroscopic samples to the parent samples as a function of $\apmag$ for both \A1689 and Coma can be fit by a \emph{selection function} of the form \citep{Wegner96}
\begin{equation}
\label{eq:selectionfunction}
S_{i} = \frac{1}{2}\left\{1-\erf\left(\frac{\apmag_{i}-\apmag_\textrm{\tiny cut}}{\sqrt{2}\delta_{\apmag}}\right)\right\}
\end{equation}
where $\erf$ is the error function, $\apmag_\textrm{\tiny cut}$ is the position of the cut (also the 50\% completeness limit) and $\delta_{\apmag}$ is the width/uncertainty of the cut (recall that Eq. \ref{eq:selectionfunction} describes a Heaviside step function at $\apmag_\textrm{\tiny cut}$ convolved with a Gaussian with $\sigma=\delta_{\apmag}$).  We fit the selection function (Eq. \ref{eq:selectionfunction}) to the spectroscopic samples of Coma and \A1689 following the approach described in \citet{Wegner96}. For \A1689, using \sex magnitudes, we find that the spectroscopic sample is consistent with selecting 60\% per magnitude bin of the parent sample down to $\apmag_\textrm{\tiny cut}=19.5$ \Mag with uncertainty $\delta_{\apmag}=0.62$ \Mag; for Coma, we find the spectroscopic data of \citet{Jorgensen96} consistent with sampling 100\% of the population down to $\apmag_\textrm{\tiny cut}=14.8$ (AB) \Mag with uncertainty $\delta_{\apmag}=0.5$ \Mag. 

We model the magnitude cuts of the two samples in a similar way to \citet{Saglia01} and \citet{LaBarbera03} by scaling the frequency distributions (i.e. the $p_{i}$) by a factor $1/f_{i}$ to account for the parameter space masked by the cut. This requires us to make an assumption about the distribution of the data \emph{along} the regression line which we assume is uniform, starting at $x\p_\n{min}=\textrm{min}(x\p_{i})$ and ending at $x\p_\n{max}=\textrm{max}(x\p_{i})$ thus having width $L=\textrm{max}(x\p_{i})-\textrm{min}(x\p_{i})$ (which is not a free parameter, but is set by the data and values of $\alpha$ and $\beta$). Therefore the probability space masked by the magnitude limit as a fraction of the total probability space is   
\begin{equation}
\label{eq:unsolved_fi_integral}
f_{i} =  \frac{1}{L\sigma_{i}\sqrt{2\pi}}\int_{x\p=x\p_\n{min}}^{x\p_\n{max}}\d x\p  \int_{y\p=y_\n{cut}}^{y\p=\infty}\d y\p  \exp\left(-\frac{{y\p}^{2}}{2\sigma_{i}^{2}}\right) 
\end{equation}
where $\sigma_{i}^{2}$ is again the total variance perpendicular to the relation and we have evaluated the integral in the rotated coordinates of the linear model, $\{x\p,y\p\}$. Eq. \ref{eq:unsolved_fi_integral} can be solved using the integral of the error function and substituting the relation $y\p = \alpha\p_\n{cut} x\p + \beta\p_\n{cut}$ where $\{\alpha\p_\n{cut},\beta\p_\n{cut}\}$ describe the magnitude cut in the rotated coordinate system. Knowing $\theta_\n{cut}=\textrm{arctan} (\alpha_\n{cut})$, it follows that 
\begin{eqnarray}
\alpha\p_\n{cut} & = & \tan(\theta_\n{cut}-\theta) \\
\beta\p_\n{cut} & = & \beta_\n{cut}[\cos \theta_\n{cut} - \tan(\theta_\n{cut}-\theta)\sin \theta_\n{cut}] - \beta\cos\theta
\end{eqnarray}
Thus Eq. \ref{eq:unsolved_fi_integral} becomes
\begin{eqnarray}
f_{i} = \left[ \frac{x\p}{2} + \left(\frac{\beta\p_\n{cut}}{\alpha\p_\n{cut}} + x\p\right) \erf\left(\frac{\alpha\p_\n{cut} x\p + \beta\p_\n{cut}}{\sqrt{2}\sigma_{i}}\right) \right.\\
+ \left.\sqrt{\frac{2}{\pi}} \frac{\sigma_{i}}{b} \exp\left(-\frac{1}{2}\left(\frac{\alpha\p_\n{cut} x\p + \beta\p_\n{cut}}{\sigma_{i}}\right)^{2}\right) \right]^{x\p_\n{max}}_{x\p_\n{min}}
\end{eqnarray}
Unlike \citeauthor{Saglia01} and \citeauthor{LaBarbera03}, we do \emph{not} assign zero probability to objects with $\apmag>\apmag_\n{cut}$, but we scale $p_{i}$ by the selection probability $S_{i}$. This accounts for the gradual nature of our magnitude cut.  

For the Coma sample, there is no other obvious selection effect and $S_{i}$ accurately describes the ratio of the spectroscopic sample to its parent distribution as a function of magnitude. However, the \A1689 data does suffer further selection effects which are not related to the magnitude cut above and should be corrected for. Firstly, only 60\% of galaxies were selected per magnitude bin on average. Furthermore, we sample more than 60\% of the most luminous galaxies due to low number statistics. Finally, we have the constraint that the data which enters the KR and FJR must have HST surface photometry, which further limits galaxies to be in the core of the cluster. Because of mass segregation, this biases the sample to more massive (luminous) galaxies. We calculate the ratio of the ideal sampling (given by $S_{i}$) to the actual sampling as a function of $\apmag$ which can be used as a weight $w_{i}$ to correct for sampling effects, as in \citet{Saglia01}. Thus, accounting for the \emph{cut} in magnitude with $S_{i}$ and $f_{i}$, and additional selection effects as a function of magnitude with weights $w_{i}$, we can describe the frequency distribution of a single data point, given the model parameters to be
\begin{equation}
p\p_{1, i} = \left[\frac{S_{1, i}}{f_{1, i}\sqrt{2\pi}\sigma_{1, i}} \exp\left(-\frac{\Delta_{1, i}^{2}}{2\sigma_{1, i}^{2}}\right)\right]^{w_{i}}
\end{equation}
The final likelihood for the model is thus
\begin{equation}
\mathcal{L}\p_\n{DLM} = \sum_{i}p\p_{1, i} + \sum_{j}p\p_{2, j}
\end{equation}
where the two sums sum over the data of the two different populations. 

Using the same MCMC set-up as before for this \emph{double linear} model allows us to measure the \emph{marginalised} offset between two scaling relations while fully accounting for magnitude cuts and selection effects.

\subsection{The Mixture Model}
\label{sec:fit:mixmodel}

We can also describe a scaling relation using a double-Gaussian mixture model (Exercise 14 in \citeauthor{HoggFitting}) in which we again have data $\{x_{i}, y_{i}\}$ with uncertainties $\{\sigma_{xi}, \sigma_{yi}\}$ that come \emph{either} from a linear relation ($y = \alpha x + \beta$, model $A$, with \emph{intrinsic} scatter $\sigma_\textrm{I}$) \emph{or} another outlier distribution (model $B$ with expectation value $Y_{b}$ and intrinsic scatter $\sigma_{b}$) that overlaps that of the linear relation; thus bad data is modelled as coming from model $B$, whereas \emph{good} data is modelled as coming from model $A$. The goal is to determine which data is good, and which bad. Say we have $N$ data points and we don't know \emph{a priori} which distribution(s) the data points are drawn from, but we assume it's either $A$ or $B$. Model $A$ is described by a normal distribution just like the linear model (\S\ref{sec:fit:linmodel}), 
\begin{equation}
p_{modelA} = \frac{1}{\sqrt{2 \pi (\sigma^{2}_\textrm{di} + \sigma^{2}_\textrm{I})}}\exp\left[-\frac{\Delta_{i}^{2}}{2(\sigma^{2}_\textrm{di}+\sigma^{2}_\textrm{I})}\right]
\end{equation}
whereas model $B$ is a simpler normal distribution
\begin{equation}
p_{\textrm{model}B} = \frac{1}{\sqrt{2 \pi (\sigma^{2}_\textrm{di}+\sigma^{2}_{b})}}\exp\left[-\frac{(y_{i} - Y_{b})^{2}}{2(\sigma^{2}_\textrm{di}+\sigma^{2}_{b})}\right]
\end{equation}
Model $B$ is described by just $Y_{b}$ and $\sigma_{b}$, rather than any linear relation: it's just a scattering of points across all $x$, centred on $Y_{b}$.

We want to be able to reject data points when constructing the likelihood. \citet{HoggFitting} explain how this is achieved by a summation over a \emph{mixture} of the linear distribution (model $A$) and the outlier distribution (model $B$)\footnote{Recall that marginalisation is in essence a weighted sum over possible outcomes, where the weights are the respective probabilities} such that the likelihood for the model is 
\begin{eqnarray}
\mathcal{L} & = & \prod_{i=1}^{N} [(1-p_{b}).p_{\textrm{model}A} + p_{b} . p_{\textrm{model}B}] \\ \nonumber
& = & \prod_{i=1}^{N}\frac{1-p_{b}}{\sqrt{2 \pi (\sigma^{2}_\textrm{di}+\sigma^{2}_\textrm{I})}}\exp\left(-\frac{\Delta_{i}^{2}}{2(\sigma^{2}_\textrm{di}+\sigma^{2}_\textrm{I})}\right) \\ \nonumber
& &+  \frac{p_{b}}{\sqrt{2 \pi (\sigma^{2}_{di}+\sigma^{2}_{b})}}\exp\left(-\frac{(y_{i} - Y_{b})^{2}}{2(\sigma^{2}_{di}+\sigma^{2}_{b})}\right) \nonumber
\end{eqnarray}
where $p_{b}$ is (a single parameter for) the probability of any given data point being bad. We again `fit' the parameters ${\theta, \beta^\prime, \sigma_\textrm{I}, p_{b}, Y_{b}, \sigma_{b}}$, via MCMC sampling. The main advantage being a less subjective approach to removing outliers: we let the data and our knowledge of its uncertainty decide what to reject and what to keep. Again, the MCMC requires reasonably accurate initial guesses for the parameters to converge rapidly onto an acceptable result and such a guess can be provided by the `maximum likelihood' estimates or by just being sensible.

\section{Acknowledgements}

We thank Inger J{\o}rgensen for proposing the observations of \A1689, selecting the spectroscopic sample and for useful comments which improved the paper. We thank Phil Marshall and Neale Gibson for interesting discussions regarding the MCMC technique in conjunction with Bayesian statistics. We also thank the anonymous referee for their helpful comments and suggestions. 

EDB was supported by the grants CPDA089220/08 and 60A02-5934/09 of
Padua University, and ASI-INAF I/009/10/0 of Italian Space Agency, and
by a grant of Accademia dei Lincei and Royal Society.  EDB
acknowledges the Sub-department of Astrophysics, Department of
Physics, University of Oxford and Christ Church College for the
hospitality while this paper was in progress.

This work was based on observations made with the NASA/ESA Hubble Space Telescope (proposal ID 9289), obtained from the data archive at the Space Telescope Science Institute. STScI is operated by the Association of Universities for Research in Astronomy, Inc. under NASA contract NAS 5-26555. This work was also based on observations obtained at the Gemini Observatory (proposal IDs GN-2001B-Q-10 \& GN-2003B-DD-3), which is operated by the 
Association of Universities for Research in Astronomy, Inc., under a cooperative agreement 
with the NSF on behalf of the Gemini partnership: the National Science Foundation (United 
States), the Science and Technology Facilities Council (United Kingdom), the 
National Research Council (Canada), CONICYT (Chile), the Australian Research Council (Australia), 
MinistŽrio da Cincia, Tecnologia e Inova‹o (Brazil) 
and Ministerio de Ciencia, Tecnolog'a e Innovaci—n Productiva (Argentina).

\bibliographystyle{mn2e}
\bibliography{A1689,HSTimaging,jabref,general}

\label{lastpage}

\end{document}